\pdfoutput=1
\documentclass[usenatbib]{mnras}
\usepackage{graphicx}
\usepackage{color}
\usepackage{aas_macros}
\usepackage{hyperref,breakurl}
\usepackage[usenames,dvipsnames]{xcolor}
\usepackage{dcolumn}
\usepackage{bm}
\usepackage{hyperref}
\usepackage{array}
\usepackage{dcolumn}

\usepackage{amsmath,amssymb,latexsym,times}

\newcommand{\galsim}{\textsc{GalSim}}
\newcommand{\wfirst}{{\slshape Roman}}

\newcommand{\changetext}[1]{\textcolor{black}{#1}}
\newcommand{\changetexttwo}[1]{\textcolor{black}{#1}}

\title[A Synthetic Roman High-Latitude Imaging Survey]{A Synthetic Roman Space Telescope High-Latitude Imaging Survey: \\
Simulation Suite and the Impact of Wavefront Errors \\on Weak Gravitational Lensing}

\author[M.~A. Troxel et al.]{M.~A. Troxel,$^1$ \thanks{E-mail: michael.troxel@duke.edu}
H. Long,$^{2,3}$
C.~M. Hirata,$^{2,3,4}$
A. Choi,$^{2,3}$
M. Jarvis,$^5$
R. Mandelbaum,$^6$
\newauthor
K. Wang,$^1$
M. Yamamoto,$^1$
S. Hemmati,$^7$
P. Capak$^{8,9}$\\
$^{1}$Department of Physics, Duke University, Durham, NC 27708, USA\\
$^{2}$Center for Cosmology and Astro-Particle Physics, The Ohio State University, 191 West Woodruff Avenue, Columbus, OH 43210, USA\\
$^{3}$Department of Physics, The Ohio State University, 191 West Woodruff Avenue, Columbus, OH 43210, USA\\
$^{4}$Department of Astronomy, The Ohio State University, 140 West 18th Avenue, Columbus, OH 43210, USA\\
$^{5}$Department of Physics and Astronomy, University of Pennsylvania, Philadelphia, PA 19104, USA\\
$^{6}$McWilliams Center for Cosmology, Department of Physics, Carnegie Mellon University, Pittsburgh, Pennsylvania 15213, USA\\
$^{7}$Jet Propulsion Laboratory, California Institute of Technology, Pasadena, CA 91109, USA\\
$^{8}$IPAC, Mail Code 314-6, California Institute of Technology, 1200 East California Boulevard, Pasadena, CA 91125, USA\\
$^{9}$Cosmic Dawn centre (DAWN), Niels Bohr Institute, University of Copenhagen, Juliane Maries vej 30, DK-2100 Copenhagen, Denmark
}

\pubyear{2019}

\begin{document}
\label{firstpage}
\pagerange{\pageref{firstpage}--\pageref{lastpage}}
\maketitle

\begin{abstract}
The {\slshape Nancy Grace Roman Space Telescope} (\wfirst) mission is expected to launch in the mid-2020s. Its weak lensing program is designed to enable unprecedented systematics control in photometric measurements, including shear recovery, point-spread function (PSF) correction, and photometric calibration. This will enable exquisite weak lensing science and allow us to adjust to and reliably contribute to the cosmological landscape after the initial years of observations from other concurrent Stage IV dark energy experiments. This potential requires equally careful planning and requirements validation as the mission prepares to enter its construction phase. We present a suite of image simulations based on \galsim\ that are used to construct a complex, synthetic \wfirst\ weak lensing survey that incorporates realistic input galaxies and stars, relevant detector non-idealities, and the current reference five-year \wfirst\ survey strategy. We present a first study to empirically validate the existing \wfirst\ weak lensing requirements flowdown using a suite of 12 matched image simulations, each representing a different perturbation to the wavefront or image motion model. These are chosen to induce a range of potential static and low- and high-frequency time-dependent PSF model errors. We analyze the measured shapes of galaxies from each of these simulations and compare them to a reference, fiducial simulation to infer the response of the shape measurement to each of these modes in the wavefront model. We then compare this to existing analytic flowdown requirements, and find general agreement between the empirically derived response and that predicted by the analytic model.
\end{abstract}


\maketitle

\section{Introduction}\label{sec:intro}

The nature of dark energy, which drives the accelerated expansion of the Universe, remains one of the most fundamental mysteries in physics twenty years after its discovery \citep{riess98,perlmutter99,detf,Frieman:2008sn,weinberg13}. 
A number of new experiments have been undertaken to probe dark energy using a variety of physical phenomena, including baryon acoustic oscillations, numbers and masses of galaxy clusters, galaxy clustering, redshift-space distortions, Type Ia supernovae, and weak gravitational lensing. 
Current-generation experiments are limited to some subset of these probes, but have already begun to expose interesting questions about the soundness of our standard cosmological model, Lambda-Cold Dark Matter (LCDM), that will require more and better data in all of these probes to resolve. 
The {\slshape Nancy Grace Roman Space Telescope} (\wfirst)\footnote{\wfirst\ was formerly named the Wide-Field Infrared Survey Telescope (WFIRST).}$^,$\footnote{\url{http://roman.gsfc.nasa.gov}} has been designed to take advantage of all of these probes to study dark energy and test general relativity with unprecedented systematic control \citep{2015arXiv150303757S,2019arXiv190205569A,2019BAAS...51c.341D}.

Weak gravitational lensing is a particularly powerful cosmological probe that is sensitive to both the expansion of the Universe and the growth of large-scale structure \citep{2001PhR...340..291B,2017arXiv171003235M}. 
In the past few years, the current generation of ground-based weak lensing experiments like the Dark Energy Survey (DES),\footnote{\url{http://www.darkenergysurvey.org/}} Hyper-Suprime Cam (HSC) survey,\footnote{\url{http://hsc.mtk.nao.ac.jp/ssp/}} and Kilo-Degree Survey (KiDS)\footnote{\url{http://kids.strw.leidenuniv.nl}} have reached levels of precision that rival the previously best possible cosmological constraints when including a free dark energy equation of state \citep{2018arXiv181206076H,2018PhRvD..98d3528T,2019PhRvL.122q1301A,2019PASJ...71...43H}. 
These surveys have spurred the development of new algorithms and methods for galaxy shape measurement and weak lensing analysis (e.g., \citealt{HuffMandelbaum2017,SheldonHuff2017,shearcat}), enhancing the potential power of weak lensing to unravel the fundamental mysteries we face in cosmology today. 

By the planned launch of \wfirst\ in the mid-2020s, we will have final results from the ongoing generation of weak lensing experiments (DES, HSC, and KiDS) and preliminary results from the Dark Energy Spectroscopic Instrument (DESI),\footnote{\url{https://www.desi.lbl.gov/}} the Large Synoptic Survey Telescope (LSST),\footnote{\url{http://www.lsst.org}} and the Euclid mission.\footnote{\url{http://sci.esa.int/euclid}} 
Faced with the unknown discovery potential of these experiments in the early 2020s, it is vital to maintain the agility of the \wfirst\ mission to respond with the best possible science, particularly in what is likely to be a systematics-dominated weak lensing field.
The process of quantifying empirically the robustness of the design requirements of the \wfirst\ mission for weak lensing in the current phase of mission development is a critical task that this paper will partly address. 
Precise control of these systematics at the statistical precision offered by current \wfirst\ mission forecasts \changetext{\citep{2020arXiv200404702E, 2020arXiv200405271E}} will enable \wfirst\ to make crucial contributions to the study of new discoveries made in the early years of LSST and Euclid, and to the resolution of any remaining disagreements between surveys.

Toward this goal, we describe in this paper a simulation framework designed to enable the empirical study of requirements flowing down from the \wfirst\ wide-field imaging survey, in particular for weak lensing. 
This simulation pipeline can incorporate a realistic simulated survey strategy, galaxy properties, and instrument effects to create a synthetic \wfirst\ wide-field imaging survey. 
We present in this paper a set of synthetic \wfirst\ imaging surveys covering approximately 6 sq.~deg.~to full five-year survey depth in one filter: a fiducial survey and 12 variations incorporating ways in which the point-spread function (PSF) could be mis-estimated. 
The simulation incorporates realistic distributions of photometric properties for galaxies and stars; complex analytic galaxy models; a simulated observing strategy for a reference five year, 2000 sq.~deg. survey; and realistic detector effects, PSF models, and WCS solutions that match current \wfirst\ design specifications. 
We use a blending-free version of this simulation to test the impact on weak lensing science of these simulated wavefront modeling errors, including static, low-, and high-frequency biases. 

We discuss the \wfirst\ weak lensing survey, the current Reference Survey structure, and the weak lensing requirements process in Sec.~\ref{sec:wfirst}. The  synthetic survey simulation suite is described generally in Sec.~\ref{sec:sim}, where we outline the simulated survey strategy, input galaxy and star catalogs, and the \galsim\ implementation of the \wfirst\ instrument used to simulate images. We discuss the specific simulation runs produced for this work to study wavefront error propagation in Sec.~\ref{sec:results} and discuss the resulting biases and how these compare to the analytic requirements flowdown in Sec.~\ref{sec:results2}. We discuss future plans for using this simulation suite in validating \wfirst\ requirements and algorithm design in Sec.~\ref{sec:future} and conclude in Sec.~\ref{sec:conclusion}. 

\section{\wfirst\ background}
\label{sec:wfirst}

We now proceed to describe requirements and the role of this suite of image simulations in verifying that the requirements flowdown is correct. We begin with a description of weak lensing in \wfirst\ that emphasizes the issues most closely tied to the image simulations (\S\ref{ss:wlprog}), and a high-level review of the requirements process in a cosmology project (\S\ref{ss:req}). There we describe where in this process we need the mapping between the wavefront error and galaxy ellipticities, $\partial e_i/\partial \psi_j$ (where $\psi_j$ denotes a Zernike mode of the wavefront error). This mapping was obtained using a simplified analytic model, calibrated by toy simulations, in the Phase A requirements flowdown; this approach is described at a high level in \S\ref{ss:dedZ}, with technical details placed in the appendices. In the rest of this paper, we will use much more advanced image simulations, based on the \galsim\ package, to estimate $\partial e_i/\partial \psi_j$.

\subsection{\wfirst\ weak lensing}
\label{ss:wlprog}

The \wfirst\ weak lensing program has undergone significant evolution over the past decade \citep{2011arXiv1108.1374G, 2012arXiv1208.4012G, 2013arXiv1305.5422S, 2015arXiv150303757S, 2018arXiv180403628D, 2019arXiv190205569A}, but the basic philosophy has not changed. The next major advance in cosmology from weak lensing will require unprecedented control of systematic errors in photometric measurements (this includes, but is not limited to, shape measurement and PSF corrections). \wfirst\ will make this measurement with a thermally controlled telescope from beyond low Earth orbit, where the PSF can be made both stable and small. The imaging observations will be carried out in multiple filters and will have a cross-linked observing strategy within each filter to enable multiple internal cross-checks in the weak lensing signal.

The current Reference Survey in the \wfirst\ Science Requirements Document (SRD)\footnote{Document reference number WFIRST-SYS-REQ-0020} envisions shape measurements in 3 filters (J129, H158, and F184), where the PSF is at least half-Nyquist sampled (i.e., pixel size $<\lambda/D$; Nyquist sampling would be $\lambda/2D$). Here F184 is the reddest filter on \wfirst, spanning 1.68--2.00 $\mu$m; it is between ground-based $H$ and $K$, and was chosen based on the thermal constraints from the previously existing telescope hardware that was transferred to the program. Photometric redshift determination requires bluer filters as well. \wfirst\ itself will do a photometric survey in the Y106 filter since there was no ground-based option that would reach the required depth. Ground-based observations will be required for the $z$ and bluer filters; the primary option for collecting these data will be LSST \citep{2009arXiv0912.0201L,2019ApJ...873..111I}.  The expected imaging depth is 26.9/26.95/26.9/26.25 mag AB in Y106/J129/H158/F184 \changetext{(5$\sigma$ point source; the limiting magnitude for the weak lensing samples is typically $\sim 2$ mag shallower and depends on source size).} The expected galaxy number density is 35 galaxies/arcmin$^2$ (H158-band, the best for shape measurement) or 50 galaxies/arcmin$^2$ (co-added bands). The Reference Survey also includes 10\% of the time devoted to medium-deep fields, which have $10\times$ the exposure time over 1\% of the overall survey area, to calibrate the properties of the source galaxies. 

The Reference Survey area is limited to 2000 deg$^2$ due to the need for internal redundancy (e.g., 2 passes over the sky in each of 4 filters means each region must be observed 8 times) and the medium-deep fields, and the need to carry out many other observing programs as well in a five-year prime mission. This area is less than considered in some previous studies. Options for larger survey area have been considered, and could include a wider layer with less redundancy (e.g., an H158-band survey overlaid with LSST data), an extended mission (\wfirst\ has no consumable cryogens, and carries propellant for at least 10 years), or both \citep{2019BAAS...51c.418E}. The actual survey -- which may look different from the Reference Survey and be informed by developments in the coming years -- will be chosen closer to launch. However, from the requirements point of view, we focus on enabling the Reference Survey.

\subsection{The requirements process}
\label{ss:req}

Every precision cosmology project has a requirements process to control both its statistical and systematic errors and ensure that the overall mission can achieve its science objectives. In the case of \wfirst, requirements on the Project (e.g., flight hardware and software or ground system support) were baselined early in the mission (the Science Requirements Document was placed under configuration control in 2018), but requirements on science analyses are more flexible and will be fixed at a later date. The statistical error requirements are usually formulated in terms of survey area, depth and image quality in each filter, cadence (for time-domain programs), etc.; their relation to the science reach of the mission is handled by forecasting tools to be described \changetext{\citep{2020arXiv200404702E, 2020arXiv200405271E}}. Systematic error control is much more difficult, and the approach may differ depending on whether a source of systematic error is {\em observational} or {\em astrophysical}. Usually, observational systematics (e.g., PSF calibration for weak lensing) can be budgeted within the systems requirements framework of a large project, whereas astrophysical systematics (e.g., baryonic feedback) are addressed through a combination of nuisance parameters, additional observations, and theory/simulation. These astrophysical systematics are important science team responsibilities but are not part of Project requirements and engineering reviews.

In general, it is important to distinguish between known systematic biases in both categories, which can be calibrated and removed from the data, and uncertainties on that calibration, which cannot be removed and must either be small enough to ignore or marginalized over in an analysis. In the description of systematic errors here, we are referring to this residual uncertainty. We focus now on the approach to observational systematic errors; our focus is on the \wfirst\ process, but note that something similar has been done for other large weak lensing programs such as LSST and Euclid \citep{euclid_srd,2016SPIE.9911E..05V,LPM-17,2018arXiv180901669T,LSE-29}.

First, one identifies a {\em data vector} that will contain the cosmological information. For setting \wfirst\ weak lensing requirements, the data vector is the concatenated list of shear power spectra and cross-power spectra ${\bf C}$ across tomographic bins. Other choices, such as including higher-order statistics, using all $3\times 2$-point information, or working in correlation function space are possible, but given the tools available at the time of Project start these would have required additional tool development that did not fit in the schedule.

Second, one identifies an {\em error metric} that summarizes the impact of a systematic error on the data vector. We have chosen the error metric $Z^2 = \Delta{\bf C}\cdot{\boldsymbol\Sigma}^{-1}\Delta{\bf C}$, where $\Delta{\bf C}$ is the bias on the data vector and ${\boldsymbol\Sigma}$ is the statistics-only covariance matrix. The metric $Z$ is essentially a metric for the ratio of the systematic to the statistical error, and this depends on the solid angle $\Omega$ covered by the survey ($Z\propto \sqrt\Omega$). One also sets a limit on the maximum allowed error $Z$; in our case, we set $Z=0.5$ at 2500 deg$^2$ (or $Z=1$ at 10,000 deg$^2$), which means that the observational systematic errors are required to be below 50\% of the statistical errors in a 2500 deg$^2$ survey and below 100\% of the statistical errors if the survey were to be extended to 10,000 deg$^2$. 

Third, we note that each category of observational systematic error contributes to $Z$. In cases where the errors are presumed independent, the $Z^2$ values can be summed (i.e., $Z$ obeys root-sum-square or RSS addition), and the ``top-level'' budget for $Z$ can be broken down into contributions from different sources. If a source of observational systematic error is parameterized by a parameter $p$ (e.g., overall shear calibration), then a requirement on knowledge of $p$ (parameterized by the $1\sigma$ uncertainty $\Delta p$) can be obtained by computing the sensitivity $d{\bf C}/dp$ and setting
\begin{equation}
\sqrt{\frac{d{\bf C}}{dp} \Delta p \cdot {\boldsymbol\Sigma}^{-1} \frac{d{\bf C}}{dp} \Delta p}
\end{equation}
equal to the allocation for $Z$ from that contribution. An important aspect of this budgeting is that, like a requirements flowdown, it is {\em hierarchical} -- a top-level requirement on observational systematics may contain an allocation for shear calibration (one of several contributions), which itself may contain a branch for PSF size (one of several contributions), which itself may contain a branch for detector non-linearity, etc. In the life cycle of a cosmology project, more detail will be filled in first on the branches that have hardware impacts, and then branches related to algorithms or simulations later on.

The details of our data vector, covariance, and systematics models are described in Appendix~\ref{app:wl-budget}. The systematic errors in the shear $\gamma$ are broken down into additive biases ($c$) and multiplicative biases ($m$) in accordance with
\begin{equation}
\gamma({\boldsymbol\theta},z;{\rm obs}) = [1+m(z)] \gamma({\boldsymbol\theta},z;{\rm true}) + c({\boldsymbol\theta},z).
\label{eq:gmod}
\end{equation}
Appendix~\ref{app:wl-budget} then allocates the systematic budget for $Z$ to residual uncertainties $\Delta c$ (in different angular bins) and $\Delta m$. One challenge is that the shear biases may be redshift-dependent. Fortunately, when we start assigning portions of the shear systematic error budget to underlying root causes, we usually know something about the redshift dependence (for example, most PSF-related errors grow with redshift because the galaxies get smaller). Therefore, we have assigned each possible redshift dependence a weighting factor $S$, which represents the ratio of what fraction (in an RSS sense) of the error budget is taken up by a systematic with a given redshift dependence, relative to a systematic that is redshift-independent with the same maximum amplitude. A redshift-independent systematic has $S=1$; due to covariance between redshift bins, it is possible to have $S>1$.

\subsection{Mapping from wavefront error to galaxy ellipticities -- analytic approach}
\label{ss:dedZ}

The requirements based on $Z$ are described in terms of shear systematics, but in order to be useful for engineering, we need to write a requirement in terms of wavefront errors. The key step to doing this is to write the derivative of the observed shear $\gamma_{\rm obs}$ with respect to the wavefront error $\psi_j$ (where $j$ denotes a Zernike mode). Because the PSF size and ellipticity are quadratic rather than linear in the wavefront error, it is necessary to take a quadratic expansion; then $\partial\gamma_{{\rm obs},i}/\partial\psi_j$ is linear in the static wavefront error $\psi_j$ \citep{2010SPIE.7731E..1EN}. Our approach is to use symmetries to categorize the possible quadratic terms, which gives 4 independent coefficients if we take the first 11 Zernike modes (up through spherical aberration). We can then use a suite of simple simulations to determine the 4 coefficients. Then -- given a limit on the static wavefront error $\lVert{\boldsymbol\psi}\rVert \le 92$ nm rms, set to achieve diffraction-limited imaging in $J$-band -- we can analytically search the space of possible static wavefront errors and find the maximum possible $|\partial\gamma_{{\rm obs},i}/\partial\psi_j|$ (units: nm$^{-1}$). This worst-case sensitivity can be used to set requirements on knowledge of the \wfirst\ wavefront.

A similar process can be used for changes in the PSF within an exposure (either line of sight motion, or wavefront jitter -- i.e., beyond the tip-tilt modes -- due to vibrations). The baseline plan for \wfirst\ will be to independently fit the line of sight motion contribution to the PSF in each exposure \citep{2012SPIE.8442E..10J}, but not the wavefront jitter. This implies a requirement on the wavefront jitter to make its contribution to the PSF negligible. This requires a computation of the derivative of $\gamma_{\rm obs}$ with respect to the second moments of the PSF (units: mas$^{-2}$); with respect to the variance or covariance of the wavefront jitter (units: nm$^{-2}$); or with respect to the covariance of wavefront jitter and line of sight motion (units: nm$^{-1}$ mas$^{-1}$).

In both of these cases, the source of bias in the shear measurement is in practice due to errors in the wavefront model leading to mis-estimation of the PSF model that is used for convolution of the galaxy model when fitting the model shape. All of these calculations are described in detail in Appendix~\ref{app:psf-stability}.

\changetext{We emphasize that while the time-dependent wavefront error and line-of-sight motion are two of the most difficult aspects of weak lensing, they are only a portion of the overall shear measurement error budget. Some other contributions related to the wavefront could come from small-scale field dependence of the wavefront due to figure errors on the fold mirrors (which are closer to an intermediate focus than a pupil) or flatness of the detectors; chromatic dependence of the wavefront; and polarization dependence of the wavefront \citep{2020MNRAS.tmp.1430L}. There are also sources associated with the calibration of the \wfirst\ detectors \citep{2020arXiv200500505M}, including but not limited to interpixel capacitance, persistence, count-rate dependent non-linearity, flat field and dark current uncertainties, and the brighter-fatter effect. In practice, as we gain additional knowledge of as-built components, new terms are added (e.g., we are currently working on adding the vertical trailing pixel effect, e.g., \citealt{2020arXiv200305978F}), so there must be margin to cover these new developments in the top-level error budget for $Z$ (the full budget is being updated and is beyond the scope of this paper). It is possible to group some of these terms together and form an intermediate level requirement on knowledge of the PSF moments, and then have, e.g., time-dependent wavefront errors as a sub-allocation, as was done for {\slshape Euclid} by \citet{2013MNRAS.431.3103C}. Given the \wfirst\ working group structure, with different groups focused on specific elements (e.g., detectors, filters, stability of the optical chain) and with the science team representatives in these groups ultimately looking after the shear bias requirements, we chose instead to treat all instrument-related systematics as sub-allocations of the shear bias requirements.}

\subsection{Limitations of the analytic approach}

The analytic approach to estimating the sensitivity to wavefront errors has some advantages: it is simple, maintains a close link to underlying physical principles, enables rapid exploration of the parameter space, and was available at an earlier stage of the project than the image simulations. However, it has some drawbacks:
\begin{list}{$\bullet$}{}
\item The analytic approach deals with single images, so it does not represent what happens when images are combined. This is especially relevant when the input images are undersampled at the native resolution of the \wfirst\ pixels. (The pixel scale is 110 mas, whereas Nyquist sampling would be $\lambda/2D = 56$, 68, or 80 mas at the average wavelength of J129, H158, or F184 bands respectively.)
\item The analytic approach computes the derivatives $\partial\gamma_{{\rm obs},i}/\partial\psi_j$ at one point in the focal plane. Therefore, it does not capture the correlations across the focal plane or tiling patterns; the distribution of systematic shear in 2-point correlation function space or in power spectrum space is not captured.
\item The analytic approach cannot be extended to include interaction of PSF errors with other aspects of the data, such as noise, detector systematics, blending/selection, etc., in the way that is possible with image simulations.
\end{list}
For these reasons, we have also estimated the mapping from wavefront error to galaxy ellipticities using pixel-level image simulations with the \galsim\ package. The version of the simulations used here is highly idealized -- for example, the matching to the ``truth catalog'' means that selection/blending effects are not realistically implemented, some detector effects were not implemented, and the input galaxies have artificially prescribed shears and do not come from a realistic large scale structure distribution. This is useful in the current study to enable us to uniquely isolate the impacts of wavefront errors on shear recover. Nevertheless, \galsim\ as a tool is extensible and could be configured to use a realistic \wfirst\ input catalog for future systematics studies.

\section{Simulation suite}\label{sec:sim}

To empirically test weak lensing requirements, methods, and algorithms in \wfirst, we have designed a synthetic survey suite that, while not entirely realistic in all object properties, contains sufficiently complex and representative objects so as to enable informative tests and preliminary algorithm development. 
This synthetic survey utilizes several external simulation and data sources, and generates \wfirst-like imaging using the \galsim\ framework and its \wfirst\ module. 
The simulation framework is generally capable of producing a full \wfirst\ HLS imaging survey in all filters matching Cycle 7 specifications.\footnote{These can be found at \url{https://wfirst.gsfc.nasa.gov/science/WFIRST_Reference_Information.html}. Note that updates to match Phase B payload design have not been incorporated into the simulation described in this paper, but this is not expected to impact the results of this paper.} 
The code is publicly available.\footnote{\url{https://github.com/matroxel/wfirst_imsim}} An example SCA image is shown in Fig.~\ref{fig:sca}. The fiducial simulation run is available for download -- this public dataset is described in App.~\ref{app:data_access}.

This approach to producing (to varying degrees) realistic, synthetic survey realizations is a common approach for weak lensing experiments, both at the catalog level \citep{2018MNRAS.480.4614M,2019arXiv190706530K} and the image level \citep{2016MNRAS.457..786S,2017MNRAS.467.1627F,2018MNRAS.481.3170M,2018MNRAS.475.4524S}. These synthetic surveys can serve as sources of calibration or characterization, validation, or increasingly as end-to-end integration tests for measurement and analysis algorithms and pipelines. Our approach here is similar to the approach being implemented in parallel by the LSST Dark Energy Science Collaboration \citep[DESC;][]{2019arXiv190706530K,dc2all}, with comparable levels of morphological complexity for weak lensing algorithm testing, but less complex true object properties. This approach is described in detail in the following subsections.

\subsection{Simulation stages}\label{stages}

The simulation is broken into several stages:

\begin{figure}
\begin{center}
\includegraphics[width=\columnwidth]{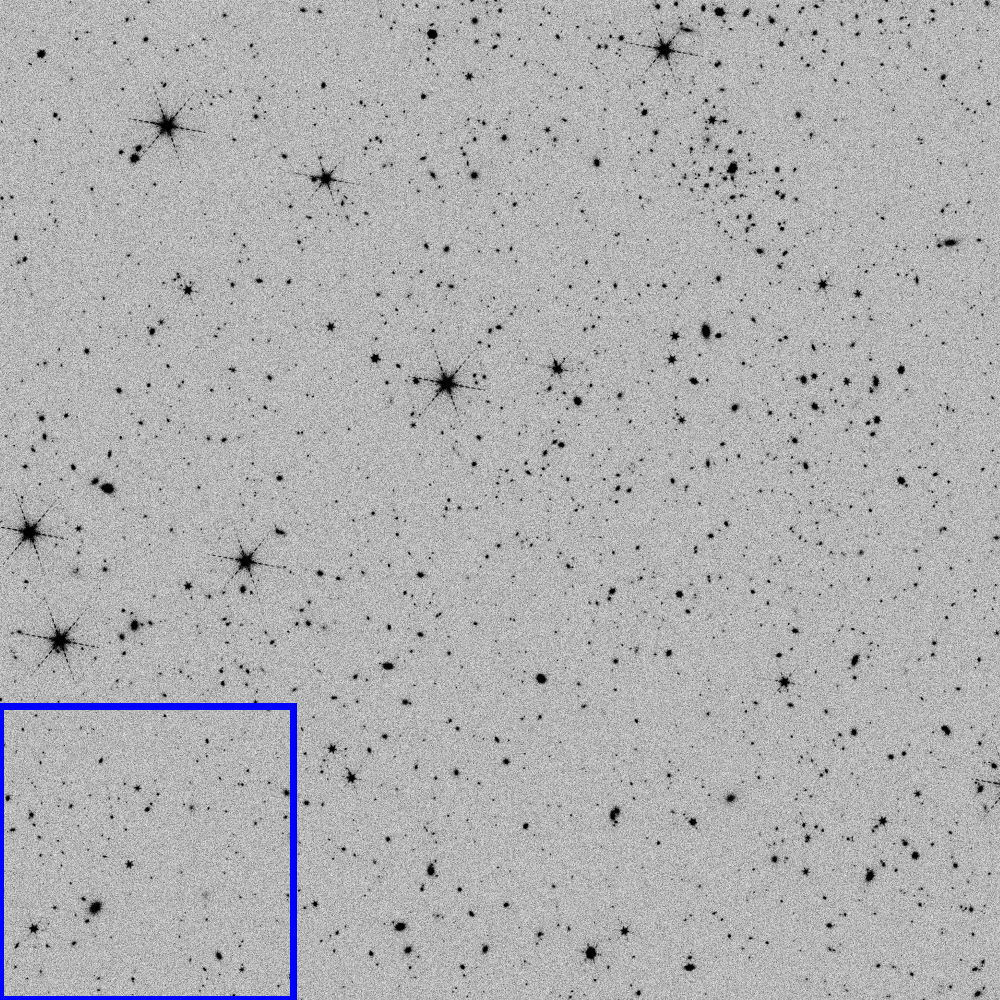}
\end{center}
\caption[]{
A simulated $\sim$140s exposure Sensor Chip Array (SCA) image, chosen for the presence of several large, bright galaxies and stars. Each SCA (HgCdTe H4RG) has a useable pixel grid of 4088$\times$4088, with a pixel scale of 0.11\arcsec. A total of 18 SCAs make up the \wfirst\ camera. For comparison, the size of the Hubble Space Telescope Wide Field Camera 3 is shown as the blue outline. Six diffraction spikes due to the \wfirst\ secondary mirror support struts are clearly visible for most stars.
\label{fig:sca}}
\end{figure}

\textbf{\textit{Truth catalog generation}} -- A truth catalog is generated from the simulated input galaxy distribution, photometric galaxy catalog, and Milky Way simulation. 
The following true object properties are assigned to each simulated galaxy:
1) The sky position in right ascension (RA) and declination (Dec) from the simulated galaxy distribution; 
2) Photometric properties (consistent Y106/J129/H158/F184 magnitudes, size, and redshift) drawn from a random object in the photometric galaxy catalog; 
3) Intrinsic ellipticity components drawn from a Gaussian distribution of width 0.27 (truncated at $\pm$0.7); 
4) A random rotation angle; 
5) The ratio of fluxes in each of the three galaxy components: a) de Vaucouleurs bulge, b) exponential disk, and c) random-walk star-forming knots (a maximum of 25\% of the flux assigned to the disk component can exist in the knots); 
6) The gravitational lensing shear applied to the object, drawn from a discrete list of $(e_1, e_2) \in \{0,\pm 0.1\}$.
Further details on the provenance of the galaxy catalogs and Milky Way simulation can be found in Secs. \ref{galcats} and \ref{starcat}, respectively. 
The true properties for all objects are saved in a single FITS table that is accessed by the following stages.

\textbf{\textit{Image generation}} -- In this stage, an empty SCA image is initialized ($4088\times4088$ pixels), and a model is built for each galaxy and star in turn, then drawn into the image. 
The galaxy models are built chromatically from the truth parameters for the object, with each component being assigned a different representative SED of types: S0 (bulge), SBa (disk), and Im (knots), respectively. 
The assigned SED is the same for all objects, since after redshifting the spectrum and applying the appropriate flux and size in each component, the model is converted to be achromatic in each passband to speed up the drawing (this is discussed further in Sec.~\ref{psf}). 
The intrinsic ellipticity, random rotation, and gravitational shear is then applied.
We model stars as point sources with the SED of Alpha Lyra.
Stars are also converted to be achromatic before drawing.
Both stars and galaxies are then convolved with the appropriate PSF for the SCA (constant across the SCA in the fiducial simulation). An example of the PSF model for an object is shown in Fig. \ref{fig:psf}, and the PSF model is discussed in more detail in Sec.~\ref{psf}. We save images of the true PSF model both at native pixel scale and oversampled by a factor of 8, in stamps of native pixel size $8\times 8$ at the position of each galaxy.

The models are drawn in dynamically-sized square stamps, the sizes of which are chosen to include at least 99.5\% of the flux.
These stamps are then added to the SCA image and saved separately (if drawing a galaxy) to provide an isolated image of each simulated galaxy to allow for tests of the impact of blending.
Objects that would have a postage stamp that overlaps the SCA image are drawn, such that light from objects in chip gaps are appropriately drawn onto the SCA, but we only save postage stamps for objects that have a centroid that falls on the SCA. 
We do not save isolated postage stamps of objects that have a stamp size of greater than 288$\times$288 pixels, but they are drawn into the images.
Finally, each isolated postage stamp is processed through the steps described in Sec.~\ref{effects} to simulate the WFIRST observatory and detectors and written to disk. This means that blended objects will be modeled differently in the isolated postage stamp and full SCA images, since some detector effects are sensitive to the total flux in nearby pixel. When all objects are added to the full SCA image, it is also processed through these steps and written to a FITS image file.

\textbf{\textit{MEDS creation}} -- We then compile the output across pointings of the isolated object stamps into MEDS (Multi-Epoch Data Structure) files.\footnote{\url{https://github.com/esheldon/meds}}
These files concatenate all exposures of unique objects to allow for fast access for object-by-object data processing (like shape measurement). 
Each MEDS file also stores for each object (and stamp) its original SCA, the object position and the stamp position within the SCA, the WCS for each stamp, the PSF model for each object, and other ancillary information and metadata.
Each MEDS file contains all objects within a $n_{\textrm{side}}=512$ Healpixel\footnote{\url{https://healpix.jpl.nasa.gov}} \citep{2005ApJ...622..759G,Zonca2019}.

\textbf{\textit{Shape measurement}} -- 
The galaxy shape is measured by jointly fitting a two-component model, de Vaucouleurs bulge and exponential disk, across all suitable exposures. 
Exposures where more than 20\% of the pixels are masked (i.e., the centroid falls too close to the edge of the SCA) are rejected. 
The model fit has 7 parameters: $e_{1,2}$, $p_{x,y}$, half-light radius, flux, and bulge flux fraction, where $e_{1,2}$ is the component of the ellipticity and $p_{x,y}$ is the pixel centroid offset. Both model components are constrained to have the same centroid, half-light radius, and shape.
The minimization is performed using the \textsc{NGMIX}\footnote{\url{https://github.com/esheldon/ngmix}} and \textsc{MOF}\footnote{\url{https://github.com/esheldon/mof}} packages \citep{2014MNRAS.444L..25S}. 
We also measure the PSF size and shape in the oversampled PSF model images using an adaptive moments method \citep{2003MNRAS.343..459H}. 
This stage writes a set of FITS files containing the galaxy and PSF measurement results and relevant truth catalog information.

\subsection{\galsim}

The images in the simulation are rendered using the \galsim\ software package \citep{Rowe15}.  
This package has been extensively tested and has been shown to yield very accurate rendered images of galaxies and stars.
Notably, the image rendering process has been shown to impart biases in the shapes of galaxies at a level much less than $10^{-3}$ for the kinds of objects we are simulating here.  

The \galsim\ package is mostly generic with respect to the telescope and observational strategy, allowing for a wide variety of options in performing the simulation.
However, it does have a sub-module (\texttt{galsim.wfirst}) that has a number of WFIRST-specific implementation details.
Some of the code in this module pre-dates this work (e.g., \cite{2016PASP..128i5001K}), but some of it was developed specifically for this project, especially updating some of the details to match Cycle 7 information, and to reflect new information from laboratory tests of persistence in \wfirst\ sensors.  
The values used for this project correspond to the \texttt{galsim.wfirst} module in \galsim\ release version 2.2.0.

\subsubsection{World coordinate system}\label{wcs}

The \texttt{galsim.wfirst} module has code to provide an estimate of the \wfirst\ WCS (world coordinate system) for each SCA given a rotation angle, date, and pointing direction.  
The WCS gives the two-dimensional mapping from $(x,y)$ coordinates on the image to RA and Dec on the sky.
The specific orientations and gaps between the sensors were updated to match Cycle 7 specifications as part of the development work for this project.
We create our scene of objects, including their surface brightness profiles, in sky coordinates (RA, Dec).
\galsim\ automatically accounts for the Jacobian of the WCS transformation when rendering the surface brightness profiles on each sensor's pixels.  
Details such as the telescope distortion and variable pixel area are correctly accounted for in this process.

\begin{figure}
\begin{center}
\includegraphics[width=\columnwidth]{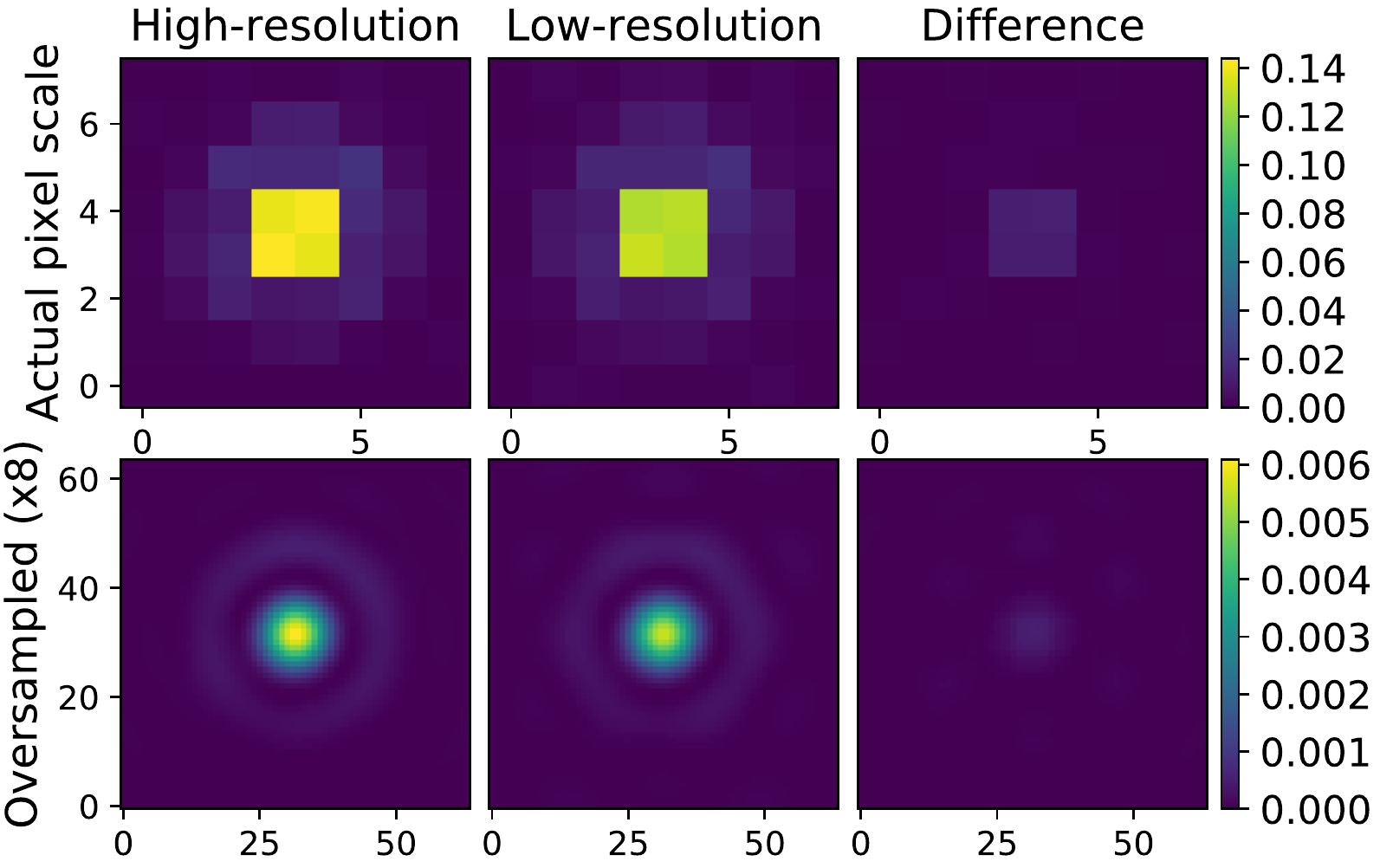}
\end{center}
\caption[]{
PSF model for SCA \#1. The top row shows the model in native pixel scale, while the bottom row is oversampled by a factor of 8. From left to right: a comparison of the high-resolution (`true') model, the low-resolution model used in the simulation, and the difference of the two models. The color bars are defined by the range of the high-resolution model. More detail about the PSF model is given in Sec.~\ref{psf}. The difference between the low- and high-resolution PSF models is negligible at the level required for the current study.
\label{fig:psf}}
\end{figure}

\subsubsection{Point-spread function}\label{psf}

For the PSF we use a model of the \wfirst\ PSF from the \texttt{galsim.wfirst} module.
While this module includes a high-resolution Cycle 7 estimate of the \wfirst\ spider pattern (i.e., the obscuration of the struts and camera in the pupil plane), we use a faster, low-resolution approximation, which gets the qualitative features correct, but has a slightly different detailed diffraction pattern. 
For the purposes of this study, we are insensitive to the differences between the two spider patterns, so we did not enable the slower, more accurate option.
The PSF uses position-dependent (Zernike) aberration polynomials \citep{noll1976}, based on an investigation of the field-dependent wavefront errors used in the original Cycle 7 documentation.
Aberrations between the tabulated positions are estimated using bilinear interpolation of the tabulated values. \changetexttwo{More details of the PSF model approximation and its implementation are described in App. \ref{app:psf_resol}.}

The wavelength-dependent features of the PSF, such as the width of the Airy diffraction pattern, and the wavelength-dependence of the aberrations, are taken at the effective wavelength of the observation bandpass.
This is an approximation, which leads to an enormous speed up in the rendering time.
However, it does omit some interesting and subtle chromatic effects as different parts of a galaxy, with different effective SEDs, would be convolved by slightly different effective PSFs.
There are plans to improve the implementation of this aspect of \galsim, but it cannot currently simulate such effects efficiently enough for our needs.

There are also plans to enable the use of WebbPSF\footnote{\url{https://webbpsf.readthedocs.io/en/stable/}} in \texttt{galsim.wfirst} to leverage the work being done on that project to simulate the \wfirst\ PSF.
The WebbPSF model is qualitatively similar to what we are using from \texttt{galsim.wfirst}, but there are slight differences.
We expect that the WebbPSF model is probably more accurate, but this will be explored in future work.

\subsubsection{Implemented detector effects}\label{effects}

Most of the development of \galsim\ has been driven by the need to render simulations of CCD images.
The HgCdTe detectors used by \wfirst\ are qualitatively similar, but there are significant differences in the physics, which lead to differences in some of the simulation steps. We discuss the implementation of some of these effects in detail below. 

For this work, each image is processed through the following stages, simulating what physically happens in the detector: 1) the Poisson background of stray light and thermal emission from the telescope is generated and a `sky' background image is created that also undergoes stages 2--9, 2) the impact of reciprocity failure is added, 3) the electron counts are quantized, 4) dark current is added to the image, 5) nonlinear response to flux is applied, 6) the effect of interpixel capacitance is applied, 7) instrument read noise is applied, 8) electron counts are converted to ADU, and 9) the ADU value is quantized. In this work, we subtract the final background image from the SCA image, simulating a perfect background subtraction algorithm.

Reciprocity failure \citep{Biesiadzinski2011} is a non-linear relationship between the voltage response in the detector to the incident flux of photons at low light levels.
The exact mechanism of this effect is unknown and hence we lack a good theoretical model. 
\galsim\ uses a power law
\begin{equation}
\frac{p}{p_\mathrm{nominal}} =
 \left(\frac{p_\mathrm{nominal}}{f_0 t_\mathrm{exp}}\right)^{\frac{\alpha}{\log(10)}}
\end{equation}
where $p_\mathrm{nominal}$ is the pixel response (in electrons) that would have occurred in the absence of reciprocity failure,
$p$ is the actual observed response due to reciprocity failure,
$f_0$ is the base flux rate (in electrons/sec) at which the nominal gain was calibrated, 
$t_\mathrm{exp}$ is the exposure time,
and $\alpha$ is taken to be $6.5 \times 10^{-3}$ for the \wfirst\ sensors.

A particularly pernicious effect present in the HgCdTe detectors is known as ``persistence'' \citep{10.1117/12.789619,2014arXiv1402.4181A,2016SPIE.9915E..0GM}.  
In a series of images taken sequentially, some small fraction of the charge accumulated in earlier exposures apparently remains in the sensor and appears in later exposures.  
The effect lasts for many minutes across multiple reset cycles.  
Therefore, for simulating the effect, we need to keep track of the precise order and time of each observation, and the electron-level (i.e., pre-read-out) images of multiple prior exposures.

The exact functional form of this effect is not very well understood, although some progress is being made in laboratory tests.  
The functional form for this effect was updated during the Cycle 7 updates, and
\galsim\ now uses a Fermi profile when the deposited flux is above the half-well level, and linear when below.
Above the half-well level, the functional form is
\begin{equation}
n_\mathrm{persist} = \frac{A \left(n/n_0\right)^a  \left(\frac{t}{1000 \mathrm{sec}}\right)^{-r}}
{ \exp(- \frac{n-n0}{dn})+ 1}
\end{equation}
where $A$, $n_0$, $a$, $r$, and $dn$ are constants estimated from laboratory measurements (and stored in the \texttt{galsim.wfirst} module).
The persistence modeling was not available when this project started, and so is not implemented in the current simulations used in this paper.

In addition to the non-linear pixel response, known as reciprocity failure, there is also a non-linearity in the conversion of accumulated charge to the measured voltage \citep{2017JInst..12C4009P,Biesiadzinski2011,10.1117/12.2314475}.
This is a different effect, which occurs at a different point in the simulation -- namely, after the application of dark current \citep{10.1117/12.790382,10.1117/12.2057308,10.1117/12.2233664} and persistence. 
\galsim\ treats this as a modification in the effective number of electrons:
\begin{equation}
n_e^\prime = n_e - 6 \times 10^{-7} n_e^2
\end{equation}
where $n_e$ is the actual number of electrons accumulated and $n_e^\prime$ is the effective number to account for the voltage response nonlinearity. The coefficient $6\times 10^{-7}$ is appropriate for one of the WFIRST development detectors measured in the lab \citep{2020PASP..132a4502C}.

Inter-pixel capacitance (IPC) \citep{2016PASP..128i5001K} essentially amounts to a convolution of the image by a $3 \times 3$ kernel in pixel coordinates.
However, the timing of the convolution is during the readout process, which means that some (but not all) of the noise has already occurred.  
Thus it cannot be treated as part of the PSF for the purpose of the simulation.  
It needs to be applied separately after the dark current and Poisson shot noise have been applied, but before the read noise.  
The IPC coefficients have been measured in the lab for \wfirst\ detectors; the values used in the \texttt{galsim.wfirst} module come from the Cycle 5 estimates. 

\begin{figure}
\begin{center}
\includegraphics[width=\columnwidth]{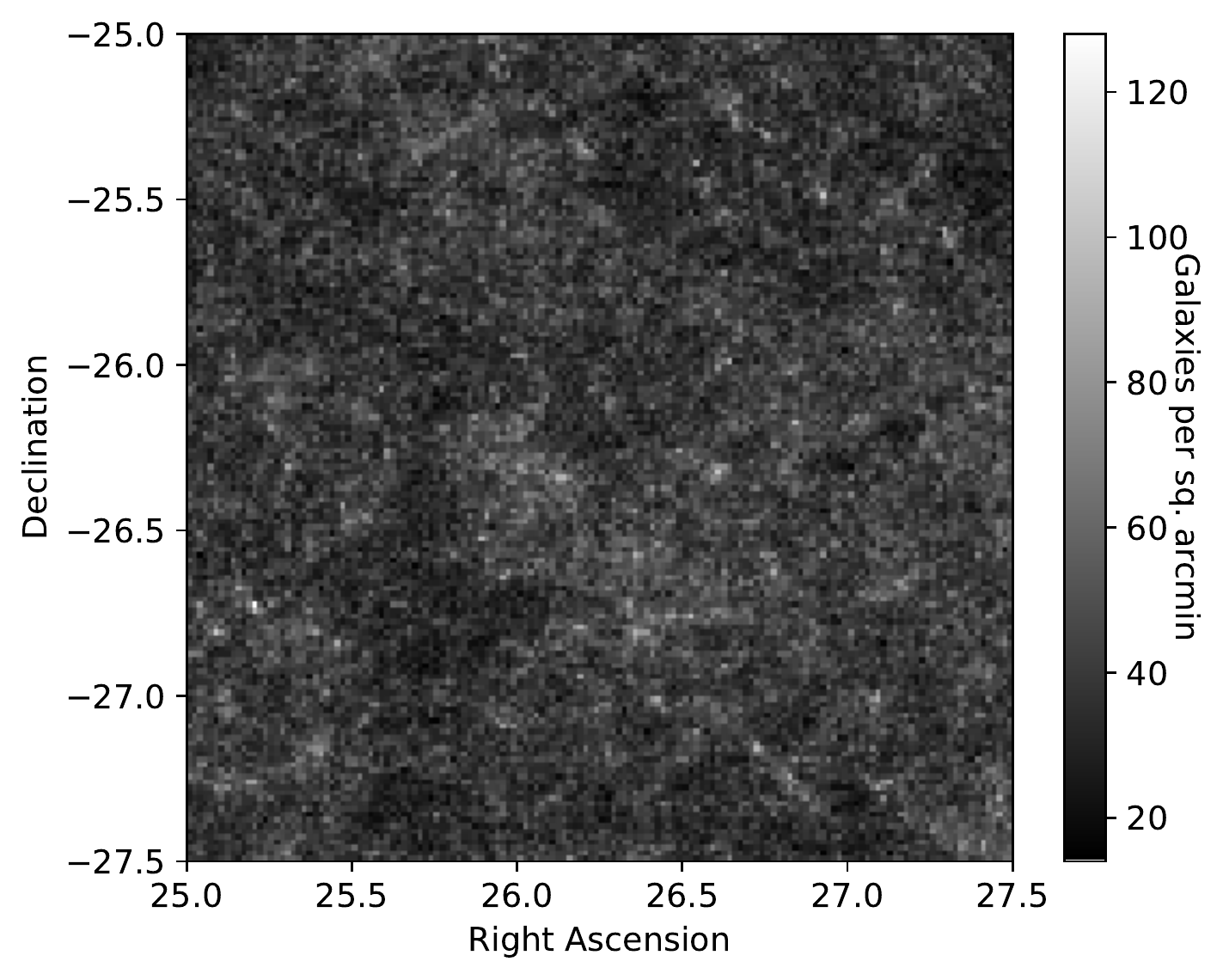}
\end{center}
\caption[]{
The distribution of simulated galaxies. The mean galaxy density is 40 arcmin$^{-2}$. 
\label{fig:galaxies}}
\end{figure}

\begin{figure*}
\begin{center}
\includegraphics[width=\textwidth]{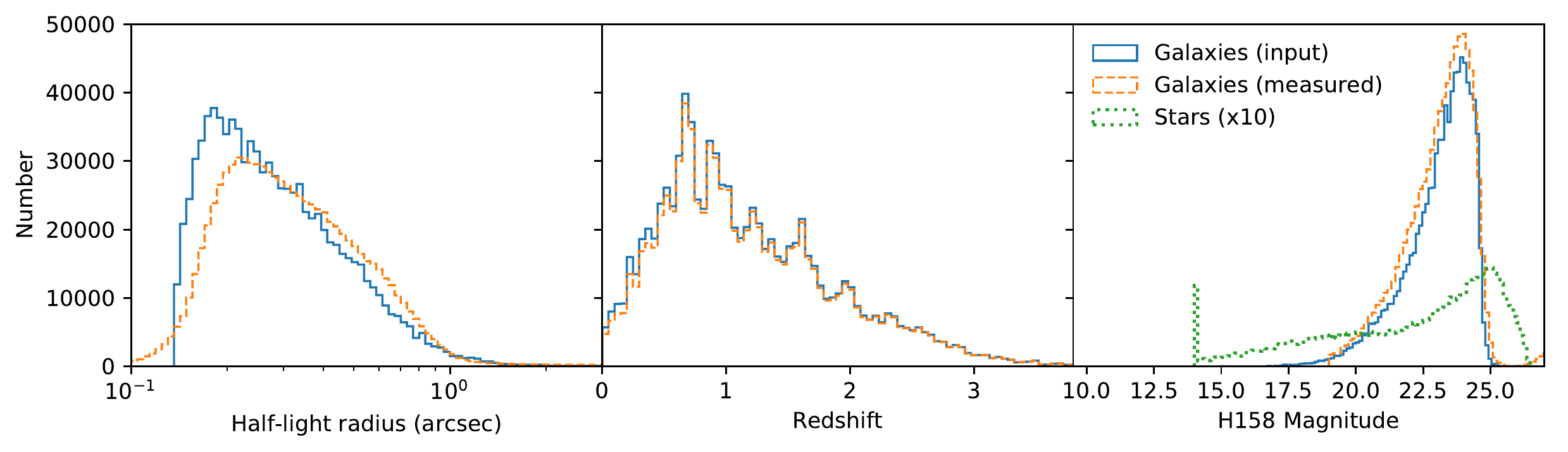}
\end{center}
\caption[]{
The true (blue) and recovered (orange) distributions of galaxy half-light radius, redshift, and H158 magnitude for galaxies, with a comparison of the magnitude distribution of stars (green). The recovered galaxy magnitude and half-light radius are the distributions inferred from the shape measurement process, while the distribution in redshift simply shows where in redshift objects do not have a valid shape fit -- mostly at low redshift, where some large objects are not used. In general, the measured size and magnitude agree well with the true values. Star magnitudes are currently capped at 14 to avoid visual artifacts in the drawn images, which has no impact on the current or most plausible weak lensing studies.
\label{fig:hist}}
\end{figure*}

\subsection{Galaxy catalogs}\label{galcats}

The input galaxy catalog is created using a simulated galaxy distribution on the sky taken from one realization of the Buzzard simulation \citep{2019arXiv190102401D,wechsler2019}, to introduce realistic galaxy clustering. 
Each galaxy is then assigned a random set of photometric properties matching a galaxy from a sample based on the Cosmic Assembly Near-infrared Deep Extragalactic Legacy Survey (CANDELS) survey that simulates the fiducial \wfirst\ weak lensing selection \citep{2019ApJ...877..117H}. We imposed selection cuts on the lensing source galaxies based on the Exposure Time Calculator \citep{2012arXiv1204.5151H}. The cuts require matched filter S/N ratio $>18$ in combined $J+H$, ellipticity error per component $<0.2$ (in the \citealt{2002AJ....123..583B} convention), and resolution factor $>0.4$ (again in the \citealt{2002AJ....123..583B} convention); note that this results in a limiting magnitude that depends on galaxy size. These cuts are also discussed in \cite{2019ApJ...877..117H}. These selections are made on the input catalog properties, which improves the efficiency of the simulation. This prevents us from exploring the impact of selection effects, but this is not important to the current work and we can use different input galaxy property distributions in future simulation runs.

The galaxy distribution, which has a mean galaxy density of approximately 40 arcmin$^{-2}$, is shown in Fig.~\ref{fig:galaxies}. 
In Fig.~\ref{fig:hist}, we show the distributions of size, redshift, and H158 magnitude in the CANDELS sample. 
We discard less than 1\% of the largest objects in the shape measurement stage, however, due to a maximum postage stamp size restriction. 
In general, the input distribution and properties of galaxies can be easily modified by configuration (i.e., specifying a different input galaxy catalog or a realistic shear field).

\subsection{Star catalog}\label{starcat}

\begin{figure}
\begin{center}
\includegraphics[width=\columnwidth]{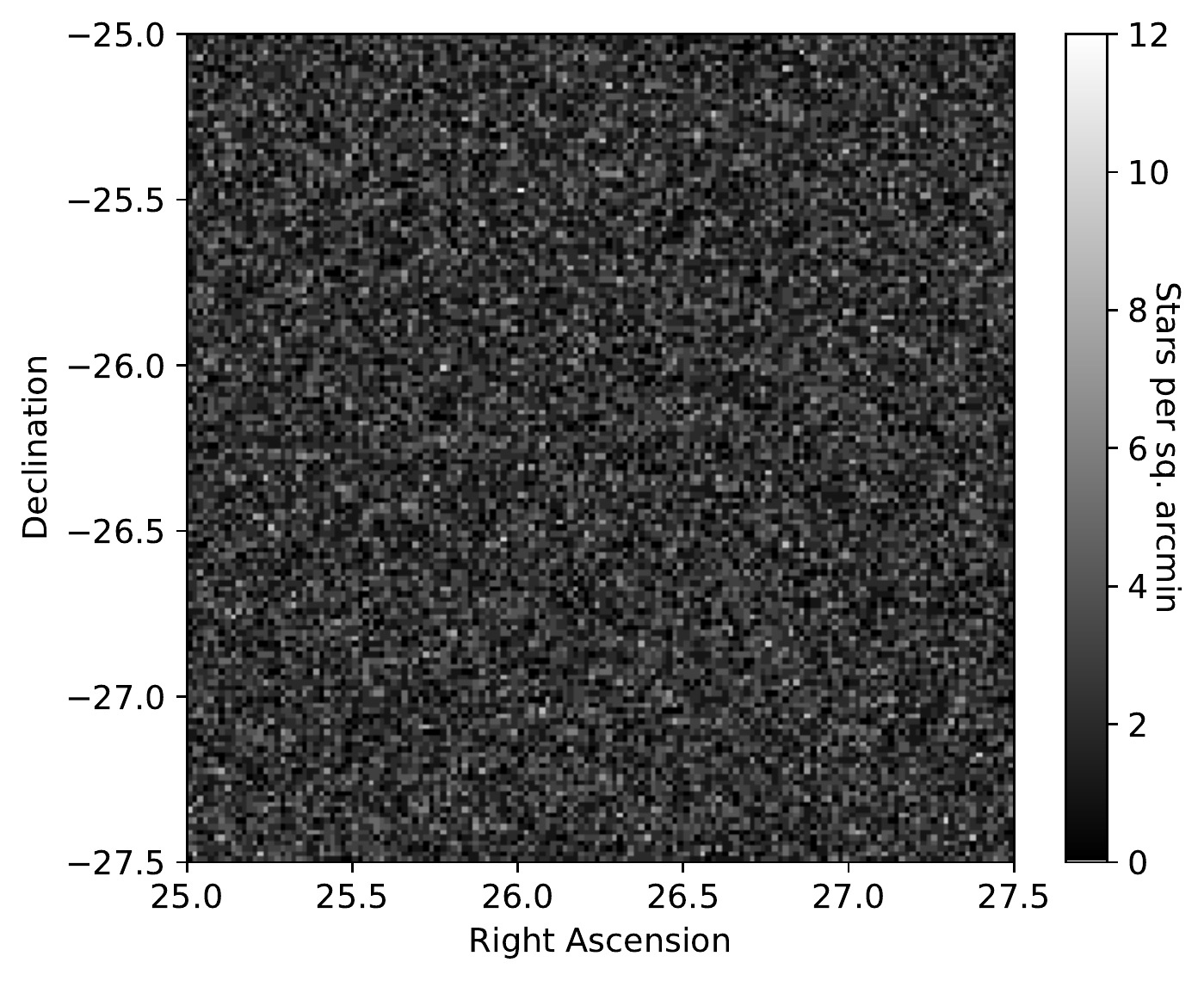}
\end{center}
\caption[]{
The distribution of simulated stars. The mean stellar density is 2.5 arcmin$^{-2}$. 
\label{fig:stars}}
\end{figure}

We simulate the positions and magnitudes of input stars in \wfirst\ passbands using the galaxy simulation Galaxia\footnote{\url{http://galaxia.sourceforge.net}} \citep{galaxia}. 
Galaxia uses an analytic model \citep{galaxia2} to simulate stars in the galaxy that includes a thin and thick disk with warp and flaring, bulge, and halo components. 
Stars are simulated to 27th magnitude in V band, extinction is added, and they are uniformly translated to \wfirst\ bandpasses using the stellar SED of Alpha Lyra derived from HST CALSPEC as packaged with \galsim.  The star distribution, which has a mean stellar density of approximately 2.5 arcmin$^{-2}$, is shown in Fig.~\ref{fig:stars}.

\subsection{Survey strategy}

\begin{figure}
\begin{center}
\includegraphics[width=\columnwidth]{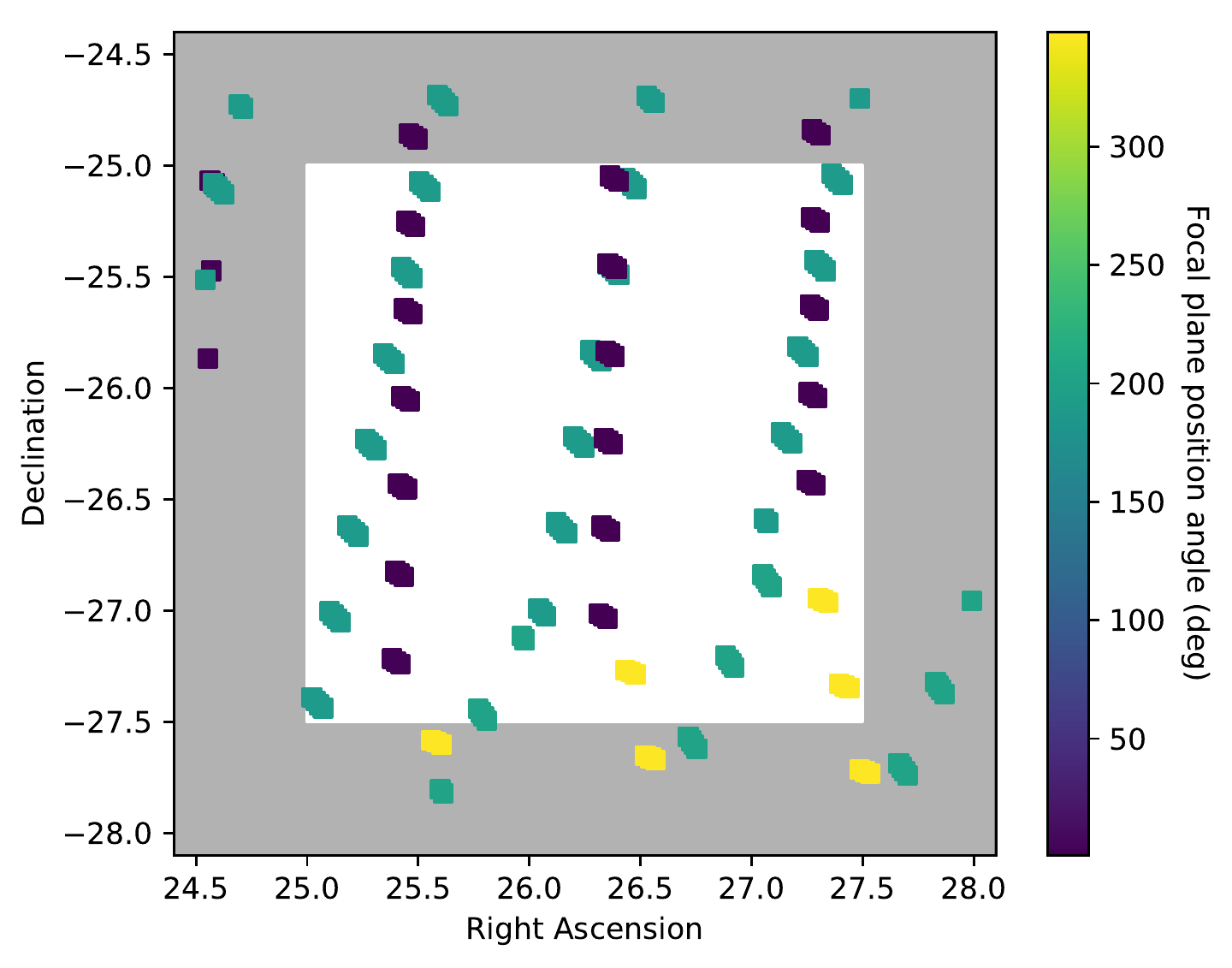}
\end{center}
\caption[]{
A visualization of the individual pointings of the telescope using the H158 filter in the five-year simulated Reference Survey that overlap the region of the sky we are simulating images for (the non-shaded region). There are a total of 189 pointings in H158 that overlap this region. Each marker is an individual pointing, whose color represents the focal plane position angle. Each cluster of pointings typically contains 3-4 very small translational dithers to cover chip gaps. The dither pattern in other filters overlaps in other directions to produce a more homogeneous coverage than is indicated in this figure. 
\label{fig:pointings}}
\end{figure}

\begin{figure}
\begin{center}
\includegraphics[width=\columnwidth]{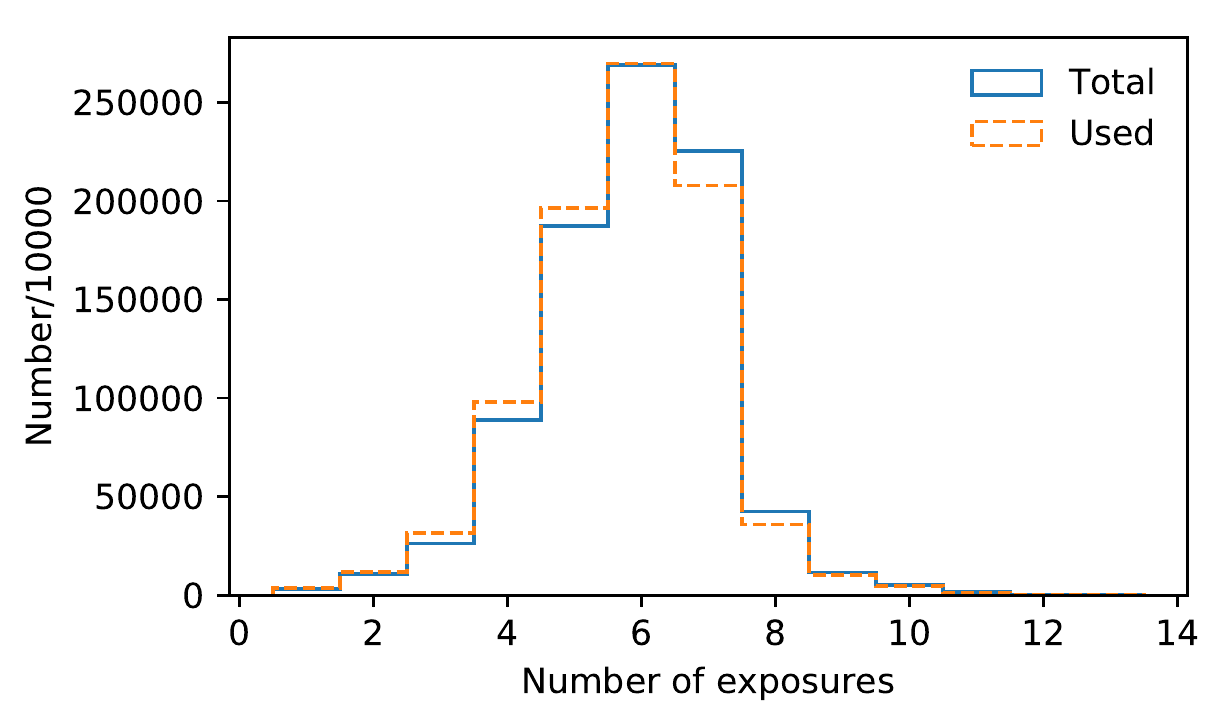}
\end{center}
\caption[]{
The number of simulated exposures per object (blue) compared to the number eventually used in the shape measurement (orange). The median number of exposures per object both simulated and used is six over this part of the five-year simulated Reference Survey.
\label{fig:hist2}}
\end{figure}

We considered a reference HLS observing strategy consisting of 2 passes in each of the 4 HLS imaging filters (plus 4 passes for the grism). To construct each pass, we take a sequence of $n$ exposures ($2\le n\le 4$, depending on the filter/grism choice), with a small diagonal step after each exposure to cover gaps between the SCAs. These steps also ensure that in each of the $n$ exposures, an image of a star or galaxy does not land on the same small chip defect, nor in the same readout channel of an SCA, and does not interact with its persistence image from the previous exposure.\footnote{Because the same set of offsets is used each time we do a field step, it is possible for all $n$ images of galaxy G to land on the $n$ persistence artifacts from a previous star S. We intend to solve this problem by introducing some pseudo-randomness in the diagonal step sizes, but this has not yet been incorporated in survey simulations.} After these exposures, we do a ``field step'' along the short axis of the field\footnote{We choose the short axis for two reasons. First, the slew times are shorter, resulting in a more efficient survey. Second, the ``arced'' layout of the focal plane means that we can make a strip with smoother edges by stepping on the short than the long axis; the resulting strips fit together much better when tiling a curved sky.} ($\sim 0.4$ degrees), and repeat the exposures. This produces a strip of observed sky; strips are tiled to cover a region on the sky. Subsequent strips are observed in opposite directions (i.e., we alternate ``up'' versus ``down''). The HLS is broken down into 8 such regions (plus deep fields), each with its own tiling. The H158 filter exposure sequence that overlaps the patch of sky simulated for this work is shown in Fig. \ref{fig:pointings} and the total number of exposures that overlap each simulated galaxy is shown in Fig. \ref{fig:hist2}.

The two passes over each region of the HLS are on grids that are rolled relative to each other. This strategy increases the number of exposures, and more importantly ensures that astronomical sources observed on one SCA have repeated observations on other SCAs. This is needed for ``ubercalibration'' internal to the HLS (e.g., \citealt{2008ApJ...674.1217P}), and will be helpful in developing a correction in the event that a few SCAs exhibit unusual behaviors (e.g., larger than normal hysteresis).

The overall survey strategy has to schedule each pass over each region, while being consistent with the needs of the other surveys and the observing constraints as well. We developed tools to do this early in \wfirst\ planning, especially since both L2 and geosynchronous orbits were under consideration, with the latter having complex Earth and Moon avoidance constraints \citep{2015arXiv150303757S}. The constraints are much more slowly varying at L2, but we still have the slowly varying Sun avoidance constraint (\wfirst\ observes between 54--126$^\circ$ degrees from the Sun), a roll angle constraint (the observatory can roll up to $\pm 15^\circ$ from the optimal orientation on the solar array; this is very important when attempting to tile a large region of the sky). Moreover there are cutouts for the microlensing seasons and -- during the middle of the reference mission -- a 30-hour supernova observing session every 5 days. This results in the need to cut each pass into shorter segments that can be observed all at once, in sequence. The strategy described here is an output from an update of the code used in \S3.10 of \cite{2015arXiv150303757S}.

\subsection{Simulation implementation for this study}

In this work, we study the impact of how a variety of biases in the PSF model propagate to shape measurement and the weak lensing signal. 
To study this, we produce a set of 13 image simulations that are identical, including noise, modulo a single PSF model change relative to the fiducial simulation in each case. 
The details of these changes and their impacts are described in more detail in Secs.~\ref{sec:results} and~\ref{sec:results2}. 
Shape measurement is then performed on the images with some PSF model bias, but using the fiducial PSF model for convolution in the galaxy shape fitter, to simulate an unknown wavefront error.

Several simplifications are employed relative to the generic synthetic survey generation described in Sec. \ref{stages} to accommodate the computational load of the many realizations of the survey we are producing. 
\begin{list}{$\bullet$}{}
\item We simulate objects in a 2.5$\times$2.5 deg$^2$ patch of the sky.
\item We only simulate pointings targeted for the H158 filter.
Since we are not simulating chromatic effects, the specific filter choice does not make a large difference in our results.
\item We use a lower-resolution version of the PSF, which significantly speeds up the convolution.
The impact of this approximation on the PSF model, in both native and oversampled pixels, can be seen in Fig. \ref{fig:psf}, but is not important for this work.
\item To better isolate the effects of PSF errors, we only utilize the isolated object postage stamps in shape measurement.
\item We do not simulate objects with photometry that would fall outside the fiducial weak lensing selection criteria.
\item We do not implement a shear calibration scheme like metacalibration \citep{SheldonHuff2017}, since we only care about changes to the recovered shape between simulation runs. Work on applying a method like metacalibration to these simulations is ongoing.
\end{list}

\begin{figure*}
\begin{center}
\includegraphics[width=\textwidth]{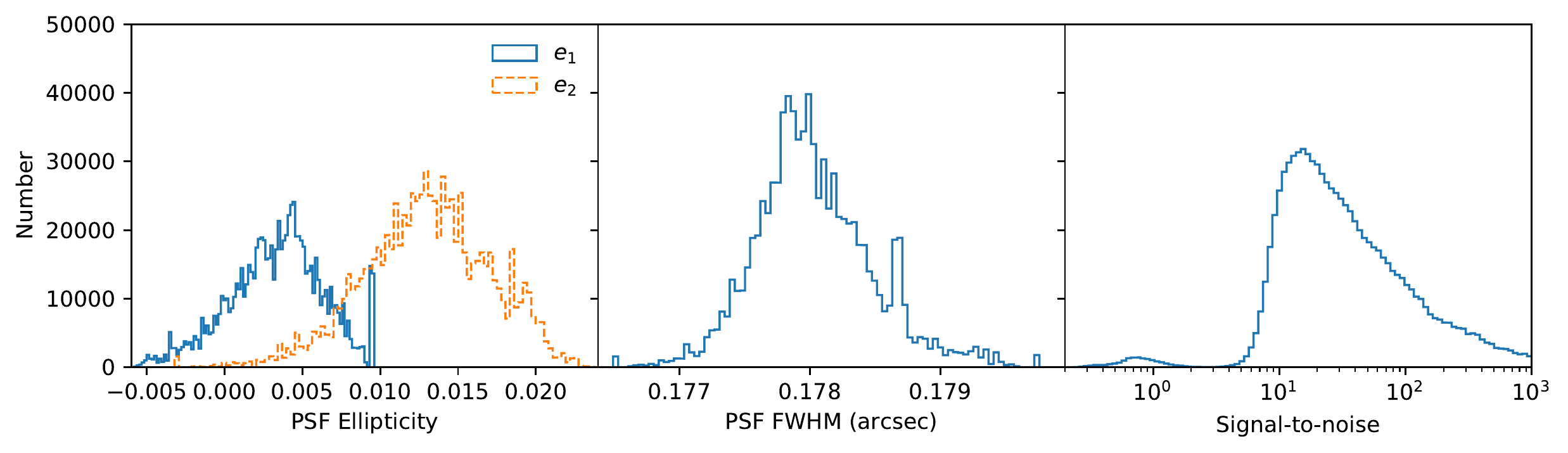}
\end{center}
\caption[]{
The measured PSF ellipticity and size, and the signal-to-noise of the galaxy measurements. The nominal signal-to-noise cut for the \wfirst\ weak lensing sample is 18, which agrees with the peak of the measured distribution. The PSF properties are measured using the oversampled version of the PSF model images (as in Fig. \ref{fig:psf}), which we determined to be the minimum required oversampling in the model to achieve unbiased size and ellipticity measurements. The typical \wfirst\ PSF tends to be more strongly elliptical in the $e_2$ direction, and the median (and mean) recovered FWHM is 0.178 arcsec.
\label{fig:hist3}}
\end{figure*}

We simulate a total of 907,170 unique galaxies and 56,128 unique stars across 189 pointings in each of the runs. The number of exposures per galaxy and the distribution of PSF properties are shown in Figs.~\ref{fig:hist2} and~\ref{fig:hist3}.

\def\arraystretch{1.4}
\setlength{\tabcolsep}{4pt}
\begin{table}
\caption{A summary of the 13 simulation runs.}
\label{table:runs}
\begin{center}
\begin{tabular}{lcccc }
\hline
\hline
Run name & PSF change & Mode & Notes \\ 
\hline
\textsc{Fiducial}              & -- & -- & -- \\
\textsc{Focus}                & $\psi_4$ & Static & --  \\
\textsc{Astig}                  & $\psi_5$ & Static &  -- \\
\textsc{Coma}                & $\psi_7$ & Static &  -- \\
\textsc{GradZ4}             & $\psi_4$ & Static & Gradient in focal plane  \\
\textsc{GradZ6}              & $\psi_6$ & Static &Gradient in focal plane    \\
\textsc{Piston} 		 & $\psi_4$ & Static  &Random per SCA   \\
\textsc{Tilt} 			 & $\psi_4$ & Static & Rand. gradient per SCA \\
\hline
\textsc{IsoJitter}		& Gaussian  & High-Freq. & Isotropic  \\
\textsc{AniJitter} 		 & Gaussian & High-Freq. & Anisotropic  \\
\textsc{RanJitter} 		 & Gaussian & High-Freq. & 15\% of pointings  \\
\hline
\textsc{OscZ4} 		 & $\psi_4$ & Low-Freq. & Time-dependent  \\
\textsc{OscZ7} 		 & $\psi_7$ & Low-Freq. & Time-dependent  \\
\hline
\hline
\end{tabular}
\end{center}
\end{table}

\begin{figure}
\begin{center}
\includegraphics[width=\columnwidth]{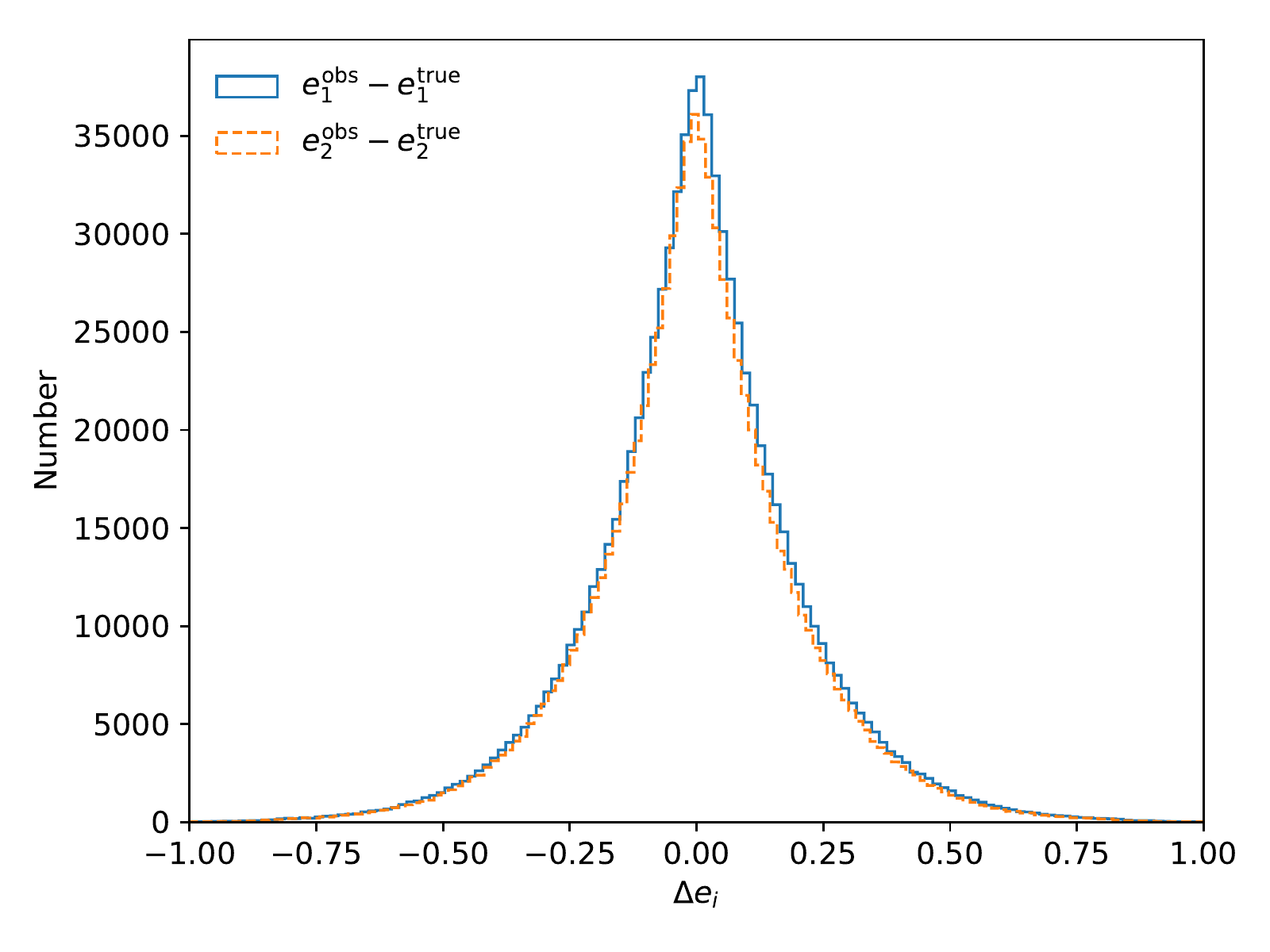}
\end{center}
\caption[]{
The difference in the measured (obs) shape relative to the true sheared intrinsic shape of the galaxies in the fiducial simulation. The inferred multiplicative and additive shear bias is discussed in Sec.~\ref{sec:results2}. 
\label{fig:shape_hist}}
\end{figure}

\section{Wavefront model errors}\label{sec:results}

In this paper, we focus on empirical tests of weak lensing requirements for wavefront model control (i.e., the PSF) in \wfirst. These are used to empirically derive the relationship between recovered shear and wavefront error modes $\partial e_i / \partial \psi_j$, which allows us to validate earlier Phase A analytic estimates of the requirements flowdown for \wfirst. In the absence of shear biases, or when comparing between runs that should have identical intrinsic shear biases, $\partial e_i / \partial \psi_j$ is equivalent to $\partial \gamma_{\mathrm{obs},i} / \partial \psi_j$.

We simulate 13 identical 2.5$\times$2.5 deg$^2$ Reference Survey cutouts: a fiducial survey that represents perfect knowledge of the PSF and 12 iterations to simulate various types of errors in the PSF reconstruction. 
These are split into three types of errors in the wavefront model: 1) static biases in the model, which are constant as a function of time, 2) high-frequency biases in the model, which correspond to rapidly changing conditions compared to the timescale of a single exposure, and 3) low-frequency biases in the model, which change over the lifetime of the mission, but can be considered static over the timescale of a single exposure. 
In each static and low-frequency mode, the (rms) amplitude of the wavefront bias corresponds to 0.005 wavelengths (a fiducial wavelength is taken to be 1293~nm), which is equivalent to approximately 6.5~nm. These PSF changes are summarized in Table~\ref{table:runs}.

\changetext{We emphasize that the purpose of these simulations is to measure the sensitivity (i.e., partial derivatives) of the shear biases $m_i$ and $c_i$ with respect to the PSF parameters. We want to do this with an area that is much less than the Reference Survey area; we choose a change in wavefront $\Delta\psi$ that is considerably greater than the expected requirement so that the partial derivative is not swamped by noise. Similar considerations apply to the pointing jitter.}

\subsection{Static biases}\label{sec:static}

We simulate seven static sources of bias in the PSF model. 
Three of these simulations include a coherent change in the PSF model Zernike coefficients, where the fiducial value is changed by 0.005 wavelengths in each of defocus $\psi_4$ (\textsc{Focus}), oblique astigmatism $\psi_5$ (\textsc{Astig}), and vertical coma $\psi_7$ (\textsc{Coma}). 
Two simulations include a coherent gradient in the defocus $\psi_4$ (\textsc{GradZ4}) and vertical astigmatism $\psi_6$ (\textsc{GradZ6}) across the focal plane with equivalent rms of 0.005 wavelengths. 
For speed, these are simulated such that the PSF is constant within a single SCA. Finally, two simulations approximate errors in the mounting of the SCAs: 1) a random vertical mounting offset of up to 0.005 wavelengths is assigned to each SCA (\textsc{Piston}), and 2) a random tilt in the $x$ or $y$ direction is assigned to each SCA  (\textsc{Tilt}), with equivalent rms of up to 0.005 wavelengths. 
These are modeled as changes in the $\psi_4$ coefficient, with the PSF being evaluated based on the object $x$--$y$ position within the SCA (i.e., each object is assigned a different PSF consistent with this random tilt of the SCA). 
Potential correlated biases in the WCS model due to these changes are ignored in this work, but should be considered in future studies of the WCS model recovery.

\subsection{High-frequency biases}\label{sec:low}

Three high-frequency resonant modes are simulated to represent residual vibrations of the telescope after orienting to a new pointing. 
These are represented by an additional convolution of the image with a Gaussian PSF. We simulate three cases: 1) an isotropic (about the pointing axis) vibration  (\textsc{IsoJitter}), 2) an anisotropic vibration (\textsc{AniJitter}), and 3) only applying this anisotropic vibration to a random 15\% of pointings (\textsc{RanJitter}). 
The additional second moments are conserved, which means $\theta_x\theta_y=\theta_{\mathrm{orig}}^2=15^2$ mas$^2$. 
For \textsc{AniJitter} and \textsc{RanJitter}, we applied a shear $e_1=\frac{\theta_x-\theta_y}{\theta_x+\theta_y}=0.3$,  with Zernike amplitude change in this case  $d\psi=\theta_x^2-\theta_y^2=297$ mas$^2$. 
In the case of \textsc{IsoJitter}, $\theta_x=\theta_y$, which leads to $d\psi=\theta_x^2+\theta_y^2=450$ mas$^2$.

\begin{figure}
\begin{center}
\includegraphics[width=\columnwidth]{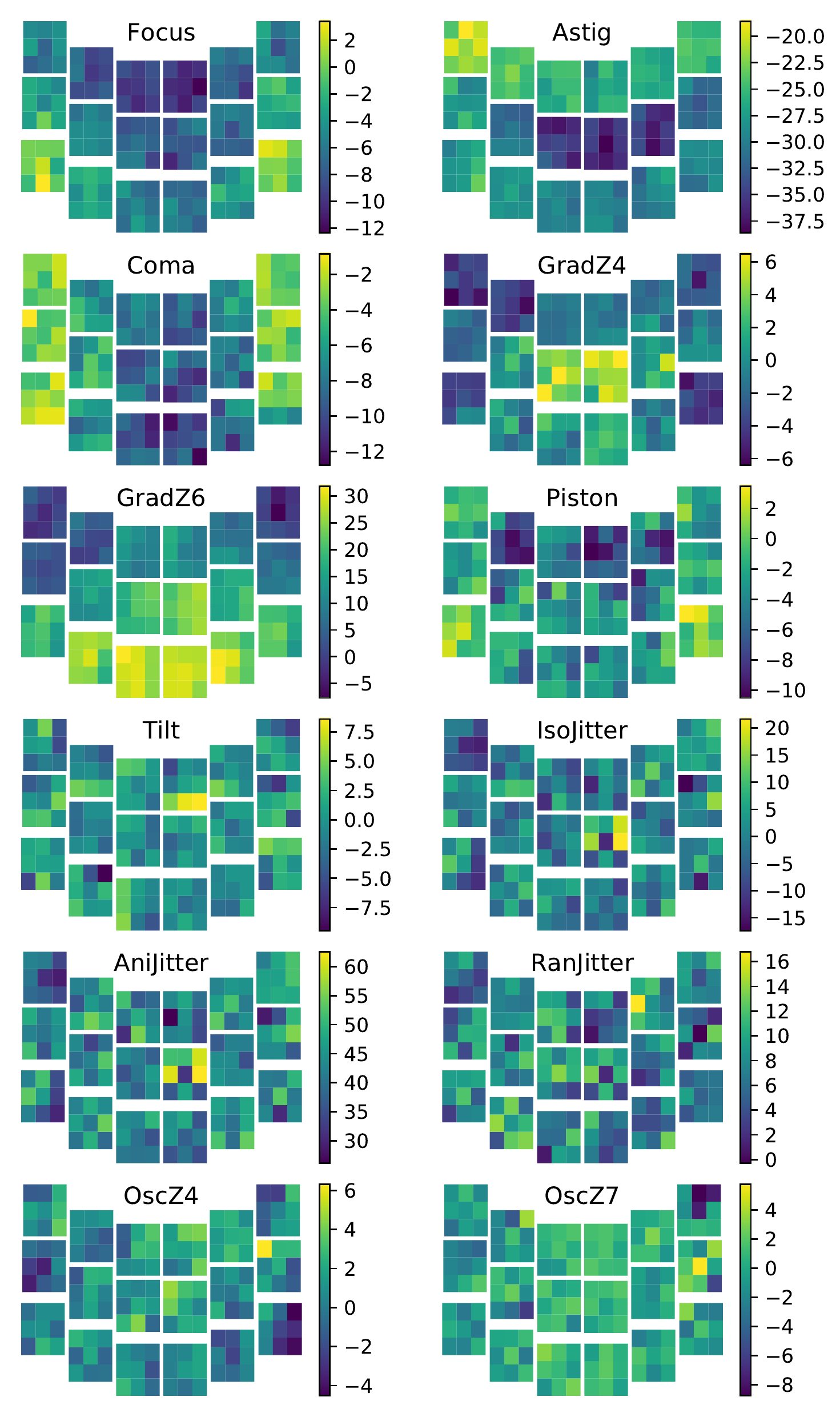}
\end{center}
\caption[]{
The binned mean difference of measured $e_1$ compared to the \textsc{Fiducial} run in the focal plane. Each color bar is in units of $1\times 10^{-4}$. Gradients across the focal plane or chips are visible for all static PSF model biases, while the mean difference is particularly large for the anisotropic jitter case, where the large mean $e_1$ difference corresponds to the direction of anisotropy in the Gaussian smearing.
\label{fig:focal_mean_e1}}
\end{figure}

\begin{figure}
\begin{center}
\includegraphics[width=\columnwidth]{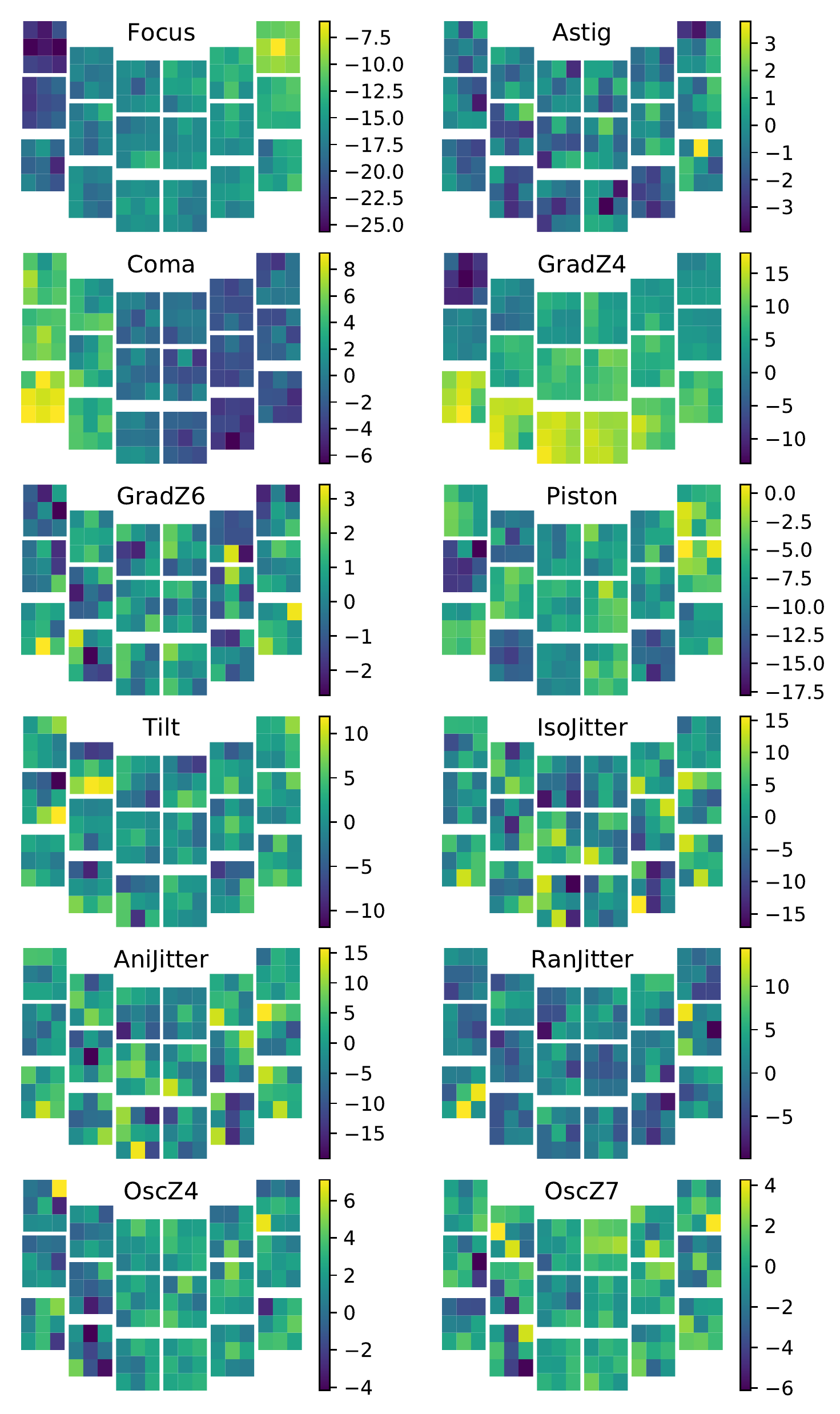}
\end{center}
\caption[]{
The binned mean difference of measured $e_2$ compared to the \textsc{Fiducial} run in the focal plane.  Each color bar is in units of $1\times 10^{-4}$. There are visible gradients for most cases of static PSF model biases.
\label{fig:focal_mean_e2}}
\end{figure}

\subsection{Low-frequency biases}\label{sec:high}

Two low-frequency biases are simulated to represent thermal drift throughout the lifetime of the mission. Thermal perturbations propagate into $\psi_4$ (\textsc{OscZ4})  and $\psi_7$ (\textsc{OscZ7}) modes.
We generate a random time-dependent function $f(t)$ with rms amplitude 0.005 wavelengths following a given power spectrum to quantify the perturbation of the Zernike coefficients over time. 
The power spectrum of thermal drift noise is taken to be a Lorenzian function 
\begin{equation}
P(\nu)=\frac{A}{1+(\nu/\nu_0)^2},
\end{equation}
with normalization factor $A$. The rms variance of $f(t)$ can be expressed as
\begin{equation}
\sigma^2 = \int_{-\infty}^{+\infty}P(\nu)d\nu=\pi A\nu_0,
\end{equation}
which leads to $A=\frac{\sigma^2}{\pi\nu_0}$. $\nu_0=\frac{1}{2\pi\tau}\approx 3.14\times10^{-4}$ Hz, with a time constant $\tau=1$ hr. This timescale is typical of thermal variations that have been seen in integrated modeling \citep[e.g.][]{2015arXiv150303757S}; we plan to use actual integrated modeling outputs for the reference observing scenario in a future version of this study.

\section{Results}\label{sec:results2}

Each simulation is analyzed in an identical way, except that shape measurement for each simulation assumes the \textsc{Fiducial} PSF model is the true model, which simulates the impact of misestimating the PSF model and introduces varying levels of bias. 
All estimates of the multiplicative and additive bias will be explored relative to the \textsc{Fiducial} simulation run, since we have not employed an absolute calibration scheme. 
This is justified to first order, since we are only interested in the relative impacts of the PSF model biases.
We find that 3\% of objects are not included in the shape measurement stage in the \textsc{Fiducial} simulation, due to being too large/bright or because too large a fraction of all cutouts are masked (fall off the edge of an SCA) -- see Sec. \ref{stages} for more details on these selections. 

Since the simulated objects have already been pre-selected as objects that should pass the fiducial \wfirst\ weak lensing selection, we are able to successfully recover a shape fit for more than 99\% of the remaining objects -- a total of 871,841 galaxies. 
We do not make an additional selection on objects that would pass the fiducial \wfirst\ shape selection based on measured properties, since we expect all objects to be within this selection if we were to simulate all remaining pointings in other bandpasses. 
The recovered multiplicative shear bias is only approximately 2\% smaller and the mean shear is unchanged if we make this selection, which removes an additional 35\% of objects, almost exclusively due to the signal-to-noise cut.

We present results for the non-\textsc{Fiducial} simulations only for objects that lie in the intersection of successful shape measurement between each simulation and the \textsc{Fiducial} simulation, to allow for 1-1 comparison of the shapes and cancellation of shape noise and sources of photon noise, which are identical in each simulation. 
We neglect the impact of selection biases here, since the intersection criteria excludes on average only 0.3\% of objects.

\subsection{Summary statistics}

The bias in an ensemble shear measurement is typically characterized in the weak limit by 
\begin{equation}
e_i^{\mathrm{obs}} = (1+m_i) e_i^{\mathrm{true}} + c_i.
\end{equation}
We find the following multiplicative and additive biases in the \textsc{Fiducial} simulation: 
\begin{align*}
m_1 &= (-7.56 \pm 0.19)\times 10^{-2}\\ 
m_2 &= (-9.40 \pm 0.19)\times 10^{-2}\\ 
c_1 &= (1.20 \pm 0.17)\times 10^{-3}\\
c_2 &= (-1.57 \pm 0.16)\times 10^{-3}.
\end{align*}
In some cases, biases are instead parameterized in terms of the PSF leakage as $e_i^{\mathrm{obs}} = (1+m_i) e_i^{\mathrm{true}} + \alpha_i e^{\mathrm{PSF}}_i + c_i$. 
The constraints on $m$ used to interpret requirements in this paper are unchanged in either parameterization. 
The difference in measured shape versus true input shape (intrinsic shape and shear) is shown in Fig.~\ref{fig:shape_hist}.

For each simulation, we compare the recovered shear to the \textsc{Fiducial} simulation in several ways. First, we calculate how the inferred values of $m$ and $c$ change from the \textsc{Fiducial} result, which is shown in Table \ref{table:bias}. More importantly, we are interested in how the inferred shear changes as a function of the induced wavefront error. This allows us to draw a direct connection to the analytic requirements predictions. We show this shear response relative to the wavefront error in Table \ref{table:partials}. Finally, we are ultimately interested in how these biases will propagate to the shear correlation function -- that is, how any coherent scale dependence of the effects will impact cosmology. 

We can study the difference in the measured ellipticity relative to the \textsc{Fiducial} simulation in both focal plane and sky coordinates. 
The mean ellipticity difference binned in the focal plane is shown in Figs.~\ref{fig:focal_mean_e1}  and~\ref{fig:focal_mean_e2}, for $e_1$ and $e_2$, respectively. 
We observe coherent, and sometimes large, biases in the mean ellipticity across the focal plane or individual chips for all static wavefront errors. 
The time-dependent wavefront errors are generally less pronounced, except in the case of a non-random anisotropic jitter which produces a very strong, coherent bias in $e_1$, the direction of the anisotropy in the smearing.

\begin{figure}
\begin{center}
\includegraphics[width=\columnwidth]{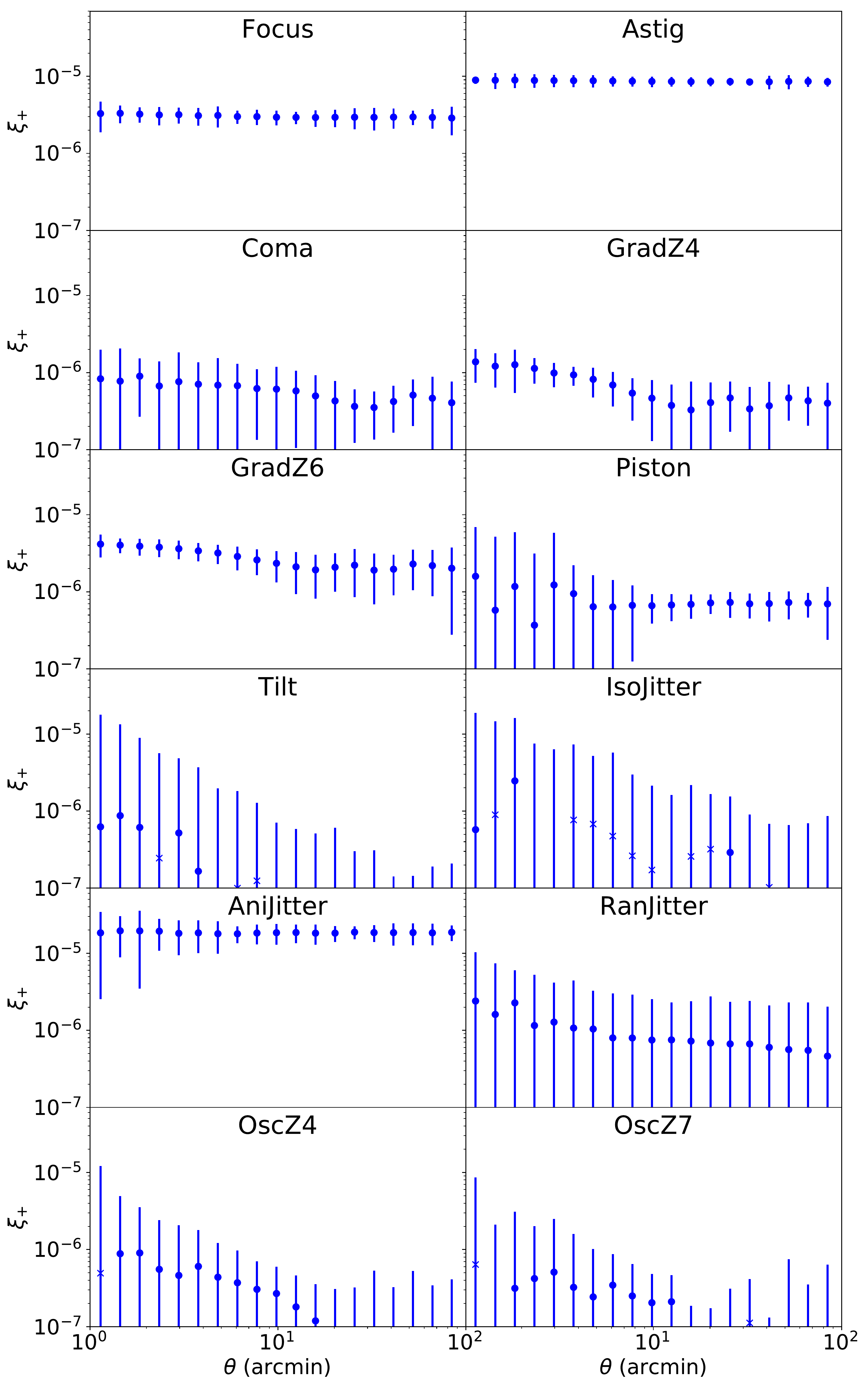}
\end{center}
\caption[]{ The correlation function $\xi_+$ of the ellipticity difference between each run, indicated by the label for each panel, and the \textsc{Fiducial} run. Negative points are shown as crosses. The $\xi_-$ values are all consistent with zero, and are not shown.
\label{fig:f2pt_corrs}}
\end{figure}

Fig.~\ref{fig:f2pt_corrs} shows the two-point correlation function $\xi_{+}$ of the ellipticity difference in sky coordinates, where
\begin{align}
\xi_{\pm} = \langle \Delta e_{+}\Delta e_{+} \rangle \pm \langle \Delta e_{\times}\Delta e_{\times} \rangle.
\end{align}
$\Delta e_{+}$ and $\Delta e_{\times}$ are the tangential and cross components, respectively, of the ellipticity difference relative to the fiducial simulation along the projected separation vector between each pair of galaxies on the sky. 
As expected for a wavefront error, the $\xi_{-}$ correlation of the differences are all consistent with zero. Like the mean shear in the focal plane, all static wavefront error cases lead to significantly non-zero $\xi_{+}$ at varying magnitudes. 
The time-dependent errors have $\xi_{+}$ consistent with zero, except for the non-random anisotropic jitter case, which shows the largest impact in $\xi_{+}$ of any case as additional smear only applies to $e_1$ component. 
The non-random anisotropic jitter, and the static focus and astigmatism errors, all produce a nearly constant $\xi_{+}$ correlation with angular scale, showing that the results are dominated by uniformly distributed $e_1$, $e_2$.

\begin{table}
\caption{Additive and multiplicative bias parameter changes in each version of the simulation relative to the \textsc{Fiducial} simulation.\ \changetext{Note that the wavefront errors injected into these simulations (6.5 nm rms) are much greater than the stability requirements; the variation simulations are designed to measure the partial derivatives of the shear biases with respect to each contribution to the PSF.}\label{table:bias}}
\begin{center}
\resizebox{\columnwidth}{!}{
\begin{tabular}{ lcccc }
\hline
\hline
Run name & $\triangle m_1\times 10^{3}$ & $\triangle m_2\times 10^{3}$ & $\triangle c_1\times 10^{4}$ & $\triangle c_2\times 10^{4}$   \\
\hline
Focus        & $4.65 \pm 0.24 $  &  $ 4.36\pm 0.23 $ &$ -5.57\pm 0.24$ & $-15.87 \pm 0.30$   \\
Astig        & $-0.23 \pm 0.24 $ &  $ 0.90\pm 0.21 $ &$ -29.14 \pm 0.30$ & $-0.55 \pm 0.20$    \\
Coma         & $-1.33 \pm 0.26 $ &  $ -1.47 \pm 0.21 $ &$ -6.66 \pm 0.26$ & $ 0.22 \pm 0.34$   \\
GradZ4       & $-2.63 \pm 0.27 $ &  $ -2.52\pm 0.23 $ &$ -0.51\pm 0.21$ & $ 6.71 \pm 0.42$  \\
GradZ6       & $0.22 \pm 0.23 $  &  $ -0.22\pm 0.23 $ &$ 14.72 \pm 0.62$ & $ 0.44\pm 0.23$ \\
Piston 		 & $-13.6 \pm 5.0 $  &  $ -15.0\pm 5.3 $ &$ -3.70 \pm 0.31$ & $-7.51 \pm 0.34$   \\
Tilt 		 & $-38.0 \pm 8.0 $  &  $ -36.6\pm 8.0 $ &$ -0.40 \pm 0.38$ & $ 0.73 \pm 0.38$   \\
\hline
IsoJitter 	 & $-40.8 \pm 8.6 $  &  $ -39.5\pm 8.5 $ &$ 0.24\pm 0.84$ & $0.43 \pm 0.91$ \\
AniJitter 	 & $-11.4 \pm 1.1 $  &  $ -11.5\pm 1.0 $ &$43.27 \pm 0.85$ & $ -0.13 \pm 0.86$ \\
RanJitter 	 & $-12.8 \pm 4.4 $  &  $ -12.6\pm 4.1 $ &$ 7.50 \pm 0.64$ & $ 0.33 \pm 0.43$   \\
\hline
OscZ4 		 & $-7.2\pm3.1 $     &  $-6.5\pm 3.1$   &$ 0.89\pm0.31$    & $2.15\pm0.49$  \\
OscZ7 	     & $-12.1\pm  6.8$   &  $-11.4\pm6.3$ &$-0.08\pm 0.31$ & $-0.17\pm 0.44$  \\
\hline
\hline
\end{tabular}}
\end{center}
\end{table}

\begin{table}
\caption{The changes of ellipticity with respect to changes in line of sight motion for jitter cases and wavefront error for the other modes.\label{table:partials}}
\begin{center}
\begin{tabular}{ lccc }
\hline
\hline
Run name & $\partial e_1/\partial\psi \times 10^{-4}$ & $\partial e_2/\partial\psi \times 10^{-4}$ & Units \\
\hline
Focus           & $-0.87  \pm 0.032$   & $-2.5  \pm 0.031$ & nm$^{-1}$ \\
Astig           & $-4.5   \pm 0.031$   & $-0.10 \pm 0.030$  & nm$^{-1}$ \\
Coma            & $-1.0   \pm 0.032$   & $0.030 \pm 0.032$  & nm$^{-1}$ \\
GradZ4          & $-0.089 \pm 0.030$   & $1.0   \pm 0.029$  & nm$^{-1}$ \\
GradZ6          & $2.2    \pm 0.029$   & $0.044 \pm 0.028$  & nm$^{-1}$ \\
Piston 		    & $-0.560  \pm 0.048$  & $-1.2  \pm 0.048$  & nm$^{-1}$ \\
Tilt 			& $-0.036 \pm 0.063$   & $0.10  \pm 0.063$  & nm$^{-1}$ \\ \hline
IsoJitter 		& $  0.0004 \pm 0.0024$     & $0.0006  \pm 0.0023$ & mas$^{-2}$ \\
AniJitter 		& $0.145   \pm 0.003$  & $-0.001 \pm 0.003$ & mas$^{-2}$ \\
RanJitter 	    & $0.0246  \pm 0.0016$  & $0.00087  \pm 0.00159$ & mas$^{-2}$ \\ \hline
OscZ4 		    & $0.13   \pm 0.039$   & $0.30  \pm 0.039$  & nm$^{-1}$ \\
OscZ7 		    & $-0.019 \pm 0.045$   & $-0.044\pm 0.044$   & nm$^{-1}$ \\
\hline
\hline
\end{tabular}
\end{center}
\end{table}

\subsection{Comparison of analytic to numerical results}

The partial derivatives $|\partial e/\partial\psi|$ of the ellipticity with respect to wavefront should be less than $\lVert {\boldsymbol\Lambda} \rVert \lVert {\boldsymbol\psi} \rVert$ (where ${\boldsymbol\Lambda}$ is the analytically derived matrix defined in Eq.~\ref{eq:LambdaDef} of Appendix~\ref{app:psf-stability}), which is $8.98\times 10^{-4}\,$nm$^{-1}$ for the H158-band (this is an RSS of the two ellipticity components). For the partial derivatives of the ellipticity with respect to the second moments of the jitter pattern, $|\partial e/\partial\psi|$ should be less than $K_{\theta\theta}$ (where $K_{\theta\theta}$ is the analytically derived sensitivity to jitter; see Eqs.~\ref{eq:K1t},\ref{eq:K2t} of Appendix~\ref{app:psf-stability}), which is $1.38\times 10^{-5}$\,mas$^{-2}$ for the H158-band. In the numerical results presented in Table~\ref{table:partials}, the largest partial derivatives are $4.5\times 10^{-4}\,$nm$^{-1}$ (for wavefront errors) and $1.45\times 10^{-5}$\,mas$^{-2}$ (for jitter). For the wavefront drift case, this is consistent with the analytic expectations. For the anisotropic jitter case, the sensitivity $|\partial e/\partial\psi|$ determined from the simulations is $5\pm 2$\%\ larger than the analytic bound (we would expect a sensitivity equal to the analytic bound since the AniJitter run is the worst-case jitter pattern: the anisotropy is in the $e_1$ component for every exposure). The $5\pm 2$\% difference may simply represent the approximations made in the analytic calculation (e.g., no treatment of undersampling/image combination, a different shape measurement algorithm, etc.).

A similar comparison is possible for the 2-point correlation functions $\xi_\pm(\theta)$ of the ellipticity changes $\Delta e$. Since we put in a wavefront change of 6.5 nm rms, the analytic prediction is that these correlation functions should be
\begin{equation}
\xi_\pm(\theta) \le (8.98\times 10^{-4}\,{\rm nm}^{-1} \times 6.5\,{\rm nm})^2 = 3.4\times 10^{-5}.
\end{equation}
As seen in Fig.~\ref{fig:f2pt_corrs}, this inequality is indeed satisfied.
A similar result can be written for the jitter cases. The most stressing case is the AniJitter case, which has $\langle\theta_x^2 - \theta_y^2\rangle = 297\,$mas$^2$ and hence should satisfy
\begin{equation}
\xi_\pm(\theta) \le (1.38\times 10^{-5}\,{\rm mas}^{-2}\times 297\,{\rm mas}^2)^2
= 1.68\times 10^{-5};
\end{equation}
The AniJitter panel in Fig.~\ref{fig:f2pt_corrs} shows a numerical result that is very close to this. The central values of $\xi_+(\theta)$ range from (1.79--1.95)$\times 10^{-5}$, which are slightly larger than the analytical estimate, although the $1\sigma$ error bars include $1.68\times 10^{-5}$. If this is not a statistical fluctuation, it is likely due to the same simplifying approximations in the analytic calculation as described above for $|\partial e/\partial\psi|$. In either case, the numerical calculation gives a result that is near the analytically estimated upper bound.

\section{Future development plans}\label{sec:future}

The simulation framework described here is a substantial step forward in the development of a significant synthetic \wfirst\ imaging survey, which incorporates a realistic set of photometric properties and distributions of galaxies and stars, complex morphological properties for galaxies, and most known detector non-idealities present in HgCdTe H4RG detectors. 
There are many advances that are still necessary, however, many of which are currently in progress. 
These include increased simulation volume for more precise end-to-end tests, increased fidelity in the simulation of the data accumulation processes within the detector simulation, and more realistic input galaxy and star information. 
Together, they will enable more precise and advanced tests of algorithm development and pipeline integration testing for the \wfirst\ weak lensing program.

In terms of advances in detector physics simulation within \texttt{galsim.wfirst}, \texttt{galsim.wfirst} now includes a model for the persistence effect in the detectors based on measurements from preliminary engineering detectors that will be implemented in future versions of the survey simulation. 
The current simulations do not include an implementation for the brighter-fatter effect. \galsim\ has an implementation of this for silicon CCDs, but not for the HgCdTe detectors used by \wfirst.  
It is possible that using the CCD implementation would be sufficiently accurate for future simulation runs, but this needs to be further investigated. 
We now also have the engineering data to implement measured realizations of the correlated noise fields derived from detector flats and darks. 
One significant difference relative to how data will be taken with the WFIRST SCAs is the lack of `up the ramp' information as charge accumulates within the SCA. 
In practice, we will have access to several linear combinations of intermediate read-outs from the SCAs, which is not currently implemented. 
Other plans for \texttt{galsim.wfirst} are also discussed in Sec.~\ref{sec:sim}. These improvements will enable us to use \galsim\ to update our knowledge requirements for detector effects (beyond the analytic estimates used during the formulation phase of the mission).  

On the mock galaxy catalog side, we have produced test runs where we interface our weak lensing survey simulation pipeline for WFIRST with the existing LSST DESC Data Challenge 2 (DC2) mock galaxy catalog \citep{2019arXiv190706530K}, CosmoDC2, to produce WFIRST imaging over the same simulated universe as is currently being used for DESC image simulations \citep{dc2all}.  
The result of this work will be described in a future paper. CosmoDC2 provides a deeper mock catalog than is currently being used, with synthetic spectra provided for each object to enable fully chromatic studies of weak lensing shape recovery. 
With planned improvements to the recovery of the near-infrared colors for objects in CosmoDC2, this will also enable a powerful joint-simulation with matched imaging as expected for both LSST and \wfirst. 
These matched simulations will enable a range of joint-processing tests at the pixel level to test combinations of ground-based imaging from LSST with space-based imaging from \wfirst.

While the current synthetic survey volume used in this paper is relatively small, due to the necessity of simulating it many times, we plan the production of much larger public simulations in the near future. 
This will include many of the improvements described above, including multi-band imaging across tens of square degrees at full \wfirst\ five-year Reference Survey depth matched to LSST imaging from DESC. 

\section{Conclusion}\label{sec:conclusion}

The \wfirst\ observatory will be an exquisite tool for the study of cosmology using weak gravitational lensing. 
Launching in the mid 2020s, it can harness a unique combination of agility in potential survey design, coupled with a unique range of capabilities and power, to clarify new discoveries and resolve disagreements between the Stage IV surveys that precede it in the early 2020s. 
To ensure we are able to take full advantage of the potential of \wfirst, we must develop the necessary tools to both validate instrument requirements and their flowdown from weak lensing cosmology and to enable pixel-level algorithm development and ultimate integration testing of our measurement pipelines.

In this paper, we have described a simulation framework to produce a realistic, synthetic \wfirst\ imaging survey populated with suitably complex objects that can serve these functions at the current level of necessary realism. 
This framework combines a simulated five-year Reference Survey, an appropriate mock galaxy and star population that would be observed by \wfirst, and a simulation of most relevant properties of the HgCdTe H4RG detectors to be integrated into the \wfirst\ camera. 
We present a set of 13 matched 2.5$\times$2.5 deg$^2$ image simulations to full depth of the reference five-year survey, each with the wavefront model perturbed in some way. These perturbations can be classified into three broad categories: static, high-, and low-frequency.
We study the galaxy shape recovery in these simulations to empirically measure the relative bias in weak gravitational lensing shear estimates due to these errors in wavefront reconstruction, in order to compare to what is anticipated from the analytical requirements flowdown that was previously developed for \wfirst.  

We present quantitative comparisons of the change in the recovered ellipticity due to these various errors in the wavefront model relative to the fiducial simulation. 
These are presented in terms of both mean shear as a function of focal plane position and the correlation function $\xi_{\pm}$ of the ellipticity difference as a function of angular separation on the sky. 
Finally, we derive the response of the change in ellipticity relative to the wavefront model error mode, which we use to evaluate differences relative to previous analytical requirements forecasts.
We find general agreement with the analytic requirements flowdown, though note that the empirical measurement of the bias induced in the non-random anisotropic jitter case is typically larger than predicted by the analytic flowdown. 
We do not consider this to be a significant concern for continued reference to the baseline, analytic requirements flowdown used by the mission, as these differences are at the 1--2$\sigma$ level, depending on the type of comparison, and thus generally consistent with the analytically predicted upper bound of the effect. 

We have outlined in Sec. \ref{sec:future} several future expansions to the validation framework described in this paper for the \wfirst\ weak lensing analysis. 
These include updates to methodology, the incorporation of new flight-candidate detector measurements, and improvements in the fidelity of the image simulations to represent the full range of both properties of objects that will be observed by \wfirst\ and the full range of non-idealities in the detector systems. 
As the \wfirst\ mission approaches its construction phase, we expect these simulations to also begin to play a substantial role as the basis for integration tests of measurement pipeline development over the next several years. 

\section*{Acknowledgements}

 We thank Dave Content, Jeff Kruk, Alice Liu, Hui Kong, and Erin Sheldon for many useful conversations\changetext{, and the anonymous referee for insightful suggestions}. 
This work supported by NASA Grant 15-WFIRST15-0008 as part of the \wfirst\ Cosmology with the High-Latitude Survey Science Investigation Team (\url{https://www.roman-hls-cosmology.space/}). 
It used resources on the CCAPP condo of the Ruby Cluster at the Ohio Supercomputing Center \citep{OhioSupercomputerCenter1987}. This research was also done using resources provided by the Open Science Grid \citep{Pordes2008,Sfiligoi2009}, which is supported by the National Science Foundation and the U.S. Department of Energy's Office of Science. Plots in this manuscript were produced partly with \textsc{Matplotlib} \citep{Hunter:2007}, and it has been prepared using NASA's Astrophysics Data System Bibliographic Services.

\section*{Data Availability}
The data underlying this article and further supporting documentation will be shared upon request to the corresponding author. Details are provided in App. \ref{app:data_access}.

\bibliographystyle{mnras}
\bibliography{short}

\appendix

\section{Overview of weak lensing systematics budgeting}
\label{app:wl-budget}

This appendix describes the requirements flowdown and error budgeting
for the weak lensing program on the \wfirst\ mission, and documents the
detailed rationale behind the summary requirements listed in the
\wfirst\ SRD. This kind of error budgeting has been performed elsewhere
in the literature \citep{2008A&A...484...67P,2013MNRAS.429..661M}, but
this document focuses on the error terms relevant to \wfirst. For
example, the PSFs are based on an obstructed pupil with low-order
aberrations rather than using generic formulae involving second
moments (some such formulae, including those used in the Joint Dark Energy Mission and
WFIRST Interim Design Reference Mission studies, were for Gaussians).

We set most systematics requirements for this mission on the basis of
having systematic errors sub-dominant to statistical errors in the weak lensing
shear power spectra or cross-power spectra (or any linear combinations
thereof). Exceptions to this policy are considered in cases
where meeting the original systematic budget becomes a cost or
complexity driver, or is not possible. Most measurement biases -- including those considered in this paper -- fall
into the ``additive'' or ``multiplicative'' forms (see
\S\ref{ss:add_mult}) and will be treated according to the formalism
therein.

\subsection{Additive and multiplicative biases}
\label{ss:add_mult}

The cosmic shear measurement is sensitive to two major types of
measurement errors. {\em Additive bias} or ``spurious shear'' $c$ is a
shear signal that is detected even when none is present. {\em
Multiplicative bias} or ``calibration bias'' $m$ is an incorrect response
to a real shear, e.g.\ a shear $\gamma$ is present in the sky but the
measurement yields 1.01$\gamma$. Normally, we think of additive biases
as resulting from mis-estimation of the PSF ellipticity (or its
variation across the sky), whereas multiplicative biases result from
mis-estimation of the size of the PSF. However, detector
nonlinearities, approximations used in the data processing/analysis
pipelines, and uncertainties about the distribution of galaxy
morphologies in the sky can also contribute to both types of
biases.  The $E$-mode
shear cross-power spectrum between two redshift bins $z_i$ and $z_j$
is modified in the presence of these biases:
\begin{equation}
C_\ell^{z_i,z_j}({\rm obs}) = (1+m_i)(1+m_j)C_\ell^{z_i,z_j}({\rm true}) + C_\ell^{c_i,c_j},
\label{eq:mod}
\end{equation}
where we write $m_i\equiv m(z_i)$ as a shorthand for the bias in bin $i$. To linear order in
the biases, the correction to the power spectrum can be written as
\begin{align}
\Delta C_\ell^{z_i,z_j} &= C_\ell^{z_i,z_j}({\rm obs}) - C_\ell^{z_i,z_j}({\rm true}) \nonumber\\
&= (m_i+m_j)C_\ell^{z_i,z_j}+C_\ell^{c_i,c_j}.
\label{eq:Delta C}
\end{align}
\changetext{If $m$ is spatially variable, there is an additional contribution \citep[e.g.][]{2012MNRAS.423.3163K, 2016MNRAS.455.3319K}:}
\begin{align}
\Delta C_\ell^{z_i,z_j} &=
\int \frac{d^2{\boldsymbol\ell}'}{(2\pi)^2} \, C_{\ell'}^{z_i,z_j} C^{\delta m_i,\delta m_j}_{{\boldsymbol\ell}-{\boldsymbol\ell}'} \cos^2 (2\varphi_{{\boldsymbol\ell},{\boldsymbol\ell}'}),
\label{eq:DeltaC-mm}
\end{align}
\changetext{where $\delta m_i$ is the fluctuation in multiplicative bias of bin $i$, and $\varphi_{{\boldsymbol\ell},{\boldsymbol\ell}'}$ is the angle between the indicated wave vectors. This is second order in the $m$-biases, so we expect it to be small compared to the first order contribution in Eq.~(\ref{eq:Delta C}), which comes from the spatially averaged part of the multiplicative bias. We will briefly discuss this spatially variable contribution again in \S\ref{ss:implement-mult}.}

\subsection{Setting requirements}

The power spectra are arranged into a vector ${\bf C}$ with
a covariance matrix ${\bf\Sigma}$. For the weak lensing power spectrum, with
$N_z$ redshift bins and $N_\ell$ angular scale bins, there are $N_\ell
N_z(N_z+1)/2$ power spectra $C_\ell^{z_i,z_j}$; hence ${\bf C}$ is a
vector of length $N_\ell N_z(N_z+1)/2$, and ${\bf\Sigma}$ is a matrix
of size $N_\ell N_z(N_z+1)/2 \times N_\ell N_z(N_z+1)/2$. A
contaminant that changes the power spectrum by $\Delta {\bf C}$ can
have its significance assessed by
\begin{equation}
Z = \sqrt{\Delta{\bf C}\cdot{\bf\Sigma}^{-1}\Delta{\bf C}},
\label{eq:alpha}
\end{equation}
which is the number of $\sigma$s at which one could distinguish the
correct power spectrum from the contaminated power spectrum. Note that
as the survey area $\Omega$ is increased, $Z$ will increase as
$\propto\Omega^{1/2}$, and hence contaminants $\Delta{\bf C}$ must be
reduced to keep them below statistical errors. If $Z=1$, then the
power spectrum is biased at the same level as the statistical
errors. We use $Z$ as a metric for contaminants, rather than
e.g. biases in $(w_0,w_a)$-space, for generality: if $Z<1$ then the
bias due to $\Delta{\bf C}$ in {\em any} cosmological parameter from
the combination of the \wfirst\ weak lensing power spectrum with {\em any} other
data set(s) from \wfirst\ or other experiments is $<1\sigma$; whereas if
one based the analysis on biases in $(w_0,w_a)$ then we would need a
separate requirement derived from every cosmological analysis planned
on \wfirst\ weak lensing data. Using $Z$ as a metric also enables us to write
requirements that do not depend on other cosmological probes (e.g.\
the \wfirst\ weak lensing systematic error budget does not change if we discover a
new way to reduce the scatter in the SN Ia Hubble diagram), which will
help to ensure the stability of our requirements going forward.

Technically the above discussion applies only to the $E$-mode of
spurious shear; we have not set a specific requirement on the
$B$-mode, which contains no cosmological information to linear order
and is used as a null test. For the latter reason, we set a
requirement on the $B$-mode that is equal to the requirement on the
$E$-mode, so that the $B$-mode null test will pass if requirements are
met. We also note that the weak lensing analysis includes a range of angular
scales, $\ell_{\rm min,tot}\le \ell \le \ell_{\rm max,tot}$;
requirements apply to sources of systematic error that affect these
scales, i.e.\ are ``in-band'' for the weak lensing measurement. The ``in-band''
qualifier is critical: as an example, pixelization errors can cause
shape measurement errors in galaxies that depend on whether the galaxy
lands on a pixel center, corner, vertical edge, or horizontal
edge. For some shape measurement methods, this error may dramatically
exceed the additive systematic error budget, but it is concentrated at
very small angular scales (multiples of $2\pi$ divided by the pixel
scale $P$, or $2\pi/P = 1.2\times 10^7$). Our requirements are set on
the portion of this power that is within (or mixes into) the band
limit, $\ell\le\ell_{\rm max,tot}$ due to e.g.\ edge effects,
selection effects, etc.

Equation~(\ref{eq:alpha}) still does not completely define a
requirement, since we have not described the redshift or scale
dependence of the spurious shear in question. Neither dependence is
expected to be trivial: errors in PSF models have a greater impact on
shape measurements for higher redshift galaxies, since they tend to be
smaller; and the angular power spectrum of PSF model errors should be
non-white in a survey strategy that ``marches'' across the sky, even
if heavily cross-linked (there may also be a characteristic scale at
the size of the field; for example, a repeating error at the $\sim
0.8\times0.4^\circ$ size of the \wfirst\ field has reciprocal lattice
frequencies at $\ell = 450$ and 900, so a large scale error in the
instrument PSF model that is ``tessellated'' as we tile the sky will
appear at these frequencies or multiples thereof). At first, we
considered assuming a particular scale and redshift dependence for the
errors, but in order to be conservative we would have to assume the
worst combination of angular and redshift dependences. Many of our
large sources of systematic error, such as PSF ellipticity due to
astigmatism, have predictable dependences (e.g.\ the systematic error
induced in galaxy shears is of the same sign in all redshift bins)
that are far from the worst case, and this could lead to
over-conservatism in the requirements. Therefore we need a more
nuanced approach to the requirements, where the allowed amplitude of
each term in the error budget is informed by the structure of the
correlations it produces.

Our approach to this problem is to write a script that accepts a
specific angular and redshift dependence (``template'') for a
systematic error, and returns the amplitude $A_0$ of the systematic
error at which we would have $Z=1$ (i.e.\ a $1\sigma$ bias on the
most-contaminated direction in power spectrum space). For cases where
the template is not known (or where we have not done the analysis),
the script is capable of searching the space of templates and finding
the most conservative choice, i.e.\ the choice that leads to the
smallest value of $A_0$. The combined results enable us to build an
error tree, where the overall top-level systematics requirement (a
limit on $Z$) can be flowed down to upper limits on each source of
systematic error. Finally, some portions of the systematic error
budget sum in quadrature (``root-sum-square'' or RSS addition) and
others linearly; in this document, we carefully account for which is
which.

\subsubsection{Data vector and covariance model}

We build our data vector for the shear power spectra and cross-spectra. We recognize that weak lensing analyses have shifted to ``$3\times 2$-point'' data vectors containing shear-shear, galaxy-shear, and galaxy-galaxy correlations, and by the time of \wfirst\ the list of standard observables may be even longer. However, for setting requirements on shape measurement, shear-shear provides the most demanding use case, and so for simplicity here we only consider shear-shear.

We use for our data vector the $N_\ell N_z(N_z+1)/2$ power spectra and
cross-spectra. Each $\ell$ is treated separately, so there are
$N_\ell=\ell_{\rm max,tot}-\ell_{\rm min,tot}$ angular bins; we use
$\ell_{\rm min,tot}=10$ and $\ell_{\rm max,tot}=3161$, thereby
covering 2.5 orders of magnitude in scale. \wfirst\ provides little
cosmological constraining power at the larger scales due both to the
finite size of its survey and due to the large cosmic variance of the
lowest multipoles. The smallest scales are generally not used in
cosmic shear analyses because the baryonic effects are severe (e.g.\
\cite{2008PhRvD..77d3507Z, 2013PhRvD..87d3509Z}). We use $N_z=15$
redshift slices, as shown in Table~\ref{tab:dNdz}, which are chosen by the Exposure Time Calculator \citep{2012arXiv1204.5151H} v17, with the Phase B exposure times ($5\times 140.25$ s in H158-band). In order to ensure
that \wfirst\ would not become systematics-limited in an extended
mission, we set the top-level requirement on systematics to $Z=1$ for
a survey of area $\Omega =10^4$ deg$^2$ (3.05 sr).

The power spectra were obtained from {\sc Class}
\citep{2011JCAP...07..034B} using the fiducial cosmology from the
{\slshape Planck} 2015 ``TT,TE,EE+lowP+lensing+ext'' results
\citep{2016A&A...594A..13P}. The shape noise contribution was added to
construct ${\bf C}^{\rm tot}$ according to
\begin{equation}
C^{{\rm tot},z_i,z_j}_\ell = C^{z_i,z_j}_\ell + \frac{\gamma_{\rm rms}^2}{\bar n_i}\delta_{ij},
\end{equation}
where $\bar n_i$ is the mean effective number density in galaxies per
steradian in redshift slice $i$, and $\gamma_{\rm rms}$ is the shape
noise expressed as an equivalent RMS shear per component; we take
$\gamma_{\rm rms} = 0.22$.

We approximate ${\boldsymbol\Sigma}$ using the usual Gaussian
covariance matrix formula,
\begin{align}
&\Sigma[C^{z_i,z_j}_\ell, C^{z_k,z_m}_{\ell'}]= \nonumber\\
& \frac{\delta_{\ell\ell'} [ C^{{\rm tot},z_i,z_k}_\ell C^{{\rm tot},z_j,z_m}_\ell + C^{{\rm tot},z_i,z_m}_\ell C^{{\rm tot},z_j,z_k}_\ell]}{(2\ell+1)f_{\rm sky}} ,
\label{eq:SGauss}
\end{align}
where $f_{\rm sky} = \Omega/(4\pi)$.  The non-Gaussian contributions
to the error covariance matrix are turned off, because since the
FoMSWG \citep{2009arXiv0901.0721A} there has been an ongoing program of
using nonlinear transformations on the data to remove them (e.g.\
\cite{2009ApJ...698L..90N, 2011ApJ...729L..11S}) and
we do not want applications of these novel statistics to \wfirst\ data
to run into systematic error limits. We also do not include astrophysical systematic errors in ${\boldsymbol\Sigma}$; we envision instead that they will be treated with nuisance parameters in the analysis. Another advantage of this is that the covariance matrix ${\boldsymbol\Sigma}$ is block diagonal in $\ell$-space (it is formally $375720\times375720$ without $\ell$-binning), which makes computations possible on a machine with limited memory. Indeed, in the Gaussian case one may write
\begin{align}
&\Delta{\bf C} \cdot {\boldsymbol\Sigma}^{-1} \Delta {\bf C}=\nonumber\\
& \sum_\ell \frac{(2\ell+1)f_{\rm sky}}2 \sum_{ijkm}
 \Delta C_\ell^{ij} [C^{{\rm tot}-1}_\ell]_{jk}
 \Delta C_\ell^{km} [C^{{\rm tot}-1}_\ell]_{mi},
\label{eq:sprodform}
\end{align}
where the matrix inverses are $N_z\times N_z$.

\begin{table}
\caption{The effective number density in each redshift bin, in units of galaxies/arcmin$^{2}$, used for setting requirements. These are {\em per bin}, i.e.\ are $dn_{\rm eff}/dz\times\Delta z$. 
\label{tab:dNdz}}
\center{
\begin{tabular}{cc|cc|cc}
\hline\hline
$z$ & $n_{\rm eff}$ & $z$ & $n_{\rm eff}$ & $z$ & $n_{\rm eff}$ \\
\hline
$0.10\pm0.10$ & 3.62 & $1.10\pm0.10$ & 3.75 & $2.10\pm0.10$ & 1.21 \\
$0.30\pm0.10$ & 2.12 & $1.30\pm0.10$ & 3.17 & $2.30\pm0.10$ & 0.95 \\
$0.50\pm0.10$ & 3.05 & $1.50\pm0.10$ & 2.52 & $2.50\pm0.10$ & 0.82 \\
$0.70\pm0.10$ & 5.90 & $1.70\pm0.10$ & 1.45 & $2.70\pm0.10$ & 0.68 \\
$0.90\pm0.10$ & 2.79 & $1.90\pm0.10$ & 1.68 & $2.90\pm0.10$ & 0.19 \\
\hline\hline
\end{tabular}
}
\end{table}

\subsubsection{Implementation: additive systematics}
\label{ss:implement-add}

\begin{table*}
\caption{The requirements for additive and multiplicative systematic errors. There are $N_{\rm band}=4$ additive error bands ranging over a total signal band from $\ell_{\rm min,tot}=10$ to $\ell_{\rm max,tot}=3161$. The fraction of the error budget allocated to each band is also indicated, as are the maximum allowed redshift-independent spurious shear ($A_0^{\rm flat}(\alpha)$, RMS per component), and the maximum scaling factors for redshift dependence, $S_{\rm max,\pm}(\alpha)$ and $S_{\rm max,+}(\alpha)$. There is only one row for the multiplicative errors, since the implementation does not contain an $\ell$ dependence; we quote a requirement on the post-calibration shear multiplicative uncertainty $\sigma_{m,\rm req't}^{\rm flat}$.
\label{tab:addbands}}
\center{
\begin{tabular}{r|rrc|c|rr}
\hline\hline
Band $\alpha$ & $\ell_{\rm min}(\alpha)$ & $\ell_{\rm max}(\alpha)$ & Allocation $Z(\alpha)$ & Sys. err. req't. & $S_{\rm max,\pm}(\alpha)$ & $S_{\rm max,+}(\alpha)$ \\ 
 & & & & $A_0^{\rm flat}(\alpha)$ or $\sigma_{m,\rm req't}^{\rm flat}$ & & \\
\hline
\multicolumn7c{Additive errors} \\
0 & 31 & 99 & 0.2596 & $7.000\times 10^{-5}$ &  8.489 &  2.782 \\
1 & 100 & 315 & 0.2539 & $9.900\times 10^{-5}$ &  5.628 &  2.041 \\
2 & 316 & 999 & 0.2575 & $1.400\times 10^{-4}$ &  3.569 &  1.509 \\
3 & 1000 & 3161 & 0.2538 & $1.900\times 10^{-4}$ &  2.119 &  1.149 \\
\hline \multicolumn7c{Multiplicative errors} \\
mult &  & & 0.4600 & $3.200\times 10^{-4}$ & 2.186 & 1.140 \\ \hline\hline
\end{tabular}
}
\end{table*}

Each additive systematic error is taken to have an angular dependence
given by some template $T_\ell$, and a redshift dependence given by a
set of weights $w_i=w(z_i)$. That is, there is a reference additive
shear $c_{\rm ref}$, with the additive shear in redshift bin $i$ given
by $c(z_i) = w_ic_{\rm ref}$. For example, a systematic error
independent of redshift bin would be specified with $w_i=1$ for all
$i$. The reference signal is taken to have a power spectrum
proportional to the template: $C_\ell^{c_{\rm ref}} = A_0^2T_\ell$,
and the template is normalized so that $c_{\rm ref}$ has variance 1
per component (from in-band fluctuations):
\begin{equation}
\sum_{\ell=\ell_{\rm min,tot}}^{\ell_{\rm max,tot}} \frac{2\ell+1}{4\pi} T_\ell = 1.
\label{eq:Tlsum}
\end{equation}
The additive cross-power spectrum is then
\begin{equation}
C_\ell^{c_i,c_j} = A_0^2 w_iw_jT_\ell,
\end{equation}
and the total RMS per component of the spurious shear in bin $i$ is $A_0|w_i|$.

The additive systematic errors can have various scale dependences. We
therefore consider a suite of $N_{\rm band}$ disjoint angular
templates that cover the shape measurement band. Each template
satisfies the normalization rule, Eq.~(\ref{eq:Tlsum}), and has
$\ell(\ell+1)T_\ell/(2\pi)=$ constant:
\begin{align}
T_\ell^{(\alpha)} = &\left[ \sum_{\ell'=\ell_{\rm min}(\alpha)}^{\ell_{\rm max}(\alpha)} \frac{2\ell'+1}{\ell'(\ell'+1)} \right]^{-1} \frac{4\pi}{\ell(\ell+1)}\nonumber\\
&\times\left\{ \begin{array}{cc} 1 & \ell_{\rm min}(\alpha)\le\ell\le\ell_{\rm max}(\alpha) \\ 0 & {\rm otherwise} \end{array} \right.\!,~\alpha=0,...N_{\rm band}-1.
\end{align}
The current bands are displayed in Table~\ref{tab:addbands}. Each band
$\alpha$ is allowed a contribution to the total error
$Z(\alpha)$. Since there are no statistical correlations between
different $\ell$s in the covariance matrix ${\bf\Sigma}$, the
$Z(\alpha)$ can be quadrature-summed (see
Eq.~\ref{eq:alpha}). However, additive systematic error is positive in
the sense that it adds rather than subtracts power; thus the power
spectrum error vectors $\Delta{\bf C}$ from two sources of additive
systematic error contributing to the same angular bin are not
orthogonal and the $Z$'s should be added linearly. Another way to
think of this is that since $Z$ is proportional to the square of the
RMS shear, $Z\propto A_0^2$, quadrature-summation of the additive
shear is equivalent to linear summation of the $Z$-values.

The allocations for each bin $Z(\alpha)$ were initially set to $\sqrt{0.25/N_{\rm band}}$, so that in an RSS sense 25\%\ of the systematic error budget is allocated to additive shear; with 4 bands this implies $Z(\alpha)=0.25$ (i.e., 6.25\% of the error budget in an RSS sense) for each band. There have been some updates of the exposure times, throughputs, and number densities since the SRD requirements were set (December 2016); we have kept the requirements the same, and updated the $Z$-values, so the latter do not exactly equal 0.25.

The construction of $Z$-values for each angular band and each additive
systematic is mathematically sufficient to build the error
budget. However, they can be difficult to conceptualize. Therefore, we
introduce some equivalent notation to describe the weak lensing error
budget. For each angular template, we introduce a limiting amplitude
$A_0^{\rm flat}(\alpha)$, defined to be the amplitude $A_0$ at which
we would saturate the requirement on $Z(\alpha)$ for bin $\alpha$ in
the case of a redshift-independent systematic $w_i=1\,\forall i$. That
is, if the additive systematics did not depend on redshift, we could
tolerate a total additive systematic shear of $A_0^{\rm flat}$ (RMS
per component) in band $\alpha$. We also introduce a scaling factor
$S[{\bf w},\alpha]$ for a systematic error
\begin{equation}
S[{\bf w},\alpha] = \frac{Z(\alpha) \,{\rm for\,this\,}w_i}{Z(\alpha)\,{\rm for\,all\,}w_i=1}
\label{eq:Swa}
\end{equation}
that depends on the redshift dependence $w_i$. An additive systematic error that is independent of redshift will have $S=1$. A systematic that is ``made worse'' by its redshift dependence will have $S>1$, and a systematic that is ``made less serious'' by its redshift dependence will have $S<1$. The requirement that the (linear) sum of $Z$s not exceed $Z(\alpha)$ thus translates into
\begin{equation}
\sum_{\rm systematics} [A(\alpha)]^2\times S[{\bf w},\alpha] \le [A_0^{\rm flat}(\alpha)]^2,
\label{eq:A-S-sum}
\end{equation}
where $A(\alpha)$ is the RMS additive shear per component due to that systematic.

In most cases, we will take the ``reference'' additive shear to be the
additive shear in the most contaminated redshift slice; in this case,
$w_i=1$ for that slice, and $|w_i|\le 1$ for the others. Under such
circumstances, we can determine a {\em worst-case scaling factor}
$S_{\rm max,\pm}(\alpha)$, which is the largest value of $S[{\bf
w},\alpha]$ for any weights satisfying the above inequality. We may
also determine a worst-case scaling factor $S_{\rm max,+}(\alpha)$
conditioned on $0\le w_i\le 1$, i.e.\ for sources of additive shear
that have the same sign in all redshift bins. The search within these
spaces is simplified by the fact that -- according to
Eq.~(\ref{eq:sprodform}) -- the contribution to $S[{\bf w},\alpha]$
considering only a single value of $\ell$ reduces to a
semi-positive-definite quadratic function of ${\bf w}$ (it is proportional to ${\bf w}^{\rm T}{\bf C}_\ell^{{\rm tot}-1}{\bf w}$, where ${\bf C}_\ell^{\rm tot}$ is $N_z\times N_z$). Therefore the
worst-case weights $\{w_i\}_{i=1}^{N_z}$ always occur at the corners
of the allowed cube in $N_z$-dimensional ${\bf w}$-space, and we can
simply search the $2^{N_z}$ corners by brute force.

\subsubsection{Implementation: multiplicative systematics}
\label{ss:implement-mult}

The implementation for \changetext{the spatial mean part of the} multiplicative systematic errors is simpler, since one can work directly with the $m_i$. Once again, we may write $m_i=mw_i$, where $m$ is the multiplicative error in the worst bin (largest absolute value) and $w_i=1$ for that bin, and $|w_i|\le 1$ for all bins; the $w_i$ thus represents the redshift dependence of the multiplicative error. Once again, we may define a scaling factor $S[{\bf w},{\rm mult}]$ for multiplicative biases analogous to Eq.~(\ref{eq:Swa}):
\begin{equation}
S[{\bf w},{\rm mult}] = \frac{Z^2({\rm mult}) \,{\rm for\,this\,}w_i}{Z^2({\rm mult})\,{\rm for\,all\,}w_i=1};
\end{equation}
this time, we define this with the $Z^2$ rather than $Z$ so that RSS addition will apply to independent multiplicative errors:
\begin{equation}
\sum_{\rm systematics} \sigma_m^2 \times S[{\bf w},{\rm mult}] \le [\sigma_{m,\rm req't}^{\rm flat}]^2,
\end{equation}
where $\sigma_{m,\rm req't}^{\rm flat}$ is the requirement on knowledge of $m$. Fundamentally, the square present here but not in Eq.~(\ref{eq:Swa}) arises because multiplicative biases in the power spectrum are proportional to $m$ but additive biases in the power spectrum are proportional to $c^2$.

The worst-case scaling factors $S_{\rm max,+}({\rm mult})$ (conditioned on $0\le w_i\le 1$) and $S_{\rm max,\pm}({\rm mult})$ (allowing either sign) can be defined analogously. In this case, since $\Delta{\bf C}$ is linear in the $m_i$ and hence $w_i$, it is actually $S^2[{\bf w},{\rm mult}]$ that is a semi-positive-definite quadratic function of ${\bf w}$ instead of $S$, but the technique of searching the corners by brute force still applies.

Once again, we initially set $Z({\rm mult})=0.5$, allocating 25\% of the systematic error in an RSS sense to multiplicative systematics; due to changes in the model since the SRD was first written, the allocation is no longer exactly 25\%. The resulting limits are quoted in Table~\ref{tab:addbands}.

\changetext{One may also consider the spatially varying multiplicative systematics, which contribute to the power spectrum via Eq.~(\ref{eq:DeltaC-mm}). As noted in \S\ref{ss:add_mult}, we expected this contribution to be small. As a simple test, we tried assessing a redshift-independent multiplicative bias, with the scale dependence $\ell(\ell+1)C_\ell^{mm}= $\,constant over the range $8\le \ell < 1442$ (i.e., from one wavelength over the survey region to a half-wavelength across an SCA), and with total variance $\sigma^2_m = \sum_\ell (2\ell+1)C_\ell^{mm}/(4\pi)$. The resulting contribution to the error budget is $Z^2 = 0.016(\sigma_m/0.01)^4$. \changetexttwo{We also tried using a Kronecker delta scale dependence, $C_\ell^{mm} = 4\pi \sigma_m^2 \delta_{\ell\ell_0}/(2\ell+1)$; by searching over all the $\ell_0$ in the range from $8\le \ell_0<1442$, we found the worst contribution to the error budget to be $Z^2 = 0.020(\sigma_m/0.01)^4$ if $\ell_0=16$. This represents an upper bound to $Z^2$ for a given $\sigma_m$.\footnote{To see why, let $\xi_\ell = (2\ell+1)C_\ell^{mm}/(4\pi\sigma_m^2)$ so that $\xi_\ell\ge 0$ and $\sum_\ell \xi_\ell = 1$. Then since $Z^2$ is a positive semi-definite function of the $C_\ell^{mm}$, it is a convex function of the $C_\ell^{mm}$, and Jensen's inequality \citep[\S3.1.8]{boyd2004} shows that $Z^2 \le \sum_\ell \xi_\ell Z^2_\ell$, where $Z^2_{\ell_0}$ is the value of $Z^2$ with the Kronecker delta scale dependence, $\xi_\ell = \delta_{\ell\ell_0}$. It follows that $Z^2$ is less than or equal to the maximum of the $Z^2_{\ell_0}$.}} It is thus clear that the requirements on spatially varying multiplicative systematics will be much looser than those on the spatially varying additive systematics, and hence it is the latter that will drive PSF and wavefront stability requirements. The allocation in terms of acceptable $Z^2$ for spatially varying multiplicative systematics the overall \wfirst\ error budget is under discussion.\footnote{We thank the anonymous referee for encouraging us to think about this more carefully.}}

\subsection{Flow-down to PSF requirements}

In order to translate a requirement on additive bias $c$ or multiplicative bias $m$ into requirements on lower-level quantities, we need to know how a given effect -- e.g.\ an error in the PSF model -- affects the shear measurement. We focus here on the additive biases, which are of interest for this paper; the multiplicative biases can be treated in the same formalism and we comment on how to do this at the end.

We need to compute $\partial\gamma_{\rm obs}(z_i)/\partial X$, where $\gamma_{\rm obs}$ is the measured shear in a region (and in redshift slice $i$) and $X$ is any quantity on which we want to set a knowledge requirement. The spurious shear in bin $i$ is taken to be
\begin{equation}
c_i = c(z_i) = \frac{\partial\gamma_{\rm obs}(z_i)}{\partial X} \Delta X,
\label{eq:X-shape}
\end{equation}
where $\Delta X = X_{\rm true} - X_{\rm model}$ is the error in
knowledge of $X$. In the context of the additive systematic errors,
the ratios of the partial derivatives $\partial\gamma_{\rm
obs}(z_i)/\partial X$ set the redshift slice dependence: if $i({\rm
max})$ is the redshift bin with the largest derivative (in absolute
value) then
\begin{equation}
w_i = \frac{\partial\gamma_{\rm obs}(z_i)/\partial X}{\partial\gamma_{\rm obs}(z_{i({\rm max})})/\partial X}
\end{equation}
and the reference additive shear is $c_{\rm ref} = c_{i({\rm max})}$.

In principle the coefficients $\partial\gamma_{\rm obs}(z_i)/\partial X$ depend on the base model for the PSF, the population of galaxies, and the shape measurement algorithm. Multiple algorithms should be used for \wfirst, but a final selection has not been made (given how rapidly the field is maturing, such a choice now would be premature). However, all practical methods of measuring shear have some basic properties in common -- if e.g.\ the true PSF has greater $e_1$ than the model (i.e.\ is elongated in the $x$-direction), then the inferred shear in that region of the sky will also have greater $c_1$, and this effect will be greater for larger PSFs or smaller galaxies. In setting requirements, we therefore chose a simple, easily understood model. This model is not, and does not need to be, an accurate description of \wfirst\ shape measurement at the few$\times 10^{-4}$ accuracy. Rather, it needs to give us estimates of $\partial\gamma_{\rm obs}(z_i)/\partial X$ early in the development of \wfirst, with the understanding that we will not update the optical stability requirements every time we have a better model for the distribution of galaxy morphologies. The very simplest choice would be to work with Gaussian PSFs and galaxies; however, our previous experience has been that the non-Gaussian tails of both PSFs and galaxies matter, and furthermore when we discuss the Zernike description of wavefront errors we have predictions for how various combinations of modes affect the ellipticities of different isophotes of the PSF. Therefore, we go one step beyond the Gaussian approximation and include in our analytic flow-down model:
\begin{list}{$\bullet$}{}
\item Our galaxies are taken to have an exponential profile, $f_{\rm circ}({\bf x}) \propto e^{-1.67834|{\bf x}|/r_{\rm eff}}$, where $r_{\rm reff}$ is the half-light radius. It can optionally be sheared by applying a {\em finite} shear $\gamma$ to arrive at the galaxy $f({\bf x})$.
\item The PSF is the Fourier transform of an
annular pupil with aberrations appearing as contributions to the
phase. The resulting ``optical'' PSF is then convolved with a detector
response that includes a tophat and charge diffusion. For HgCdTe
detectors, we take the charge diffusion length to be 2.94 $\mu$m rms
per axis \citep{2007PASP..119..466B}.\footnote{This was measured on an H2RG. At the time we had to fix this for Phase A requirements
flow-down, we did not have a measurement on the H4RG. We now know the charge diffusion for \wfirst\ detectors is smaller than this number \changetext{\citep{2020arXiv200500505M}}, but it makes a small enough difference that we have not re-done the requirements flowdown.} Other effects that are likely significant for \wfirst\ analyses, such as inter-pixel capacitance, the brighter-fatter effect, or polarization, are not included. We turned the spider off. The main effect of the spider is the production of 12 diffraction spikes, but it is the core of the PSF that matters most for shape measurement and has the greatest change as one adjusts the Zernike coefficients. The spider further leads to an asymmetric pupil, i.e.\ with odd-order modes in the decomposition of the amplitude, but this has no appreciable effects on the relation of ellipticity to low-order Zernike modes.\footnote{It is known that an odd-order mode in the phase can
mix with other asymmetric phase modes to produce PSF ellipticity,
e.g. if one introduces a large trefoil $t$ then the ellipticity
develops a linear term in coma, proportional to $tc^\ast$
\citep{2010SPIE.7731E..1EN}. However, an amplitude feature with 3-fold
or other odd symmetry, such as the spider, does not lead to such an
effect.}
\item We use as our measure of ellipticity the 2-component ellipticity $e_I$ of the observed image $I = f \star P$ (where $f$ is the galaxy, $P$ is the PSF, and $\star$ denotes convolution). The ellipticity is determined according to the adaptive moment algorithm of \cite{2002AJ....123..583B}, \S3.1.
\end{list}

The observed 2-component ellipticity $e_I$ of the galaxy is related to
the shear by a $2\times 2$ responsivity matrix
\begin{equation}
R_{ij} = \frac{\partial e_{I,i}}{\partial\gamma_j} = {\cal R}\delta_{ij} + R^{\rm aniso}_{ij},
\end{equation}
which we have decomposed into an isotropic part ${\cal R}$ and a
traceless matrix $R^{\rm aniso}$ characterizing the anisotropic part
of the responsivity. The inverse of the responsivity matrix relates a
bias in the galaxy ellipticities to a bias in the shear:
\begin{equation}
c_i = \sum_{j=1}^2 [{\bf R}^{-1}]_{ij} \frac{\partial e_{I,j}}{\partial X} \Delta X.
\end{equation}
Since the isotropic part of the responsivity dominates except for
extreme PSF ellipticity, anisotropic noise correlations, etc., we take
the isotropic part and write
\begin{equation}
\frac{\partial\gamma_{{\rm obs},i}(z_k)}{\partial X} = \left\langle {\cal R}^{-1} \frac{\partial e_{I,i}}{\partial X} \right\rangle,
\end{equation}
where the average is taken over the source galaxies in that redshift
bin. The various partial derivatives are easily computed as finite
differences of the galaxy simulation and ellipticity measurement
process.

As this model is intended to be simple, the average is taken only over
the distribution of source sizes $r_{\rm eff}$ -- we do not include
the intrinsic source ellipticity or a distribution of Sersi\'c
indices.

The main difference that occurs with the multiplicative systematic errors is that when one changes the PSF size, one must look at the change in responsivity, i.e., $m = \partial \ln {\cal R}/\partial X$.

\section{Requirements on wavefront stability for the PSF calibration}
\label{app:psf-stability}

The determination of the PSF in imaging mode will be based on an empirical (principal components or more advanced version thereof) approach (these methods have a long history in weak lensing -- see, e.g., \citealt{2004astro.ph.12234J, 2007PASP..119.1403J} -- but a large amount of work will be required to adapt them to \wfirst), or on physical fitting of the optical model \citep{2012SPIE.8442E..10J} with empirical corrections. Central to both of these approaches is that we must limit the number of possible principal components in the data by limiting the number of properties of the PSF that vary from one image to another. The \wfirst\ approach is to keep the PSF stable during an exposure so that no parameters are needed to describe time dependence of the PSF during an exposure. We make one exception to this policy for image motion, since at the \wfirst\ weak lensing level of precision this is unavoidable. Thus the requirement is for the optics + image motion PSF to be the convolution of the optics PSF with a kernel coming from the image motion, with small residuals. Here ``small'' means that the residual error must fit within the overall error budget for PSF (or shear) errors.

We note that since \wfirst\ detectors can be read non-destructively, and 6 sub-exposures will be sent to the ground, that one could imagine building a time-dependent PSF from these sub-exposures. We have chosen {\em not} to set a looser requirement based on this expectation, since we plan to calibrate detector non-linearity using the consistency of the sub-exposures; this approach does not work if the PSF is varying in an uncontrolled way.

Requirements are derived for the two major sources of wavefront change: slow drifts induced by, e.g. thermal variations (\S\ref{as:a-drift}), and jitter induced by e.g.\ vibrations from the reaction wheels (\S\ref{as:a-wfjitter}).

\subsection{Wavefront drift}
\label{as:a-drift}

\subsubsection{Flowdown methodology}
\label{as:drift-method}

In general, we suppose that there is a vector of parameters ${\bf p}$
that determines the PSF in each exposure (including its field
dependence). Some of these are associated with the equilibrium
wavefront -- this is the subject of this section -- whereas others are
associated with image motion, jitter, detector properties, etc. The
amplitudes $\psi_i({\boldsymbol\theta})$ of each Zernike component of
the wavefront error -- which depend on field position
${\boldsymbol\theta}$ -- are functions of these parameters, and will
each have their own time dependence
$\psi_i({\boldsymbol\theta};t)$. This induces a time dependence in the
PSF $G({\bf x};{\boldsymbol\theta};t)$, and hence in the observed
shear $\gamma_{\rm obs}$ for an object.

We may write the amplitudes $\psi_i$ at a given position as a vector
${\boldsymbol\psi}$ of length $N_{\rm Zern}$, where $N_{\rm Zern}$ is
the number of Zernike coefficients kept. We normalize the Zernike
modes to unit RMS, so that $|{\boldsymbol\psi}({\boldsymbol\theta})|$
is the RMS wavefront error at position ${\boldsymbol\theta}$. That is,
we write the wavefront error at pupil position ${\boldsymbol\eta}$ and
field position ${\boldsymbol\theta}$ as
\begin{align}
&\psi({\boldsymbol\eta};{\boldsymbol\theta}) = \nonumber\\
&\sum_{n=2}^\infty
\sum_{m} \sqrt{n+1}\,\psi_{nm}({\boldsymbol\theta}) R_n^m(\rho)
\times\left\{\begin{array}{ccc} 1 & & m=0 \\ \sqrt2\,\cos m\varphi & &
m>0 \\ \sqrt2\,\sin m\varphi & & m<0 \end{array} \right.,
\end{align}
where $\rho$ is the radius of the pupil position normalized to 1 at
the edge, and $\varphi$ is the polar angle in the pupil plane, $m$ is
summed over integers with the same parity as $n$ (both odd or both
even) and $|m|\le n$ (so that there are $n+1$ terms in the $m$-sum),
and $R_n^m$ is the Zernike polynomial with normalization
$R_n^m(1)=1$. The factor of $\sqrt{n+1}$ and (sometimes) $\sqrt2$
guarantee the unit normalization of the RMS over the unit disc.

If the wavefront is drifting over time, then to first order in the
drift rate we may write
\begin{equation}
\psi_i({\boldsymbol\theta};t) = \psi_i({\boldsymbol\theta};t_0) +
\dot\psi_i({\boldsymbol\theta}) (t-t_0),
\end{equation}
where $t_0$ is the central epoch chosen and $-\frac12\Delta t < t-t_0
< \frac12\Delta t$. Again to linear order in $t-t_0$, the PSF that is
determined by a least-squares fit with uniform weighting in time will
have an expectation value that is $G({\bf
x};{\boldsymbol\theta};t_0)$. There is then a corresponding error in
the shear in a given redshift bin $z_k$:
\begin{equation}
c_{k,i}(t) = \sum_j \frac{\partial \gamma_{{\rm obs},i}(z_k)}{\partial\psi_j} \dot\psi_j({\boldsymbol\theta}) (t-t_0),
\end{equation}
where in this equation $k$ denotes a redshift bin and $i$ denotes a
component. Taking just the most strongly affected (in the sense of
$|c|$) redshift bin to start as the reference, we see that
\begin{equation}
|c_{\rm ref}(t)| \le \left\lVert \frac{\partial \gamma_{{\rm obs,ref},i}}{\partial\psi_j} \right\rVert |\dot{\boldsymbol\psi}({\boldsymbol\theta})| |t-t_0|,
\end{equation}
where $\lVert~\rVert$ denotes an operator norm (i.e.\ the maximum singular value of the $2\times N_{\rm Zern}$ matrix). The variance of $c$ {\em per component} (i.e.\ divided by 2) is
\begin{equation}
A^2 \equiv \frac12\langle |c_{\rm ref}|^2 \rangle \le \frac12 \left[ \left\lVert \frac{\partial \gamma_{{\rm obs,ref},i}}{\partial\psi_j} \right\rVert |\dot{\boldsymbol\psi}({\boldsymbol\theta})| \right]^2 \langle(t-t_0)^2\rangle;
\end{equation}
the last expectation value is $\frac1{12}\Delta t^2$ with the average taken over a uniform interval, leading to
\begin{equation}
A \le \frac1{\sqrt{24}} \left\lVert \frac{\partial \gamma_{{\rm obs,ref},i}}{\partial\psi_j} \right\rVert |\dot{\boldsymbol\psi}({\boldsymbol\theta})| \Delta t.
\label{eq:ADt}
\end{equation}
Thus from a requirement on $A$, a determination of the matrix
$\partial \gamma_{{\rm obs,ref},i}/\partial\psi_j$, and an interval of
time $\Delta t$, we can set a requirement on the wavefront drift rate
$|\dot{\boldsymbol\psi}|$. The matrix $\partial \gamma_{{\rm
obs,ref},i}/\partial\psi_j$ depends on the static aberration pattern
and its determination is described below. The interval $\Delta t$ for
PSF fitting is a free parameter, and the wavefront drift rate
requirement is tighter if $\Delta t$ is increased. This must be traded
against the {\em statistical} error in the PSF solution, where the
target precision is easier to achieve if the time baseline $\Delta t$
used in fitting the model is increased.

\subsubsection{Sensitivity matrix}
\label{as:drift-sens}

From Eq.~(\ref{eq:ADt}), we see that a key step is to compute the
sensitivity matrix $\partial \gamma_{{\rm
obs,ref},i}/\partial\psi_j$. Unfortunately, this matrix depends on the
specific combination of static wavefront errors, because
${\boldsymbol\gamma}_{\rm obs,ref}$ is not a linear function of
${\boldsymbol\psi}$. Indeed, due to symmetries the possible form of
${\boldsymbol\gamma}_{\rm obs,ref}$ is restricted, with the result
that $\partial \gamma_{{\rm obs,ref},i}/\partial\psi_j$ may be
suppressed at zero wavefront error (${\boldsymbol\psi}=0$) and be much
larger in the realistic case where ${\boldsymbol\psi}\neq 0$ (e.g., 
\citealt{2010SPIE.7731E..1EN}). 
\changetext{We do not know {\em a priori} what the static wavefront error will be -- we have set a requirement of $|{\boldsymbol\psi}|<92\,$nm, but until we have the as-built observatory we do not know how this will be distributed among the Zernikes. We address this by expanding the sensitivity matrix $\partial \gamma_{{\rm obs,ref},i}/\partial\psi_j$ to linear order in ${\boldsymbol\psi}$, which means we need $\gamma_{{\rm obs,ref},i}$ Taylor expanded to quadratic order in ${\boldsymbol\psi}$ around ${\boldsymbol\psi}=0$. A consequence of this is that if the static wavefront error ${\boldsymbol\psi}$ is larger, then the sensitivity matrix is also larger and the stability requirements are tighter.
Then we} search the
entire space of possible wavefront errors ${\boldsymbol\psi}$ --
bounded by the top-level requirement that $|{\boldsymbol\psi}|<92\,$nm
-- to find the place where the operator norm is maximized.

The \changetext{fact} that the PSF inverts (i.e.\ preserves ellipticity and
hence spurious shear) under
${\boldsymbol\psi}\rightarrow-{\boldsymbol\psi}$ implies that
${\boldsymbol\gamma}_{\rm obs,ref}$ is an even function of
${\boldsymbol\psi}$ (this statement remains true even for an
asymmetric pupil, due e.g.\ to the spider). For a circularly symmetric
pupil \changetext{(i.e., an annular pupil, so allowing for a secondary obstruction, but not accounting for the offset of the secondary obstruction when using an off-axis portion of the field, nor for the spider arms)}, we find the further restrictions that
\begin{align}
\gamma_{\rm obs,ref\,1} =& C_{fa} \psi_{20} \psi_{22} + C_{sa} \psi_{40} \psi_{22}
+ C_{cc} (\psi_{31}^2-\psi_{3-1}^2) \nonumber\\
&+ C_{ct} (\psi_{31}\psi_{33} + \psi_{3-1}\psi_{3-3})
+ ... \label{eq:temp.314159}
\end{align}
and
\begin{align}
\gamma_{\rm obs,ref\,2} =& C_{fa} \psi_{20} \psi_{2-2} + C_{sa} \psi_{40} \psi_{2-2}
+ 2C_{cc} \psi_{31} \psi_{3-1} \nonumber\\
&+ C_{ct} (\psi_{31}\psi_{3-3} - \psi_{3-1}\psi_{33})
+ ...\,,
\end{align}
where we have taken the lowest-order aberrations (focus, astigmatism,
coma, trefoil, and spherical) as these dominate the wavefront
stability budget. With the wavefront error vector written in this
order, ${\boldsymbol\psi} = (\psi_{20}; \psi_{22},\psi_{2-2};
\psi_{31},\psi_{3-1}; \psi_{33},\psi_{3-3}; \psi_{40})$, we find a
sensitivity matrix
\begin{align}
{\bf M}^{\rm T} &= \left[ \frac{\partial \gamma_{{\rm obs,ref},i}}{\partial\psi_j} \right]^{\rm T}\nonumber\\
&= \left( \begin{array}{cc}
C_{fa}\psi_{22}  & C_{fa}\psi_{2-2} \\
C_{fa}\psi_{20}+C_{sa}\psi_{40} & 0 \\
0 & C_{fa}\psi_{20}+C_{sa}\psi_{40} \\
2C_{cc}\psi_{31}+C_{ct}\psi_{33} & 2C_{cc}\psi_{3-1}+C_{ct}\psi_{3-3} \\
-2C_{cc}\psi_{3-1}+C_{ct}\psi_{3-3} & 2C_{cc}\psi_{31}-C_{ct}\psi_{33} \\
C_{ct}\psi_{31} & -C_{ct}\psi_{3-1} \\
C_{ct}\psi_{3-1} & C_{ct}\psi_{31} \\
C_{sa}\psi_{22} & C_{sa}\psi_{2-2}
\end{array} \right)
\label{eq:TheMatrix}
\end{align}
(we show the transpose here for ease of display; the operator norm is the same). \changetext{The real pupil is not circularly symmetric, however as noted above ${\boldsymbol\gamma}_{\rm obs,ref}$ remains an even function of ${\boldsymbol\psi}$ even for a pupil of general asymmetry; thus ${\bf M}$ remains an odd function of ${\boldsymbol\psi}$ and the leading term is the linear term. The consequence of asymmetry of the pupil is that the coefficients in Eq.~(\ref{eq:TheMatrix}) may be slightly different in different directions, e.g., in $J$-band in the $z=1.0-1.2$ bin, using the full pupil for SCA \#1, we find that the top row of ${\bf M}^{\rm T}$ is $(8.273\psi_{22}+0.002\psi_{2-2}, 0.002\psi_{22}+8.289\psi_{2-2})\,\mu$m$^{-2}$, versus $8.337(\psi_{22},\psi_{2-2})\,\mu$m$^{-2}$ for the circularly symmetric (annular) pupil. This corresponds to a difference in sensitivity of 0.2\% (8.273 vs.\ 8.289) between the two astigmatism modes, and a 0.8\% (8.273 vs.\ 8.337) change in sensitivity relative to the annular pupil case. This in principle changes our requirements by 0.8\% (or 0.9\%, which is the maximum over any of the bands and redshift bins); in practice, given the necessary margin factors, it does not make sense to track stability requirements at the $<1\%$ level. This statement about the {\em sensitivity} holds even though the annular pupil is missing some important features of the real PSF (particularly diffraction spikes).}

We want a limit on the maximum singular value of
Eq.~(\ref{eq:TheMatrix}), subject to a limit on
$|{\boldsymbol\psi}|$. To do so, let us first consider writing the
singular value decomposition ${\bf M} = {\bf UDV}^{\rm T}$, where
${\bf U}$ is a $2\times 2$ orthogonal matrix, ${\bf D}$ has 2 diagonal
non-negative entries in non-increasing order ($D_{11}\ge D_{22}$) and
is otherwise zeroes (and has dimension $2\times N_{\rm Zern}$), and
${\bf V}$ is $N_{\rm Zern}\times N_{\rm Zern}$. Here ${\bf U}$ is
simply a rotation of the shear derivative, and due to circular
symmetry can be set to the identity by rotating the entire aberration
pattern. Thus without loss of generality we can consider cases where
${\bf U}$ is the identity, and then
\begin{equation}
\lVert{\bf M} \rVert = \sqrt{ \sum_j \left(\frac{\partial \gamma_{{\rm obs,ref},1}}{\partial\psi_j} \right)^2 }
= \sqrt{ {\boldsymbol\psi}^{\rm T} {\boldsymbol\Lambda}{\boldsymbol\psi} }
\le \lVert{\boldsymbol\Lambda} \rVert |{\boldsymbol\psi}|,
\end{equation}
where we used the fact that ${\bf M}$ is a linear function of ${\boldsymbol\psi}$ and defined the matrix ${\boldsymbol\Lambda}$ to be the matrix of derivatives of the first row of ${\bf M}$:
\begin{equation}
{\boldsymbol\Lambda} = \left( \begin{array}{cccccccc}
0 & C_{fa} & 0 & 0 & 0 & 0 & 0 & 0 \\
C_{fa} & 0 & 0 & 0 & 0 & 0 & 0 & C_{sa} \\
0 & 0 & 0 & 0 & 0 & 0 & 0 & 0 \\
0 & 0 & 0 & 2C_{cc} & 0 & C_{ct} & 0 & 0 \\
0 & 0 & 0 & 0 & -2C_{cc} & 0 & C_{ct} & 0 \\
0 & 0 & 0 & C_{ct} & 0 & 0 & 0 & 0 \\
0 & 0 & 0 & 0 & C_{ct} & 0 & 0 & 0 \\
0 & C_{sa} & 0 & 0 & 0 & 0 & 0 & 0
\end{array} \right),
\label{eq:LambdaDef}
\end{equation}
which has norm
\begin{equation}
\lVert {\boldsymbol\Lambda} \rVert = \max \left\{
\sqrt{C_{fa}^2 + C_{sa}^2}, ~|C_{cc}|+\sqrt{C_{cc}^2+C_{ct}^2}
\right\}.
\label{eq:LambdaNorm}
\end{equation}
(There are both even-aberration and odd-aberration sectors of this
matrix; the operator norm is determined by whichever has greater
leverage on the spurious shear. In most cases, the even sector -- the
first term -- is dominant.) We show the derived coefficients in Table~\ref{tab:wfej}.

We are not quite done because we have not specified the redshift or
scale dependence of this systematic. Since $C_{fa}$ is usually
dominant, we adopt its redshift dependence to determine the weights
$w(z_i)$, with the last bin as the reference bin because it is the
most heavily contaminated -- the galaxies are smallest in that bin and
the $C$-coefficients are largest. However, the weights $w(z_i)$
obtained from $C_{ct}$ (the next largest coefficient) are only
slightly different.

We find that in J129, H158, and F184 bands, the operator norms are $\lVert{\boldsymbol\Lambda}\rVert = 1.14\times 10^{-5}$, $9.76\times 10^{-6}$, and $8.46\times 10^{-6}$ nm$^{-2}$, respectively. Using the J129-band limit, which has the worst total contamination, we find a limit on the total wavefront error of
$|{\boldsymbol\psi}({\boldsymbol\theta})|<92\,$nm, we find
\begin{equation}
A \le \frac1{\sqrt{24}}\lVert{\boldsymbol\Lambda} \rVert |{\boldsymbol\psi}| |\dot{\boldsymbol\psi}({\boldsymbol\theta})| \Delta t
= 2.14\times 10^{-4}\, {\rm nm}^{-1}\times |\dot{\boldsymbol\psi}({\boldsymbol\theta})| \Delta t.
\end{equation}

The current wavefront drift sub-allocation is that $\Delta t$ = 1 exposure
(140 s) and $|\dot{\boldsymbol\psi}({\boldsymbol\theta})| \Delta
t<0.37\,$nm, which produces a spurious shear of $7.91\times 10^{-5}$,
RMS per component. However the $S$-factor for the
focus$\times$astigmatism mode is 0.5488 in the worst angular bin, so the
implied spurious shear is $AS^{1/2} = 5.86\times 10^{-5}$. The
requirements in Table~\ref{tab:addbands} give a top-level error of
$2.65\times 10^{-4}$; 4.9\%\ of the additive shear systematic error budget, in
an RSS sense, is currently being taken up by wavefront drift. For a sub-allocation of 1 nm (instead of 0.37 nm), this would be 36\% of the additive shear systematic error budget.

\subsection{Wavefront jitter}
\label{as:a-wfjitter}

The wavefront jitter is handled by a similar calculation to the
wavefront drift. The principal difference is that we are now
interested in the spurious shear from a PSF that is the superposition
of many instantaneous PSFs with different wavefronts. Moreover, the
PSFs can have different line-of-sight positions, so instead of simply
considering the covariance matrix of the Zernike amplitudes, we must
also consider the line-of-sight motion (parameterized by $\theta_x$
and $\theta_y$). The spurious shear thus depends on the full
covariance matrix of the Zernike amplitudes ${\boldsymbol\psi}$ and
the line-of-sight motion ${\boldsymbol\theta}$. Of this covariance
matrix, the ``line-of-sight block'' ${\rm
Cov}({\boldsymbol\theta},{\boldsymbol\theta})$ corresponds to simple
image motion, and is not related to wavefront jitter. \changetext{The PSF modeling procedure for \wfirst\ will explicitly allow for image motion to be fit separately in each exposure \citep{2012SPIE.8442E..10J}, so the presence of ${\rm
Cov}({\boldsymbol\theta},{\boldsymbol\theta})$ does not represent a bias in fitting the PSF.} On the other
hand, the blocks ${\rm Cov}({\boldsymbol\theta},{\boldsymbol\psi})$
and ${\rm Cov}({\boldsymbol\psi},{\boldsymbol\psi})$ involve the
wavefront jitter, and their effects on ${\boldsymbol\gamma}_{\rm
obs,ref}$ must be treated here. \changetext{(Due to the large number of parameters, we cannot fit a full covariance matrix of all the Zernikes in each exposure.)}

We can then write the matrix of second derivatives:
\begin{align}
\gamma_{{\rm obs},i}(z_k) &= \gamma_{{\rm obs},i}(z_k)|_{\rm no~wf~jitter}\nonumber\\
&+ \sum_{aj} K^{\rm LOS,WFE}_{iaj}(z_k) {\rm Cov}({\theta}_a,{\psi}_j)\nonumber\\
&+ \frac12 \sum_{jj'} K^{\rm WFE,WFE}_{ija}(z_k) {\rm Cov}({\psi}_j,{\psi}_{j'}).
\label{eq:gamma-S-Sens}
\end{align}
The matrix ${\bf K}^{\rm WFE,WFE}$, describing how much small
high-frequency vibrations of the wavefront impact the shear, has a
dependence on redshift bin $z_k$, shear component $i$, and the Zernike
modes $j$ and $j'$. The matrix ${\bf K}^{\rm LOS,WFE}$ describes the
effects of correlations between LOS motion and wavefront jitter.
Although not strictly necessary for an analysis of wavefront jitter, we do also compute the sensitivity to line-of-sight motion, which implies a term $K^{\rm LOS,LOS}_{iab}(z_k) {\rm Cov}({\theta}_a,{\theta}_b)$ in Eq.~(\ref{eq:gamma-S-Sens}).

The matrix ${\bf K}$ in principle varies with the wavefront error, but
since it is a second derivative it is nonzero even for zero
aberrations. One option is to take this leading term (i.e.\ ${\bf K}$
evaluated at ${\boldsymbol\psi}=0$) to set requirements. Another would
be to also include the linear dependences on ${\boldsymbol\psi}$; this
would be necessary if we were to separately write requirements on the
individual Zernike modes, since due to symmetries some entries in
${\bf K}$ are exactly zero in the unaberrated case\changetext{, but we do not expect this to be necessary when setting overall limits on wavefront jitter (we verify this explicitly below)}.

Following the methodology of \S\ref{as:drift-sens}, and again
exploiting the symmetries of the problem and suppressing the $z_k$
index, we find that at ${\boldsymbol\psi}=0$, the terms involving the
covariance of line-of-sight motion and wavefront jitter are
\begin{equation}
K_{1aj}^{\rm LOS,WFE} = \left( \begin{array}{cccccccc}
0 & 0 & 0 & K_{\theta c} & 0 & K_{\theta t} & 0 & 0 \\
0 & 0 & 0 & 0 & -K_{\theta c} & 0 & K_{\theta t} & 0
\end{array} \right)
\label{eq:temp.078524}
\end{equation}
and
\begin{equation}
K_{2aj}^{\rm LOS,WFE} = \left( \begin{array}{cccccccc}
0 & 0 & 0 & 0 & K_{\theta c} & 0 & K_{\theta t} & 0 \\
0 & 0 & 0 & K_{\theta c} & 0 & -K_{\theta t} & 0 & 0
\end{array} \right),
\end{equation}
where the two rows are $a=1,2$ and the eight columns are the low-order
Zernikes. Similarly, for the wavefront jitter variance, we have
\begin{equation}
\setlength\arraycolsep{1.5pt}
K_{1jj'}^{\rm WFE,WFE} = \left( \begin{array}{cccccccc}
0 & K_{fa} & 0 & 0 & 0 & 0 & 0 & 0 \\
K_{fa} & 0 & 0 & 0 & 0 & 0 & 0 & K_{sa} \\
0 & 0 & 0 & 0 & 0 & 0 & 0 & 0 \\
0 & 0 & 0 & 2K_{cc} & 0 & K_{ct} & 0 & 0 \\
0 & 0 & 0 & 0 & -2K_{cc} & 0 & K_{ct} & 0 \\
0 & 0 & 0 & K_{ct} & 0 & 0 & 0 & 0 \\
0 & 0 & 0 & 0 & K_{ct} & 0 & 0 & 0 \\
0 & K_{sa} & 0 & 0 & 0 & 0 & 0 & 0 \\
\end{array} \right)
\end{equation}
and
\begin{equation}
\setlength\arraycolsep{1.5pt}
K_{2jj'}^{\rm WFE,WFE} = \left( \begin{array}{cccccccc}
0 & 0 & K_{fa} & 0 & 0 & 0 & 0 & 0 \\
0 & 0 & 0 & 0 & 0 & 0 & 0 & 0 \\
K_{fa} & 0 & 0 & 0 & 0 & 0 & 0 & K_{sa} \\
0 & 0 & 0 & 0 & 2K_{cc} & 0 & K_{ct} & 0 \\
0 & 0 & 0 & 2K_{cc} & 0 & -K_{ct} & 0 & 0 \\
0 & 0 & 0 & 0 & -K_{ct} & 0 & 0 & 0 \\
0 & 0 & 0 & K_{ct} & 0 & 0 & 0 & 0 \\
0 & 0 & K_{sa} & 0 & 0 & 0 & 0 & 0 \\
\end{array} \right).
\end{equation}
Finally, for the line-of-sight-motion only, we have
\begin{equation}
K_{1ab}^{\rm LOS,LOS} = \left( \begin{array}{cc} K_{\theta\theta} & 0 \\ 0 & -K_{\theta\theta} \end{array} \right)
\label{eq:K1t}
\end{equation}
and
\begin{equation}
K_{2ab}^{\rm LOS,LOS} = \left( \begin{array}{cc} 0 & K_{\theta\theta} \\ K_{\theta\theta} & 0 \end{array} \right).
\label{eq:K2t}
\end{equation}
Image simulations are required to determine the specific values of
$K_{\theta c}$, $K_{\theta t}$, $K_{fa}$, $K_{sa}$, $K_{cc}$, and
$K_{ct}$. These depend on the galaxy sizes, and hence indirectly on
redshift slice $z_k$. The coefficients in the ``worst'' redshift slice
are shown in Table~\ref{tab:wfej}.

Once again, the maximum value of the apparent shear induced by
wavefront error can be determined from the eigenvalues of the ${\bf
K}$ matrices, the RMS wavefront jitter, and the line of sight motion
per axis. We note that the RMS wavefront jitter is
\begin{equation}
\sigma_{\rm wfe-jitter} = \sqrt{\sum_j {\rm Var} \psi_j},
\end{equation}
and that covariances between WFE jitter and LOS jitter are limited by the rule that the covariance matrix be positive-definite (in particular, the correlation coefficients cannot exceed 1). This implies the limits
\begin{align}
&\left|\sum_{aj} K^{\rm LOS,WFE}_{1aj}(z_k) {\rm
Cov}({\theta}_a,{\psi}_j)\right| \nonumber\\
&\le \sqrt{K_{\theta c}^2 + K_{\theta
t}^2}~\, \sigma_{\rm los-jitter} \sigma_{\rm wfe-jitter}
\label{eq:g1-wfej.corr}
\end{align}
and
\begin{align}
&\left|\frac12 \sum_{jj'} K^{\rm WFE,WFE}_{ij\changetext{j'}}(z_k) {\rm
Cov}({\psi}_j,{\psi}_{j'})\right| \nonumber\\
&\le \frac12 \max\left(
\sqrt{K_{fa}^2+K_{sa}^2}, |K_{cc}|+\sqrt{K_{cc}^2+K_{ct}^2}
\right)\,\sigma_{\rm wfe-jitter}^2,
\label{eq:g1-wfej.var}
\end{align}
where $\sigma_{\rm los-jitter}$ is the RMS line-of-sight jitter per
axis, i.e.\ we set ${\rm Cov}(\theta_a,\theta_b) = \sigma_{\rm
los-jitter}^2\delta_{ab}$. (Note that only the jitter contributes
here: the controlled motion of the line of sight does not correlate
with the wavefront jitter since it is not in the same frequency band.)
The sum of Eqs.~(\ref{eq:g1-wfej.corr}) and (\ref{eq:g1-wfej.var})
represents a bound on the RMS spurious shear in the $\gamma_1$
component (a similar bound applies to $\gamma_2$).

\changetext{We also computed ${\bf K}$ for the case of the full pupil (including spider) and one realization of the static wavefront error at the center of SCA \#1. \footnote{\url{https://roman.gsfc.nasa.gov/science/Roman\_Reference\_Information.html}} In this case, the sparseness pattern of ${\bf K}$ changes, and Eq.~(\ref{eq:g1-wfej.corr}) changes by replacing $\sqrt{K_{\theta c}^2 + K_{\theta t}^2}$ with the maximum singular value of ${\bf K}^{\rm LOS,WFE}_{1aj}$. Over the 3 bands (J, H, and F184), and the 15 redshift bins, we found a maximum change of 1.9\% in the coefficient of Eq.~(\ref{eq:g1-wfej.corr}) for the $\gamma_1$ component, and 3.1\% for the $\gamma_2$ component; in all cases, the sensitivity actually went down. We therefore conclude that for the purposes of setting requirements, computing ${\bf K}$ by expansion around the unaberrated annular pupil was a sufficient approximation.}

\begin{table*}
\center{\caption{\label{tab:wfej}The coefficients of spurious shear at
${\boldsymbol\psi}=0$, appearing in
Eqs.~(\ref{eq:temp.314159}--\ref{eq:LambdaNorm}) for the top half of the table (wavefront drift) and
Eqs.~(\ref{eq:temp.078524}--\ref{eq:g1-wfej.var}) for the bottom half (wavefront jitter). Coefficients are
shown for the worst (most contaminated) redshift bin. This redshift
bin and the $S$-factor are shown in the two right-most columns for the
J129 band (which is always the most sensitive to wavefront jitter
because it has the shortest wavelength). The $S$-factor is shown for
the worst angular bin, which is always the smallest scales
($3.0<\log_{10}\ell<3.5$). The $C$ and $K$ coefficients are the same for the ``even'' aberrations but different for the ``odd'' aberrations.}
\begin{tabular}{c|rrr|c|cc}
\hline
\hline
Band: & J129 & H158 & F184 & Units & Worst $z$-bin & Worst $S$-factor \\
\hline
\multicolumn7c{Wavefront drift coefficients} \\
\hline
$C_{fa}$ & 10.60 & 9.22 & 8.06 & $10^{-6}$ mas$^{-1}$ nm$^{-1}$ & 2.8--3.0 & 0.549 \\
$C_{cc}$ & 2.36 & 2.03 & 1.81 & $10^{-6}$ mas$^{-1}$ nm$^{-1}$ & 2.8--3.0 & 0.612 \\
$C_{ct}$ & 6.23 & 5.41 & 4.71 & $10^{-6}$ mas$^{-1}$ nm$^{-1}$ & 2.8--3.0 & 0.558 \\
$C_{sa}$ & 4.14 & 3.21 & 2.59 & $10^{-6}$ mas$^{-1}$ nm$^{-1}$ & 2.8--3.0 & 0.672 \\
\hline
 $\lVert{\boldsymbol\Lambda}\rVert$ & 11.38 & 9.76 & 8.46 & $10^{-6}$ mas$^{-1}$ nm$^{-1}$ \\
\hline
\multicolumn7c{Wavefront jitter coefficients} \\
\hline
$K_{\theta c}$ & 7.69 & 6.16 & 4.83 & $10^{-6}$ mas$^{-1}$ nm$^{-1}$ & 2.8--3.0 & 0.592 \\
$K_{\theta t}$ & 3.29 & 3.78 & 4.08 & $10^{-6}$ mas$^{-1}$ nm$^{-1}$ & 1.8--2.0 & 0.438 \\
\hline
$K_{fa}$ & 10.60 & 9.22 & 8.06 & $10^{-6}$ nm$^{-2}$ & 2.8--3.0 & 0.549 \\
$K_{cc}$ & 3.52 & 2.76 & 2.23 & $10^{-6}$ nm$^{-2}$ & 2.8--3.0 & 0.648 \\
$K_{ct}$ & 5.52 & 4.74 & 4.10 & $10^{-6}$ nm$^{-2}$ & 2.8--3.0 & 0.568 \\
$K_{sa}$ & 4.14 & 3.21 & 2.59 & $10^{-6}$ nm$^{-2}$ & 2.8--3.0 & 0.672 \\
\hline
$K_{\theta\theta}$ & 14.32 & 13.77 & 13.34 & $10^{-6}$ mas$^{-2}$ & 2.8--3.0 & 0.504 \\
\hline
\hline
\end{tabular}
}
\end{table*}

The RMS spurious shear per component in J-band, weighted by the worst scaling factor $S = 0.672$ (which accounts for redshift dependence), is then
\begin{align}
\gamma_{\rm rms}\sqrt{S N_{\rm ind}} &\le 6.85\times 10^{-6}\,{\rm
nm}^{-1}\,{\rm mas}^{-1}\,\sigma_{\rm los-jitter}\sigma_{\rm
wfe-jitter} \nonumber\\
&+ 4.67\times 10^{-6} \,{\rm nm}^{-2}\,\sigma_{\rm
wfe-jitter}^2.
\label{eq:wfej-grms}
\end{align}
This should be compared to the requirement of $2.65\times 10^{-4}$. The line-of-sight jitter is required to be $\sigma_{\rm los-jitter}=12$ mas rms per axis, and the observing strategy has two passes at epochs separated by many months so we take $N_{\rm ind}=2$. From Eq.~(\ref{eq:wfej-grms}), we find that the entire error budget would be taken up for $\sigma_{\rm wfe-jitter}=3.76$ nm rms. If $\sigma_{\rm wfe-jitter}=1$ nm rms, then $\gamma_{\rm rms}S^{1/2}=6.14\times 10^{-5}$, and 5.4\% of the error budget is used (in an RSS sense).

\section{Simulation data access}
\label{app:data_access}

The simulated data for the \textsc{Fiducial} run used in this paper form a 2.5$\times$2.5 deg$^2$, full-depth synthetic \wfirst\ Reference Survey in the H158-band, which is suitable for a variety of uses in testing algorithms to apply to \wfirst\ weak lensing data. The simulated dataset will be available for download via a public shared Globus endpoint following publication of this paper. This endpoint directory includes the following sub-directories:

\begin{itemize}
\item \textsc{images}: A set of FITS images for each SCA in each pointing. 
\item \textsc{meds}: A set of MEDS files that contain cutouts of each object in each exposure, which do not include any neighboring galaxy light. These MEDS files were used for the analysis in this paper. 
\item \textsc{truth}: A FITS catalog of true object properties and a FITS catalog containing information about where each object appeared in each SCA in each pointing. The true centroid of the object in SCA pixel coordinates is offset by 0.5 pixels in $x$ and $y$ relative to the positions recorded in the second FITS catalog, which must be corrected when doing precision operations with the images like shape measurement. This correction is not needed if using the pre-made MEDS files.
\end{itemize}

\section{Performance statistics}
\label{app:performance}

Each stage of the image simulation suite can be trivially parallelized. Disk I/O is not a practical issue in the image generation when running at scale across 4-5k jobs. Within each image generation job (simulation of a complete SCA), the drawing of each object is currently also trivially parallelized across available threads. This will likely change as the simulation of the detectors becomes more realistic, though. In both modes, the image generation stage achieves about 95\% CPU utilization. Stages that process the image output into different data formats are less efficient, and the generation of the MEDS files is generally limited by disk I/O and remote data transfer. Typical timing (per thread), memory usage, and resulting data sizes are provided in Table \ref{table:performance}. The total data volume is 66GB of FITS images, 412GB of MEDS files, and $<$1GB of catalog data, for a total of about 478GB. These values will scale approximately linearly with the area of sky simulated and the density of objects.

\begin{table}
\center{\caption{\label{table:performance} Performance information for the major image simulation suite stages. Each tile is 0.013 deg$^2$ on the sky. Each image contains about 2255 galaxies and 140 stars. For the image simulation created in this paper, with an area of 6.25 deg$^2$, these numbers correspond to a total CPU time cost per simulation realization of about 7,500 CPU hours for image generation and 9,000 CPU hours total. The time required for shape measurement is expected to decrease by at least an order of magnitude, since the current measurement algorithm employed is very slow by current standards, but most of the CPU cost is still in generating the images, which is unlikely to be reduced in the future. 
}
\begin{tabular}{cccc}
\hline
\hline
Benchmark & Image generation & MEDS creation& Shape  \\
  &  (per SCA image) & (per tile ) & measurement \\
\hline
CPU time & 180 min. & 10-15 min. & 5 sec.  \\
Memory  & 2-4GB & 1-2GB & $<$1GB  \\
Data size  & 25MB & 1GB & $<$1MB  \\
\hline
\hline
\end{tabular}
}
\end{table}

\section{PSF model \changetexttwo{approximation}}
\label{app:psf_resol}

\begin{figure}
\begin{center}
\includegraphics[width=\columnwidth]{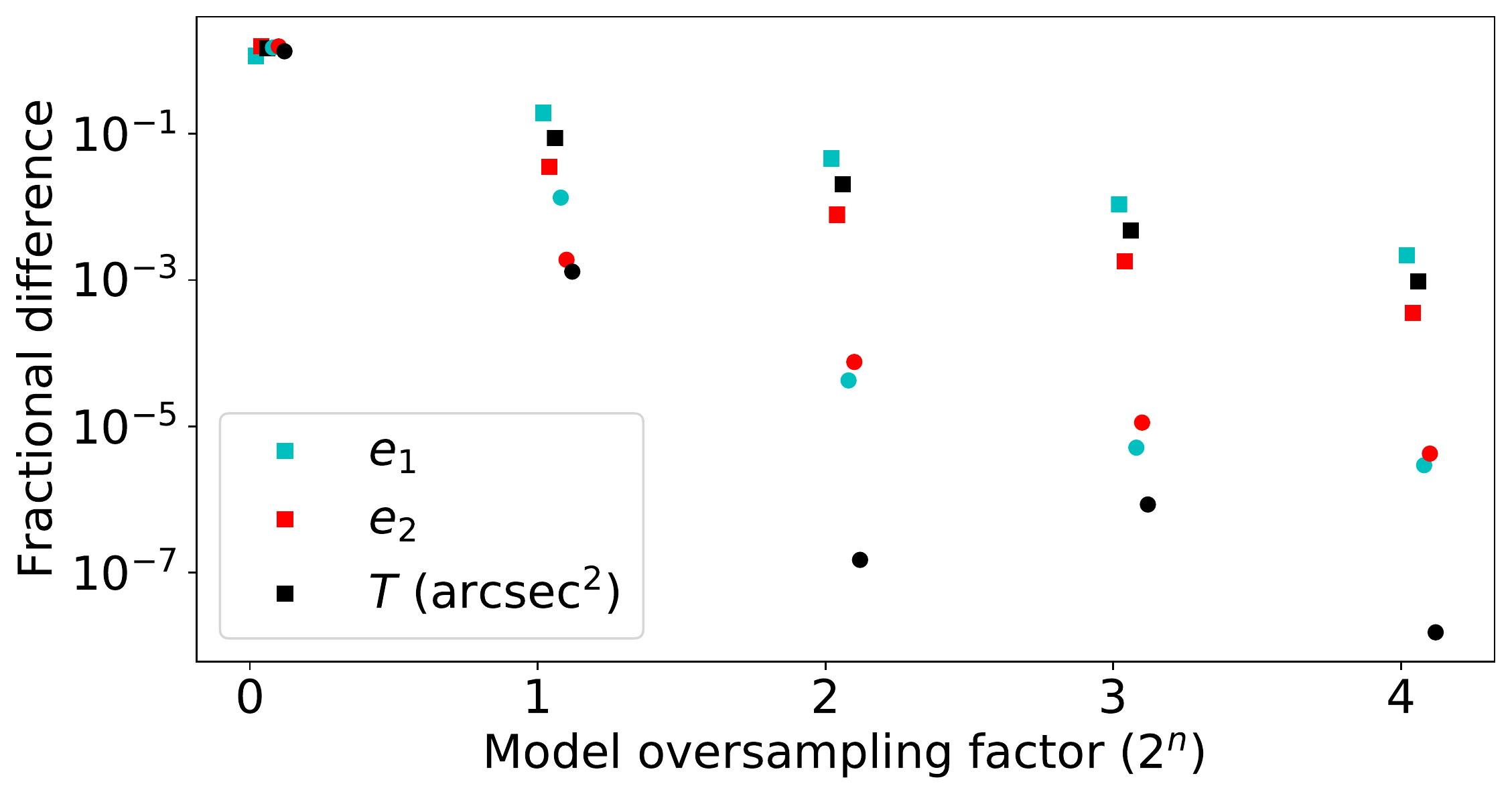}
\end{center}
\caption[]{
The fractional error in the recovered PSF ellipticity ($e_1$ and $e_2$) and size ($T$) for various factors of PSF model pixel scale oversampling factors of 1, 2, 4, 8, and 16 relative to the native \wfirst\ pixel scale, in cutouts of size $8\times 8$ native pixels. The fractional error is measured relative to a `true' PSF model, which is represented by a cutout of native pixel size $64\times64$ pixels with a resolution that is oversampled by a factor of 32. \changetext{Results both with (square) and without (circle) the pixel convolution are included. The points are artificially offset horizontally for clarity.}
\label{fig:psf_oversampling}}
\end{figure}

For this set of image simulations, we have saved the PSF model at the position of galaxies in two resolutions: the native pixel scale and at a pixel scale that is smaller by a factor of 8 to enable unbiased measurements of the PSF model size and ellipticity, which is undersampled in the native pixel resolution. Measurements of PSF properties use this oversampled PSF model image. The motivation for choosing a pixel grid of $8\times8$ pixels and an oversampling factor of 8 was to recover the true model ellipticity ($e_1$ and $e_2$) and size ($T$) to better than 0.1\%. We show in Fig. \ref{fig:psf_oversampling} the fractional difference in the PSF size and shape measured with various oversampling factors relative to a `true', high-resolution PSF image. This choice has no impact on the measurement of galaxy shapes as implemented in this paper. \changetext{Measurements of PSF ellipticity and size are performed using an adaptive moments method (e.g., \cite{2003MNRAS.343..459H}). In all cases we compare results from a fast approximation of the PSF model used in these simulations, which is shown in Fig. \ref{fig:psf}, and not the PSF derived from the real pupil plane image. Future studies do include PSF models inferred from down-sampled versions of the full pupil plane image, which are more accurate.}

\changetexttwo{This fast approximation to the true PSF model uses six radially-oriented struts to create a generic pupil plane image with the struts and a central obscuration. The resolution of this image is set such that the pixelization effects are significantly smaller than the strut width, so should be negligible. Since the 6 struts in the Roman pupil plane are all at slightly different angles, this leads to 12 diffraction spikes, rather than 6, so visually the “approximate struts” PSF (with only 6 spikes) is noticeably different from the correct appearance outside of the PSF core. In addition, a bug in how the PSF model rotates with the observatory roll angle was discovered prior to publication, which will be corrected in future simulations. This acts to enhance any biases observed in this suite of simulations, due to the PSF model not averaging as much as it should across multiple exposures, but otherwise does not change the primary conclusions.}
 
\section{\changetext{Error metrics for setting requirements}}
\label{app:reqpar}

\changetext{In this paper, we have set requirements using the error metric $Z$, which is based on the error in the data vector (length $N_{\rm data}$) relative to its covariance matrix (see \S\ref{ss:req}). This is one of several possible choices. There have also been suggestions to choose an error metric more directly related to the cosmological parameters, since these rather than the data vector are the ultimate science result from the mission. There are several ways to implement this idea. Examples are:
\begin{list}{$\bullet$}{}
\item A similar error metric, $r$, defined in the space of cosmological parameters ${\boldsymbol\theta}$ (length $N_{\rm cosmo}$): $r = \sqrt{\Delta{\boldsymbol\theta} \cdot {\boldsymbol\Sigma}_{\boldsymbol\theta}^{-1} \Delta{\boldsymbol\theta}}$, where ${\boldsymbol\Sigma}_{\boldsymbol\theta}$ is the cosmological parameter covariance matrix, and $\Delta{\boldsymbol\theta}$ is the bias in the cosmological parameters. This metric asks not about the data vector, but whether the bias in the cosmological parameters is within the $1\sigma$, $2\sigma$, etc.\ error ellipsoid. If this is done, one presumably fits not just the cosmological parameters to the data vector, but also a set of nuisance parameters ${\boldsymbol\nu}$ (length $N_{\rm nuis}$).
\item One might use a similar error metric, but only care about some of the cosmological parameters -- e.g., the LSST DESC Science Requirements Document uses the space $(w_0,w_a)$ \citep{2018arXiv180901669T}, and \citet{2013MNRAS.429..661M} considers the bias on the single parameter $w$ (in which case the error metric is ``bias divided by sigma'' or $\sqrt r$). This is mathematically the same problem as using $Z_{\rm cosmo}$, but one treats other parameters such as $\Omega_{\rm m}$ or $\sigma_8$ as nuisance parameters. \citet{2020A&A...635A.139E} presents ``bias over sigma'' for $w_0$ and $w_a$ separately, which can be thought of as two versions of the $r$ metric but each in a different 1-dimensional space.
\item One could instead treat a systematic error in the data vector as an effective increase in its covariance matrix ${\boldsymbol\Sigma}$: ${\boldsymbol\Sigma} \rightarrow {\boldsymbol\Sigma} + \Delta{\bf C}\,\Delta{\bf C}^{\rm T}$, where $\Delta{\bf C}$ is the bias in the data vector. (One thinks of this as a ``systematics contribution to the data covariance.'') Then one can ask by what factor the ellipsoid volume in cosmological parameter space is increased: $f_{\rm deg} = \sqrt{|{\boldsymbol\Sigma}_{{\boldsymbol\theta},\rm new}|/|{\boldsymbol\Sigma}_{{\boldsymbol\theta},\rm old}|}$. Note that if the cosmological parameter space is $(w_0,w_a)$ (and other parameters are taken as nuisance parameters), this is equivalent to the degradation factor in Dark Energy Task Force Figure of Merit \citep{detf}.
\end{list}
The error metrics $Z$ and $r$ satisfy $Z,r\ge 0$, whereas $f_{\rm deg}\ge 1$, with equality holding for no degradation.}

\changetext{Our objective in this appendix is to explore the mathematical properties of these different metrics ($Z$, $r$, and $f_{\rm deg}$), and their robustness under different circumstances. Do these error metrics add linearly, by RSS addition, or in some other way? What happens when we introduce a new nuisance parameter (which may be astrophysical)? What if we shorten the data vector with a scale cut (which is really a special case of adding nuisance parameters)? What if we combine \wfirst\ weak lensing with an external data set?
The $Z$ metric -- while it is the most conservative and leads to the tightest requirements -- has the advantage of obeying straightforward rules under these operations, which make it well-suited to error budgeting when these tighter requirements can be met. The other metrics might still be considered as a fall-back option under some circumstances if the $Z$ metric cannot reasonably be met, or at the requirements verification stage if one is focused on a particular set of analyses \citep[e.g.][]{2020A&A...635A.139E}.}

\changetext{We note that the $f_{\rm deg}$ error metric treats the systematic error in a probabilistic sense, whereas the others as written treat the error as deterministic but unknown. If we take the probabilistic point of view instead for $Z$ or $r$, we should instead consider the RMS error metrics, $Z_{\rm rms} = \langle Z^{2}\rangle^{1/2}$ or $r_{\rm rms} = \langle r^2\rangle^{1/2}$. This approach is particularly useful in error budgeting when adding systematic contributions that are ``independent'' (indeed, only in the probabilistic point of view does considering systematics to be ``independent'' make sense).}

\changetext{The major results from this appendix are shown in Table~\ref{tab:error-metrics}. Some aspects of these error metrics have been considered before. Section 4.1 of \citet{2013MNRAS.429..661M} discusses at length the determinstic vs.\ probabilistic interpretation of $r$. Appendix B2 of \citet{2018arXiv180901669T} has a discussion of the comparison of $Z$ and $r$.}

\changetext{In this appendix, we use the symbol ${\bf A}\succ 0$ to denote that the symmetric matrix ${\bf A}$ is positive definite (or ${\bf A}\succeq 0$ to indicate semipositive definite) and ${\bf A}\succ{\bf B}$ to denote that ${\bf A}-{\bf B}$ is positive definite. We limit our analysis to the Fisher matrix approximation, where the covariance matrix does not depend on the data vector, and the derivatives of the theory data vector with respect to the parameters are constants. We make extensive use of the fact that all of these error metrics are invariant under general invertible linear transformations of the data vector, GL$(N_{\rm data})$, and of the cosmological parameter space, GL$(N_{\rm cosmo})$. Note that these are general linear transformations, not just rotations. Many results here are easiest to understand and prove using particular choices of basis. For example, general linear transformations allow one to use a basis where ${\boldsymbol\Sigma}$ is the $N_{\rm data}\times N_{\rm data}$ identity matrix ${\mathbb I}_{N_{\rm data}}$.}

\begin{table*}
\caption{\label{tab:error-metrics}\changetext{The rules governing the systematic error metrics considered here: $Z$, which is the size of the systematic error relative to the statistical error in the data vector; $r$, which is the size of the systematic error relative to the statistical error in the cosmological parameter vector; and $f_{\rm deg}$, which is the increase in volume in parameter space when the systematic error is included in the covariance matrix. For $Z$ and $r$, one may consider the systematic error in either a deterministic sense or a probabilistic sense (in which case we take the RMS). Table entries include ``No rule'' if there is no rule for addition, ``N/A'' for not applicable, ``Triangle ineq.'' for triangle inequality addition, ``RSS'' for RSS addition, and ``$\le$RSS'' when RSS addition provides an upper bound.}}
\begin{tabular}{l|cc|cc|c}
\hline\hline
Error metric & \multicolumn2c{Data vector} & \multicolumn2c{Cosmological parameter} & Volume ratio \\
& \multicolumn2c{space norms} & \multicolumn2c{space norms} & \\
 & $Z$ & $Z_{\rm rms}$ & $r$ & $r_{\rm rms}$ & $f_{\rm deg}$ \\
\hline
Relations & $Z\le r$ & $Z_{\rm rms}\le r_{\rm rms}$ & & & $f_{\rm deg} \le \left( 1 + \frac{r^2}{N_{\rm cosmo}} \right)^{N_{\rm cosmo}/2}$ \\
\hline
Addition of errors (deterministic) & Triangle ineq. & N/A & Triangle ineq. & N/A & No rule \\
Addition of errors (probabilistic, & N/A & RSS & N/A & RSS & No rule \\
~~independent) & & & & & \\
\hline
Adding nuisance parameter & No effect & No effect & No rule & No rule & No rule \\
\hline
Combining independent data sets (A+B): & & & & & \\
~~~~~-- general & Triangle ineq. & $\le$RSS & No rule & No rule & No rule \\
~~~~~-- if no common nuisance pars. & Triangle ineq. & $\le$RSS & Triangle ineq. & $\le$RSS &
 $f_{\rm deg}^{\rm(A+B)} \le f_{\rm deg}^{\rm(A)} f_{\rm deg}^{\rm(B)}$ \\
 & & & & & (equality only if $f_{\rm deg}^{\rm(A)} = f_{\rm deg}^{\rm(B)}=1$) \\
\hline\hline
\end{tabular}
\end{table*}

\subsection{\changetext{Relations among the error metrics}}

\changetext{Here we show that $Z$ is in some sense the ``most conservative'' of the error metrics, followed by $r$, and then $f_{\rm deg}$. This is essentially because $Z$ counts all systematic errors in the data vector, whereas $r$ counts only those that have a projection onto the cosmological parameters. Finally, $f_{\rm deg}$ treats systematic errors as a contribution to the covariance and hence allows them to be marginalized out.}

\subsubsection{\changetext{Comparison of $Z$ and $r$}}

\changetext{Let's suppose that the cosmological parameters that we fit, ${\boldsymbol\theta}$, have some dependence on the data vector ${\bf C}$: there is a Jacobian, ${\bf R} = \partial {\boldsymbol\theta}/\partial{\bf C}$ (matrix dimension: $N_{\rm cosmo}\times N_{\rm data}$, with $N_{\rm cosmo}\le N_{\rm data}$). Then
\begin{equation}
{\boldsymbol\Sigma}_{\boldsymbol\theta} = {\bf R}{\boldsymbol\Sigma}{\bf R}^{\rm T},
\end{equation}
and so
\begin{equation}
r^2 = \Delta{\bf C}^{\rm T}{\bf R}^{\rm T} ({\bf R}{\boldsymbol\Sigma}{\bf R}^{\rm T})^{-1} {\bf R}\Delta{\bf C}.
\end{equation}
If we go to a basis where ${\boldsymbol\Sigma} = {\mathbb I}_{N_{\rm data}}$, and do a singular value decomposition of ${\bf R} = {\bf UDV}^{\rm T}$, where ${\bf U}$ and ${\bf V}$ are orthogonal and ${\bf D}$ has $N_{\rm cosmo}$ diagonal entries, then we find $r^2 = \sum_{i=1}^{N_{\rm cosmo}} ({\bf V}^{\rm T}\Delta{\bf C})_i^2$, whereas $Z^2 = \sum_{i=1}^{N_{\rm data}} ({\bf V}^{\rm T}\Delta{\bf C})_i^2$. This shows that
\begin{equation}
Z \le r.
\label{eq:ZZC}
\end{equation}
It follows that this relation holds in the probabilistic sense as well, $Z_{\rm rms}\le r_{\rm rms}$.
}

\subsubsection{\changetext{Comparison of $Z_{\rm cosmo}$ and $f_{\rm deg}$}}
\label{ss:czf}

\changetext{
We now consider $f_{\rm deg}$. First, let's consider what happens {\em without} nuisance parameters. If ${\bf J}$ is the $N_{\rm data} \times N_{\rm cosmo}$ matrix of partial derivatives of the theory data vector with respect to the parameters $\partial{\bf C}_{\rm th}/\partial{\boldsymbol\theta}$, with a partition, then standard Fisher matrix formulae tell us that the covariance matrix of the cosmological+nuisance parameters, ${\boldsymbol\Sigma}_{\boldsymbol\theta}$ is given by
\begin{equation}
{\boldsymbol\Sigma}_{\boldsymbol\theta} = ({\bf J}^{\rm T}{\boldsymbol\Sigma}^{-1}{\bf J})^{-1}.
\end{equation}
If ${\boldsymbol\Sigma}$ is increased by the addition of a systematics contribution,  ${\boldsymbol\Sigma} \rightarrow {\boldsymbol\Sigma} +\Delta{\boldsymbol\Sigma}$, then 
\begin{equation}
f_{\rm deg} = \sqrt{ \frac{ |{\bf J}^{\rm T}{\boldsymbol\Sigma}^{-1}{\bf J} |}
{| {\bf J}^{\rm T}({\boldsymbol\Sigma}+\Delta{\boldsymbol\Sigma})^{-1}{\bf J} | } }.
\label{eq:fdeg-.5}
\end{equation}
This equation simplifies if we do a general linear transformation on the data vector to make ${\boldsymbol\Sigma} = {\mathbb I}_{N_{\rm data}}$. This does not uniquely define the choice of basis for the data vector space; we may further do a singular value decomposition of ${\bf J} = {\bf UDV}^{\rm T}$, and then do a rotation in data vector space to set ${\bf V} = {\mathbb I}_{N_{\rm data}}$, a rotation in cosmological parameter space to set ${\bf U} = {\mathbb I}_{N_{\rm data}}$, and a rescaling of the $i$th basis vector in cosmological parameter space by $D_{ii}$ to set ${\bf D}$ to 1's down the diagonal. This makes ${\bf J}$ equal to 1's down the diagonal and 0's elsewhere. In this basis, we have
\begin{equation}
f_{\rm deg} = \left\{ \det [({\mathbb I}_{N_{\rm data}} + \Delta{\boldsymbol\Sigma})^{-1}]_{N_{\rm cosmo}\,\rm block} \right\}^{-1/2},
\end{equation}
where the subscript ``$N_{\rm cosmo}\,\rm block$'' means that we take the upper-left $N_{\rm cosmo}\times N_{\rm cosmo}$ block of the $N_{\rm data}\times N_{\rm data}$ matrix.}

\changetext{Now we want a bound on $f_{\rm deg}$ in terms of $r$. In the basis we have chosen here, the error on the parameters is related to the error in the data vector by
\begin{equation}
\Delta{\boldsymbol\theta} = ({\bf J}^{\rm T}{\boldsymbol\Sigma}^{-1}{\bf J})^{-1} {\bf J}^{\rm T}{\boldsymbol\Sigma}^{-1}\Delta{\bf C},
\end{equation}
so $\Delta\theta_i = \Delta C_i$ for $1\le i\le N_{\rm cosmo}$ and then
\begin{equation}
r^2 = \sum_{i=1}^{N_{\rm cosmo}} \Delta\theta_i^2 = \sum_{i=1}^{N_{\rm cosmo}} \Delta C_i^2
= {\rm Tr} [\Delta{\boldsymbol\Sigma}]_{N_{\rm cosmo}\,\rm block}.
\end{equation}
(Every step in this equality is valid in either a deterministic or probabilistic sense; in the latter case, the left-hand side becomes $r_{\rm rms}^2$.)
It follows trivially from the block inversion formula that for a positive definite matrix, the determinant of the block of an inverse is greater than or equal to the determinant of the inverse of the block. Therefore, if $\{\lambda_i\}_{i=1}^{N_{\rm cosmo}}$ are the eigenvalues of $[\Delta{\boldsymbol\Sigma}]_{N_{\rm cosmo}\,\rm block}$, then
\begin{equation}
f_{\rm deg} \le \sqrt{ \prod_{i=1}^{N_{\rm cosmo}} (1+\lambda_i)}
~~~{\rm and} ~~
r^2 = \sum_{i=1}^{N_{\rm cosmo}} \lambda_i.
\end{equation}
We may write the formula for $f_{\rm deg}$ in the form
\begin{equation}
2\ln f_{\rm deg} = \sum_{i=1}^{N_{\rm cosmo}} \ln (1+\lambda_i),
\end{equation}
where the last term has a strictly negative second derivative. This means that for a fixed $\sum_{i=1}^{N_{\rm cosmo}} \lambda_i = r^2$, the maximum value of $2\ln f_{\rm deg}$ is obtained when all of the $\lambda_i$ are the same: $\lambda_i = r^2/{N_{\rm cosmo}}$. Therefore, we have
\begin{equation}
f_{\rm deg} \le \left( 1 + \frac{r^2}{N_{\rm cosmo}} \right)^{N_{\rm cosmo}/2}.
\label{eq:fdeg-ineq}
\end{equation}
}

\changetext{This relation still applies if we have nuisance parameters, since inclusion of a nuisance parameter can be thought of as a modification to ${\boldsymbol\Sigma}$: ${\boldsymbol\Sigma} \rightarrow {\boldsymbol\Sigma} + \sigma_p^2 (\partial{\bf C}_{\rm th}/\partial p)(\partial{\bf C}_{\rm th}/\partial p)^{\rm T}$.}

\subsection{\changetext{Addition of error terms}}
\label{ss:addition-error}

\changetext{A common problem in error budgeting is addition of error terms -- when there are two or more sources of error that must be included in the budget, how should they be added? And what can we say when those two sources of error are independent?}

\changetext{This problem is simplest for $Z$ and $r$, because they are the ordinary vector lengths of $\Delta{\bf C}$ in ${\mathbb R}^{N_{\rm data}}$ and $\Delta{\boldsymbol\theta}$ in ${\mathbb R}^{N_{\rm cosmo}}$ respectively, if we use the bases where ${\boldsymbol\Sigma}$ and ${\boldsymbol\Sigma}_{\boldsymbol\theta}$ are the identity. Therefore they satisfy the ``usual'' rules of error addition:
\begin{list}{$\bullet$}{}
\item If we take the deterministic point of view, then $Z$ and $r$ obey the triangle inequality: if there is one source of error (``A'') that produces a bias in the data vector $\Delta{\bf C}^{(\rm A)}$ and another that produces a bias $\Delta{\bf C}^{(\rm B)}$, then always the error metrics satisfy $Z^{(\rm A+B)} \le Z^{(\rm A)} + Z^{(\rm B)}$.
\item If we take the probabilistic point of view, then $Z_{\rm rms}$ and $r_{\rm rms}$ obey root-sum-square (RSS) addition with two {\em independent} contributions A and B: $Z^{(\rm A+B)\,2}_{\rm rms} = Z^{(\rm A)\,2}_{\rm rms} + Z^{(\rm B)\,2}_{\rm rms}$.
\end{list}}

\changetext{In contrast, $f_{\rm deg}$ does not obey any simple error addition rule. A simple counterexample to any proposed inequality can be found with $N_{\rm cosmo}=1$, $N_{\rm data}=2$, and matrices in the language of \S\ref{ss:czf}:
\begin{equation}
{\bf J} = \left( \begin{array}c 1 \\ 0 \end{array} \right) ~~{\rm and}~~
{\boldsymbol\Sigma} = \left( \begin{array}{cc} 1 & 0 \\ 0 & 1 \end{array} \right),
\end{equation}
with systematic error contributions A and B:
\begin{equation}
\Delta{\bf C}^{(\rm A)} =  \left( \begin{array}c \alpha \\ \beta \end{array} \right) ~~{\rm and}~~
\Delta{\bf C}^{(\rm B)} =  \left( \begin{array}c \alpha \\ -\beta \end{array} \right).
\end{equation}
We take these contributions to be additions to the data covariance matrix: ${\boldsymbol\Sigma}$ gets an extra contribution of $\Delta{\bf C}^{(\rm A)}\Delta{\bf C}^{(\rm A)\,T}$, $\Delta{\bf C}^{(\rm B)}\Delta{\bf C}^{(\rm B)\,T}$, or $\Delta{\bf C}^{(\rm A)}\Delta{\bf C}^{(\rm A)\,T} + \Delta{\bf C}^{(\rm B)}\Delta{\bf C}^{(\rm B)\,T}$. Treating the combination of A and B by adding their contributions to the covariance matrix is a direct consequence of supposing them to be independent. However, in this case one can evaluate Eq.~(\ref{eq:fdeg-.5}) and find the degradation factors:
\begin{equation}
f_{\rm deg}^{(\rm A)} = f_{\rm deg}^{(\rm B)} = \sqrt{\frac{ 1 + \alpha^2 + \beta^2 }{ 1 + \beta^2 }},
\end{equation}
whereas
\begin{equation}
f_{\rm deg}^{(\rm A+B)} = \sqrt{1 + 2\alpha^2}.
\end{equation}
It is clear that in this case, by making $\alpha$ large enough, we can make $f_{\rm deg}^{(\rm A+B)}$ as large as we want; but then at fixed $\alpha$, by making $\beta$ large enough, we can make $f_{\rm deg}^{(\rm A)}$ and $f_{\rm deg}^{(\rm B)}$ as close to 1 as we want. In other words, using the $f_{\rm deg}$ metric, two systematic error contributions A and B, which are individually arbitrarily small can combine to make an arbitrarily large contribution A+B.}

\changetext{This means that if one builds an error budget using $f_{\rm deg}$ as a metric, one must repeat the full Fisher matrix analysis each time a new term is added and cannot rely on each contribution to the error budget staying within an ``allocation'' (e.g., stating $f_{\rm deg} < 1.02$ or $<2\%$ loss of figure of merit for a particular systematic). Keeping this level of coordination across different contributions to error budgets is a challenge in a large interdisciplinary team, but it can be considered particularly in cases where error budgets based on $r$ or $Z$ become cost, schedule, or engineering risk drivers.}

\subsection{\changetext{Introduction of new nuisance parameters}}
\label{ss:app-nuisance}

\changetext{We now consider what happens when new nuisance parameters are added. This is simplest in the case of $Z$: since it is built entirely in data vector space, $Z$ does not depend on how the data vector is used and is unaffected when nuisance parameters are added.}

\changetext{It turns out that no simple rule occurs for $r$. To see this, let's consider the simple case of 1 cosmological parameter $\theta$ and 1 nuisance parameter $\nu$. We also approximate the theory data vector as a linear function of the parameters over the range of interest, i.e., $\partial{\bf C}_{\rm th}/\partial({\boldsymbol\theta},{\boldsymbol\nu})$ constant. The $\chi^2$ surface for these parameters will generally be of the form
\begin{equation}
\chi^2 = \left( \begin{array}{cc} \theta & \nu \end{array} \right) {\bf A}  \left( \begin{array}c \theta \\ \nu \end{array} \right)
- 2 (b_\theta \theta + b_\nu\nu) + c,
\end{equation}
where ${\bf A}$ is the $2\times 2$ inverse covariance matrix, and ${\bf b}$ is a vector. A change in the data vector will lead to a change in $\Delta{\boldsymbol b}$, the slope (it does not affect ${\bf A}$ if the derivatives are constant, and we do not care about $\Delta c$). The bias in the parameters if we include the nuisance parameter is
\begin{equation}
\left( \begin{array}c \theta^{(+)} \\ \nu^{(+)} \end{array} \right) = {\bf A}^{-1}\Delta{\boldsymbol b},
\end{equation}
so
\begin{equation}
\Delta\theta^{(+)} = \frac{A_{\nu\nu}\,\Delta b_\theta + A_{\theta\nu}\,\Delta b_\nu}{A_{\theta\theta} A_{\nu\nu} - A_{\theta\nu}^2}.
\end{equation}
If the nuisance parameter is not added, then we look at the $\chi^2$ surface with $\nu$ fixed to 0, in which case the bias in $\theta$ is
\begin{equation}
\Delta\theta^{(-)} = \frac{\Delta b_\theta}{A_{\theta\theta}}.
\end{equation}
The corresponding values of $r^2 = \Delta\theta^2 / {\rm Var}(\theta)$ are:
\begin{equation}
r^{(+)}{^2} = \frac{(A_{\nu\nu}\,\Delta b_\theta + A_{\theta\nu}\,\Delta b_\nu)^2}{A_{\nu\nu}(A_{\theta\theta} A_{\nu\nu} - A_{\theta\nu}^2)}
\end{equation}
and
\begin{equation}
r^{(-)}{^2} = \frac{(\Delta b_\theta)^2}{A_{\theta\theta}}.
\end{equation}
It is now apparent that there is no inequality relating $r^{(+)}$ to $r^{(-)}$: as long as $A_{\theta\nu}\neq 0$, any value of the ordered pair $(\Delta b_\theta, A_{\nu\nu}\,\Delta b_\theta + A_{\theta\nu}\,\Delta b_\nu)\in{\mathbb R}^2$ is possible, and therefore knowledge of $r^{(-)}$ by itself provides no constraint on $r^{(+)}$ and vice versa. So the $r$ error metric may go up, go down, or stay the same when nuisance parameters are added.
}

\changetext{Finally, we show that there is also no inequality relating $f_{\rm deg}^{(+)}$ (with a nuisance parameter) to $f_{\rm deg}^{(-)}$ (without it). Let us consider the same example above with 1 cosmological parameter and a single nuisance parameter. The degradation factor with and without the extra nuisance parameter is
\begin{equation}
f_{\rm deg}^{(+)} = \sqrt{ \frac{
(A_{\theta\theta}^{\rm(old)} A_{\nu\nu}^{\rm(old)} - [A_{\theta\nu}^{\rm(old)}]^2)/A_{\nu\nu}^{\rm(old)}
}{
(A_{\theta\theta}^{\rm(new)} A_{\nu\nu}^{\rm(new)} - [A_{\theta\nu}^{\rm(new)}]^2)/A_{\nu\nu}^{\rm(new)}
}}
\end{equation}
and
\begin{equation}
f_{\rm deg}^{(-)} = \sqrt{ \frac{A_{\theta\theta}^{\rm(old)}}{A_{\theta\theta}^{\rm(new)}} },
\end{equation}
where ${\bf A}^{\rm(old)}$ or ${\bf A}^{\rm(new)}$ is the old (without degradation) or new (with degradation) inverse covariance matrix. These matrices are given by
\begin{equation}
{\bf A}^{\rm(old)} = {\bf J}{\boldsymbol\Sigma}^{-1}{\bf J}^{\rm T}
~~{\rm and}~~
{\bf A}^{\rm(new)} = {\bf J}({\boldsymbol\Sigma} + \Delta{\bf C}\,\Delta{\bf C}^{\rm T})^{-1}{\bf J}^{\rm T},
\end{equation}
where ${\bf J}$ is the $N_{\rm data}\times 2$ matrix of partial derivatives of the data vector. For the sake of constructing an example that shows there is no inequality relating $f_{\rm deg}^{(+)}$ to $f_{\rm deg}^{(-)}$, we take $N_{\rm data}=2$ and ${\bf J}$ to be the identity. If we take ${\boldsymbol\Sigma}$ to be the $2\times 2$ identity and $\Delta{\bf C} = (\alpha~~\beta)^{\rm T}$, then
\begin{equation}
f_{\rm deg}^{(+)} = \sqrt{1+\alpha^2} ~~{\rm and}~~ f_{\rm deg}^{(-)} = \sqrt{\frac{1+\alpha^2+\beta^2}{1+\beta^2}}.
\end{equation}
We can see that any configuration with $1 \le f_{\rm deg}^{(-)} \le f_{\rm deg}^{(+)}$ can be obtained by choosing $\alpha$ (to get the desired $f_{\rm deg}^{(+)}$) and then $\beta$ (to get the desired $f_{\rm deg}^{(-)}$). However, if alternatively we taken
\begin{equation}
{\boldsymbol\Sigma} = \left( \begin{array}{cc} 1 & \rho \\ \rho & 1 \end{array} \right)
~~~{\rm and}~~~
\Delta{\bf C} = \left( \begin{array}c \gamma \\ 0 \end{array} \right)
\end{equation}
with $|\rho|<1$, then
\begin{equation}
f_{\rm deg}^{(+)} = \sqrt{1+\gamma^2} ~~{\rm and}~~ f_{\rm deg}^{(-)} = \sqrt{\frac{1-\rho^2+\gamma^2}{1-\rho^2}}.
\end{equation}
This time, we can choose any configuration with $1\le f_{\rm deg}^{(+)}\le f_{\rm deg}^{(-)}$ by choosing $\gamma$ (to get the desired $f_{\rm deg}^{(+)}$) and then $\rho$ (to get the desired $f_{\rm deg}^{(-)}$). Between these two examples, we see that there is no relation between $f_{\rm deg}^{(+)}$ and $f_{\rm deg}^{(-)}$: any behaviour is possible when we add a nuisance parameter.}

\subsection{\changetext{Combination with external data sets}}

\changetext{Finally, we consider combinations with an external data set. In the context of \wfirst\ weak lensing, this might be another experiment (e.g., cosmic microwave background) or within \wfirst\ (e.g., the supernova survey). We assume that the data vector for this external data set is independent.}

\changetext{The $Z$ metric undergoes RSS addition between data sets, i.e., if we have data sets A and B, then $Z^2({\rm A+B)} = Z^2({\rm A}) + Z^2({\rm B})$, because the covariance matrix ${\boldsymbol\Sigma}$ of the combined data set A+B is block diagonal. This is true both for the deterministic version of the metric and the probabilistic version, $Z_{\rm rms}$.}

\changetext{What happens to the $r$ and $f_{\rm deg}$ metrics is more complicated. If there are shared nuisance parameters between the experiments, then in general there is no rule on how $Z_{\rm cosmo}$ and $f_{\rm deg}$ change when an external data set is added, because one could consider the limiting case where the external data set effectively fixes a nuisance parameter, and then one has the situation described in \S\ref{ss:app-nuisance}. We note that if $f_{\rm deg}$ is defined in terms of $(w_0,w_a)$ space, and other cosmological parameters such as $\Omega_{\rm m}$ and $H_0$ are treated as nuisance parameters, then almost all practical cases of combined data sets will have shared nuisance parameters.}

\changetext{If there are {\em no} shared nuisance parameters, then it is possible to say more about $r$ and $f_{\rm deg}$. Let's consider $r$ first. When two data sets A and B are combined, and we have Gaussian likelihoods, the bias in parameters in the combined data set is
\begin{equation}
\Delta{\boldsymbol\theta}^{(\rm A+B)} =
[{\boldsymbol\Sigma}_{\boldsymbol\theta}^{\rm(A)}{^{-1}}
+{\boldsymbol\Sigma}_{\boldsymbol\theta}^{\rm(B)}{^{-1}}]^{-1}
[{\boldsymbol\Sigma}_{\boldsymbol\theta}^{\rm(A)} \Delta{\boldsymbol\theta}^{(\rm A)} +
{\boldsymbol\Sigma}_{\boldsymbol\theta}^{\rm(B)} \Delta{\boldsymbol\theta}^{(\rm B)}]
\end{equation}
and the covariance matrix is $[{\boldsymbol\Sigma}_{\boldsymbol\theta}^{\rm(A)}{^{-1}} + {\boldsymbol\Sigma}_{\boldsymbol\theta}^{\rm(B)}{^{-1}}]^{-1}$. Then
\begin{eqnarray}
r^{\rm(A+B)}{^2}\!\!\!\! &=& \!\!\!\!
\Delta{\boldsymbol\theta}^{(\rm A)}{^{\rm T}} {\boldsymbol\Sigma}_{\boldsymbol\theta}^{\rm(A)} 
 [{\boldsymbol\Sigma}_{\boldsymbol\theta}^{\rm(A)}{^{-1}}
+{\boldsymbol\Sigma}_{\boldsymbol\theta}^{\rm(B)}{^{-1}}]^{-1}
{\boldsymbol\Sigma}_{\boldsymbol\theta}^{\rm(A)} \Delta{\boldsymbol\theta}^{(\rm A)}
\nonumber \\
&& \!\!\!\!
+
\Delta{\boldsymbol\theta}^{(\rm B)}{^{\rm T}} {\boldsymbol\Sigma}_{\boldsymbol\theta}^{\rm(B)} 
 [{\boldsymbol\Sigma}_{\boldsymbol\theta}^{\rm(A)}{^{-1}}
+{\boldsymbol\Sigma}_{\boldsymbol\theta}^{\rm(B)}{^{-1}}]^{-1}
{\boldsymbol\Sigma}_{\boldsymbol\theta}^{\rm(B)} \Delta{\boldsymbol\theta}^{(\rm B)}
\nonumber \\
&& \!\!\!\!
+ 2
\Delta{\boldsymbol\theta}^{(\rm A)}{^{\rm T}} {\boldsymbol\Sigma}_{\boldsymbol\theta}^{\rm(A)} 
 [{\boldsymbol\Sigma}_{\boldsymbol\theta}^{\rm(A)}{^{-1}}
+{\boldsymbol\Sigma}_{\boldsymbol\theta}^{\rm(B)}{^{-1}}]^{-1}
{\boldsymbol\Sigma}_{\boldsymbol\theta}^{\rm(B)} \Delta{\boldsymbol\theta}^{(\rm B)}.
\nonumber \\ &&
\end{eqnarray}
Since ${\boldsymbol\Sigma}_{\boldsymbol\theta}^{\rm(A)} \succeq [{\boldsymbol\Sigma}_{\boldsymbol\theta}^{\rm(A)}{^{-1}}
+{\boldsymbol\Sigma}_{\boldsymbol\theta}^{\rm(B)}{^{-1}}]^{-1}$, we know that the first term is $\le \Delta{\boldsymbol\theta}^{(\rm A)}{^{\rm T}} {\boldsymbol\Sigma}_{\boldsymbol\theta}^{\rm(A)}{^{-1}}\Delta{\boldsymbol\theta}^{(\rm A)} = r^{\rm(A)}{^2}$.
Similarly, the second term is $\le r^{\rm(B)}{^2}$. The Cauchy-Schwarz inequality then shows that the third term is less than or equal to twice the geometric mean of the first two, i.e., $\le 2r^{\rm(A)} r^{\rm(B)}$. So ti follows that $r$ obeys linear addition:
\begin{equation}
r^{\rm(A+B)} \le r^{\rm(A)} + r^{\rm(B)}.
\end{equation}
If one treats the systematic errors in the two data sets as independent in the probabilistic sense, then when taking an average the third term drops out and $r_{\rm rms}$ obeys RSS addition (with an inequality):
\begin{equation}
r_{\rm rms}^{\rm(A+B)}{^2} \le r_{\rm rms}^{\rm(A)}{^2} + r_{\rm rms}^{\rm(B)}{^2}.
\end{equation}
}

\changetext{We may also consider $f_{\rm deg}$; for two independent probes A and B and no common nuisance parameters, we have
\begin{equation}
f_{\rm deg}^{\rm(A+B)} = \sqrt{ \frac{\det ({\bf F}^{\rm(A,old)} + {\bf F}^{\rm(B,old)})}{\det ({\bf F}^{\rm(A,new)} + {\bf F}^{\rm(B,new)})} },
\end{equation}
where ``old'' and ``new'' indicate the Fisher matrices without and with the systematic, respectively (and after the nuisance parameters have been marginalized out). We can understand what happens if we turn this into an integral:
\begin{equation}
\ln f_{\rm deg}^{\rm(A+B)} = \frac12 \int_0^1 \frac{d}{d x} \ln \det[{\bf F}^{\rm(A)}(x) + {\bf F}^{\rm(B)}(x)]\,d x,
\label{eq:2lnf}
\end{equation}
with $x=0$ corresponding to the ``new'' Fisher matrix, $x=1$ corresponding to the old:
\begin{equation}
{\bf F}^{\rm(A)}(x) = x {\bf F}^{\rm(A,old)} + (1-x) {\bf F}^{\rm(A,new)}.
\end{equation}
Since ${\bf F}^{\rm(A,new)}\preceq {\bf F}^{\rm(A,old)}$, we have $d{\bf F}^{\rm(A)}(x)/dx \succeq 0$. The integral in Eq.~(\ref{eq:2lnf}) then represents the degradation of parameter constraint volume as we introduce the systematic. Then the derivative of a log determinant satisfies
\begin{eqnarray}
\ln f_{\rm deg}^{\rm(A+B)} &=& \frac12 \int_0^1 {\rm Tr} \Bigl\{ [{\bf F}^{\rm(A)}(x) + {\bf F}^{\rm(B)}(x)]^{-1}
\nonumber \\ && \times
\left[ \frac{d{\bf F}^{\rm(A)}(x)}{dx} + \frac{d{\bf F}^{\rm(B)}(x)}{dx} \right]
\Bigr\}
\,d x.
\end{eqnarray}
Now for any ${\bf A}\succeq0$ and ${\bf B}\succ {\bf D} \succ 0$, we may write ${\bf A} = \sum_{i=1}^{N_{\rm cosmo}} {\bf uu}^{\rm T}$ and then ${\rm Tr}({\bf BA}) = \sum_{i=1}^{N_{\rm cosmo}} {\bf u}^{\rm T}{\bf Bu}$. It follows that ${\rm Tr}({\bf BA})\ge {\rm Tr}({\bf DA})$ (with equality only for ${\bf A}=0$). Applying this to $[{\bf F}^{\rm(A)}(x) + {\bf F}^{\rm(B)}(x)]^{-1} \prec [{\bf F}^{\rm(A)}(x)]^{-1}$ and similarly for B, we show that
\begin{eqnarray}
\ln f_{\rm deg}^{\rm(A+B)} \!\!\!\! &\le&\!\!\!\! \frac12 \int_0^1 {\rm Tr} \Bigl\{ [{\bf F}^{\rm(A)}(x)]^{-1}\frac{d{\bf F}^{\rm(A)}(x)}{dx} \Bigr\} \,d x
\nonumber \\ &&\!\!\!\!
+ \frac12 \int_0^1 {\rm Tr} \Bigl\{ [{\bf F}^{\rm(B)}(x)]^{-1}\frac{d{\bf F}^{\rm(B)}(x)}{dx} \Bigr\} \,d x~~~~
\nonumber \\
&=& \ln f_{\rm deg}^{\rm(A)} + \ln f_{\rm deg}^{\rm(B)}.
\end{eqnarray}
(Equality holds only when $d{\bf F}^{\rm(A)}/dx = d{\bf F}^{\rm(B)}=0$.) We see that for no shared nuisance parameters,
\begin{equation}
f_{\rm deg}^{\rm(A+B)} \le f_{\rm deg}^{\rm(A)} f_{\rm deg}^{\rm(B)},
\end{equation}
with equality only in the case where the systematic has no effect on the Fisher matrices, i.e., when $f_{\rm deg}^{\rm(A+B)} = f_{\rm deg}^{\rm(A)} = f_{\rm deg}^{\rm(B)} = 1$.
}

\changetext{A particular example showing that this inequality cannot be strengthened is in the space of 2 cosmological parameters:
\begin{equation}
{\bf F}^{\rm(A,old)} = \left( \begin{array}{cc} 1 & 0 \\ 0 & \epsilon \end{array} \right),
~~
{\bf F}^{\rm(B,old)} = \left( \begin{array}{cc} \epsilon & 0 \\ 0 & 1 \end{array} \right),
\nonumber
\end{equation}
\begin{equation}
{\bf F}^{\rm(A,new)} = \left( \begin{array}{cc} \alpha^{-2} & 0 \\ 0 & \epsilon \end{array} \right),
~~{\rm and}~~
{\bf F}^{\rm(B,new)} = \left( \begin{array}{cc} \epsilon & 0 \\ 0 & \beta^{-2} \end{array} \right),
\end{equation}
with $\alpha,\beta\ge 1$. By taking the limit of $\epsilon\rightarrow 0$, we see that $f_{\rm deg}^{\rm(A)}\rightarrow\alpha$, $f_{\rm deg}^{\rm(B)}\rightarrow\beta$, and $f_{\rm deg}^{\rm(A+B)}\rightarrow \alpha\beta$.
}

\bsp	
\label{lastpage}
\end{document}